\documentclass[aps,pre,twocolumn,10pt,superscriptaddress,nofootinbib,floatfix,longbibliography]{revtex4-2}

\usepackage[T1]{fontenc}
\usepackage[utf8]{inputenc}
\usepackage{amsmath,amssymb,amsfonts,bm,mathtools}
\usepackage{graphicx}
\usepackage{xcolor}
\usepackage{booktabs}
\usepackage{array}
\usepackage{hyperref}
\usepackage{enumitem}
\usepackage{tikz}
\usetikzlibrary{arrows.meta,positioning,calc,shapes.geometric}

\hypersetup{colorlinks=true,linkcolor=blue!45!black,citecolor=blue!45!black,urlcolor=blue!45!black}

\newcommand{\dd}{\mathrm d}
\newcommand{\ii}{\mathrm i}
\newcommand{\Order}{\mathcal O}
\newcommand{\avg}[1]{\langle #1\rangle}
\newcommand{\grad}{\bm{\nabla}}
\newcommand{\lap}{\nabla^2}
\newcommand{\phit}{\widetilde\phi}
\newcommand{\Dphi}{\Delta_\phi}
\newcommand{\eps}{\varepsilon}
\newcommand{\Tr}{\operatorname{Tr}}
\newcommand{\Span}{\operatorname{span}}
\newcommand{\EOM}{\mathcal E}
\newcommand{\Id}{\mathsf I}
\newcommand{\vphi}{\varphi}
\newcommand{\aeq}{a_{\rm eq}}
\newcommand{\ac}{a_{\rm ch}}
\newcommand{\aJ}{a_J}
\newcommand{\yth}{y_{\Theta_1}}
\newcommand{\yJ}{y_J}
\newcommand{\ych}{y_{\rm ch}}
\newcommand{\kinS}{\mathtt{s}}
\newcommand{\kinT}{\mathtt{t}}
\newcommand{\kinV}{\mathtt{v}}
\newcommand{\LPAA}{LPA$'$+A}

\tikzset{
  diagline/.style={line width=0.55pt},
  response/.style={diagline,-{Latex[length=1.7mm,width=1.2mm]}},
  correlation/.style={diagline,double,double distance=1.25pt},
  passivevertex/.style={circle,draw=black,fill=black,inner sep=1.45pt},
  noisevertex/.style={rectangle,draw=red!80!black,fill=red!10,
    minimum size=4.1pt,inner sep=1pt},
  lambdavertex/.style={diamond,draw=blue!75!black,fill=blue!8,
    minimum size=5.2pt,inner sep=1pt},
  xivertex/.style={regular polygon,regular polygon sides=3,
    draw=cyan!60!black,fill=cyan!10,minimum size=5.8pt,inner sep=0.5pt},
  currentvertex/.style={regular polygon,regular polygon sides=6,
    draw=violet!80!black,fill=violet!10,minimum size=6.0pt,inner sep=0.5pt},
  activevertex/.style={regular polygon,regular polygon sides=5,
    draw=black,fill=white,minimum size=10.0pt,inner sep=0.9pt},
  regulatorinsertion/.style={circle,draw=black,fill=white,inner sep=1.2pt}
}

\newcommand{\passiveicon}{\tikz[baseline=-0.55ex]{\node[passivevertex] {};}}
\newcommand{\noiseicon}{\tikz[baseline=-0.55ex]{\node[noisevertex] {};}}
\newcommand{\lambdaicon}{\tikz[baseline=-0.55ex]{\node[lambdavertex] {};}}
\newcommand{\xiicon}{\tikz[baseline=-0.55ex]{\node[xivertex] {};}}
\newcommand{\currenticon}{\tikz[baseline=-0.55ex]{\node[currentvertex] {};}}

\begin{document}

\title{Nonequilibrium corrections to conserved Ising criticality in scalar
active matter: Ward identities, spectrum, and long crossovers}

\author{Piotr Zdybel}
\email{pzdybel@ippt.pan.pl}
\thanks{ORCID: \href{https://orcid.org/0000-0001-7484-1425}{0000-0001-7484-1425}}
\affiliation{Department of Biosystems and Soft Matter,
Institute of Fundamental Technological Research,
Polish Academy of Sciences,
Pawi\'nskiego 5B, 02-106 Warsaw, Poland}

\date{\today}

\begin{abstract}
We identify the slowest-decaying nonequilibrium perturbations near the conserved Ising critical point in three dimensions and determine how they affect finite-size observables.  We consider two distinct classes of nonequilibrium perturbation: a field-dependent noise-to-mobility ratio $\Theta(\phi)=D(\phi)/M(\phi)$, and the gradient activity of Active Model~B+.  Starting from the Martin--Siggia--Rose--Janssen--De~Dominicis (MSRJD) action, we compute the linearized flow with the functional renormalization group (FRG). The transport flow is block triangular, and its leading odd eigenvalue is $y_{\Theta_1}=-\Delta_\phi$. Here $d$ is the spatial dimension, $\eta$ the anomalous dimension, and $\Delta_\phi=(d-2+\eta)/2$ the field scaling dimension.  The gradient sector contains one direction that preserves detailed balance. Projecting it out leaves two genuinely nonequilibrium modes, one chemical and one current-like. Two smooth regulators give $y_{\Theta_1}\simeq-0.52$, $y_J\simeq-0.56$, and
$y_{\rm ch}\simeq-0.89$.  A translation Ward identity represents the current operator by the divergence of the stress tensor. Together with conservation and It\^o causality, this forbids chemical operators from generating the current mode at any loop order, so the nonequilibrium stability matrix is triangular.  An independent two-loop test in $d=4-\varepsilon$ finds no additional contact counterterm for the current operator.  Within the FRG truncation the two slow modes are separated only by the anomalous dimension, $y_J-y_{\Theta_1}=-\eta$. Beyond these calculations the relation holds only if neither sector develops a further contact anomaly.  Under this condition, using the three-dimensional Ising reference value
$\eta_{\rm ref}=0.0362978(20)$ [Kos~\textit{et al.}, 2016] gives
$y_{\Theta_1,\rm ref}\simeq-0.5181$ and
$y_{J,\rm ref}\simeq-0.5544$.  Because the two exponents nearly coincide, a single-power fit returns an amplitude-dependent apparent exponent, and the crossover length can exceed any accessible system size.  We derive the corresponding finite-size scaling rules. Odd block observables respond linearly to activity, whereas even observables receive only quadratic active corrections.
\end{abstract}

\maketitle

\section{Introduction}
\label{sec:intro}

Active matter is intrinsically out of equilibrium because microscopic
energy consumption sustains motion and currents that are not constrained by
detailed balance~\cite{Marchetti2013,CatesTailleur2015,CatesNardini2025}.
A paradigmatic
example is motility-induced phase separation (MIPS), where repulsive active
particles separate into dense and dilute phases as their motility decreases
with local density~\cite{TailleurCates2008,CatesTailleur2013,Stenhammar2013}.
At long wavelengths the order parameter is naturally a conserved scalar
density, so the passive reference theory is the conserved Cahn--Hilliard
or Model-B dynamics~\cite{CahnHilliard1958,HohenbergHalperin1977,
HalperinHohenbergMa1976,Tauber2014}.

The status of Ising criticality in active scalar systems is subtle.
Several numerical studies find behavior compatible with the Ising
universality class, while other studies emphasize visible deviations,
strong corrections to scaling, or long crossover regimes
\cite{Siebert2018,PartridgeLee2019,Maggi2021,Dittrich2021,Speck2022}.
We therefore ask a different question from whether a new infrared fixed point exists: \emph{which nonequilibrium perturbations of
the passive conserved Wilson--Fisher fixed point decay most slowly, and how
do they appear in finite systems?}

Two classes of perturbations are particularly natural.  First, the local
mobility $M(\phi)$ and conserved-noise amplitude $D(\phi)$ may depend on the
field.  Their equilibrium relation is most economically tested by the local
noise-to-mobility ratio
\begin{equation*}
 \Theta(\phi)\equiv\frac{D(\phi)}{M(\phi)} ,
\end{equation*}
which equals the temperature for constant equilibrium transport.  A
nonconstant $\Theta(\phi)$ is therefore a local
fluctuation--dissipation violation~\cite{Cugliandolo2011,Aron2016}.  A
field-dependent mobility is known to leave its own imprint on the ordering
kinetics of scalar active matter~\cite{DeLuca2024}.
Second, scalar active field theories
contain gradient nonlinearities.  Active Model B introduces a
nonequilibrium chemical-potential term, while Active Model B+ adds a
non-gradient current and captures phenomena such as reversed Ostwald
ripening and microphase separation
\cite{Wittkowski2014,Nardini2017,Tjhung2018,CaballeroCates2018}.
Recent perturbative and functional renormalization-group (FRG) studies have
analyzed the fixed-point
structure generated by these gradient couplings
\cite{CaballeroCates2020,PapanikolaouSpeck2024,
FejosSzepYamamoto2026}; the FRG has also exposed genuinely
nonequilibrium universality classes in more general active fluids
\cite{JentschLee2023}, and a related dynamic renormalization-group analysis
identifies new universality classes in a critical active Ising
model~\cite{WongLee2025}.

Those recent studies ask how the full nonlinear active theory flows and
whether additional fixed points exist.  Our target is different and more local in theory space. We study the \emph{linearized nonequilibrium spectrum around the passive conserved Wilson--Fisher fixed point}. We retain the complete Active-Model-B+ gradient basis, evaluate all sectors within one field-theoretic framework, and use Ward identities to separate a kinematic cancellation from the genuinely dynamic content of the calculation.  This yields correction-to-scaling exponents, operator mixing, and finite-size observables that the global-flow analyses of Refs.~\cite{PapanikolaouSpeck2024,FejosSzepYamamoto2026} do not fix.  The calculation requires three ingredients.  First, the stochastic model
must be formulated carefully because multiplicative conserved noise and
gradient activity generate different
Martin--Siggia--Rose--Janssen--De~Dominicis (MSRJD) vertices.  Second, the local
transport sector and the gradient sector must be projected at different
orders in the external momentum.  Third, the active gradient basis must be
complete: the two chemical couplings $(\Lambda,\Xi)$ and the independent
current coupling $\zeta_J$ mix under coarse graining, while the combination
$\Xi=2\Lambda$, $\zeta_J=0$ is the equilibrium tangent.

Our main results are as follows.  We define the local perturbation
$\vartheta(\phi)$ by $\Theta(\phi)=\Theta_0[1+\vartheta(\phi)]$, where
$\Theta_0$ is the constant passive value.  At the passive fixed point the
linearized FRG system for the pair
(potential, transport perturbation) is block triangular: $\vartheta$ feeds
into the potential flow, but the flow of $\vartheta$ is proportional to
$\vartheta$ itself and receives no feedback.  The transport spectrum is
therefore governed by a closed linear operator.
Within the functional truncation its leading odd eigenfunction is
$\vartheta_1\propto\phi$.  We denote the spatial dimension by $d$, the
anomalous dimension by $\eta$, and the scaling dimension of the physical
field by $\Dphi$.  We use renormalization-group (RG) eigenvalues $y_i$ such that a scaling-field
amplitude changes as $u_i\mapsto b^{y_i}u_i$ when lengths are increased by
$b>1$; negative $y_i$ therefore describe decaying, irrelevant perturbations.
Here ``odd'' refers to Ising field inversion.  An asterisk will denote the
passive Wilson--Fisher fixed point.  The corresponding local eigenvalue is
\begin{equation}
 \yth=-\Dphi=-\frac{d-2+\eta}{2}.
 \label{eq:introYtheta}
\end{equation}
To isolate departures from equilibrium, we treat two gradient perturbations as equivalent whenever they differ only by the detailed-balance-preserving direction.  The resulting
quotient coordinates use the dimensionless couplings (denoted by bars) of
Eq.~\eqref{eq:dimlessActive}:
\begin{equation}
 \aeq=\bar\Lambda,\qquad
 \ac=\bar\Xi-2\bar\Lambda,\qquad
 \aJ=\bar\zeta_J.
 \label{eq:introQuot}
\end{equation}
The genuine nonequilibrium block is triangular: chemical perturbations never generate the current mode, whereas a current perturbation does acquire a chemical component.  We denote the RG
eigenvalues of the current-like and chemical quotient modes by $\yJ$ and
$\ych$, respectively.  In three dimensions our two smooth-regulator
calculations give
\begin{equation*}
 \yth\simeq-0.52,\qquad
 \yJ\simeq-0.56,\qquad
 \ych\simeq-0.89 .
\end{equation*}

\begin{figure}[t]
\centering
\includegraphics[width=\columnwidth]{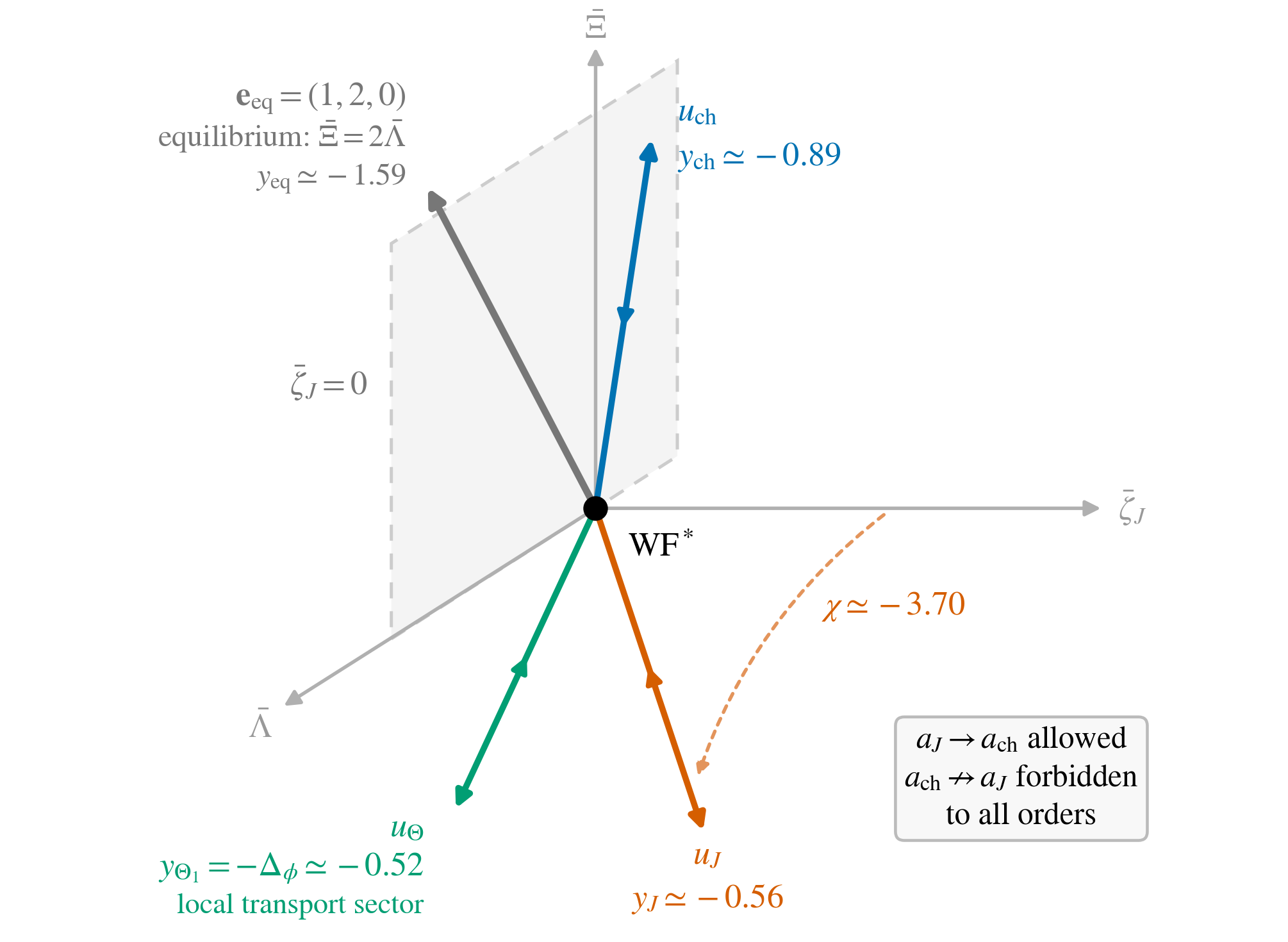}
\caption{Perturbation space around the passive conserved Wilson--Fisher
fixed point (WF$^*$, filled circle) in $d=3$.  Thin gray axes are the
gradient-active couplings $(\bar\Lambda,\bar\Xi,\bar\zeta_J)$; the shaded patch is the
chemical plane $\bar\zeta_J=0$.  The thick gray ray is the equilibrium tangent
$\mathbf e_{\rm eq}=(1,2,0)$, along which $\bar\Xi=2\bar\Lambda$ and detailed
balance is preserved.  Colored rays represent the three retained nonequilibrium eigenmodes; their
amplitudes are denoted by $u_\Theta,u_{\rm ch},u_J$, respectively.
Their eigenvalues are given in Table~\ref{tab:spectrum}; inward
arrowheads indicate that all three are irrelevant.  Quotienting by the
equilibrium direction gives the $(a_{\rm ch},a_J)$ plane of
Eq.~\eqref{eq:introQuot}.  The directions of $u_{\rm ch}$ and $u_J$ follow
from Eqs.~\eqref{eq:MQE} and \eqref{eq:chi}; the dotted arc shows the
chemical component $\chi\simeq-3.70$ of the current-like eigenvector.  The
angular position of $u_\Theta$ is conventional because the local-transport
sector is independent of the gradient sector in the present projections.
The box records the one-way mixing of quotient coordinates -- current into chemical, but not the converse -- rather than the diagonal eigenfields, and the all-orders triangular structure of Sec.~\ref{sec:ward}.}
\label{fig:theoryspace}
\end{figure}

The simple form of the current eigenvalue is not accidental. It follows from an operator identity.  Let $K$ denote
the square-gradient stiffness, $U(\phi)$ the local static potential, and
$T_{ij}$ the renormalized spatial stress tensor.  Translation invariance
implies, within the derivative sector retained here,
\begin{equation*}
 K(\nabla^2\phi)\partial_i\phi
 =
 \partial_iU+\partial_jT_{ij},
\end{equation*}
so the non-gradient current is represented, modulo chemical gradients, by
the divergence of the stress tensor.  Conservation and It\^o causality then make this triangularity exact to all loop orders: chemical operators cannot generate the genuine current quotient at the passive fixed point.  We also
carry out the first nontrivial perturbative test of an additional current
renormalization.  We set $\eps=4-d$ and denote by $c_{J,\rm ct}$ the coefficient
of a possible order-$\eps^2$ contact anomalous dimension.  At two loops
it vanishes,
\begin{equation*}
 c_{J,\rm ct}=0,
\end{equation*}
so
\begin{equation*}
 y_J=
 \frac{2-d}{2}-\frac{3\eta}{2}+\Order(\eps^3)
 \quad (d=4-\eps).
\end{equation*}
Within the FRG truncation the local odd mode obeys Eq.~\eqref{eq:introYtheta},
hence
\begin{equation*}
 \yJ-\yth=-\eta .
\end{equation*}
The small three-dimensional Ising reference value
$\eta_{\rm ref}\simeq0.0363$~\cite{Kos2016} suggests that this near degeneracy persists beyond the truncation, provided neither sector develops an additional contact anomaly.  We show that this produces unusually large crossover scales and makes
single-power finite-size fits potentially misleading.
Figure~\ref{fig:theoryspace} summarizes the resulting geometry: the
three-dimensional gradient-active sector splits into the equilibrium tangent
and a two-dimensional genuine nonequilibrium quotient, and the three retained odd
nonequilibrium directions flow into the passive fixed point.

The paper is organized as follows.  Section~\ref{sec:model} defines the
stochastic model and derives the MSRJD action.  Section~\ref{sec:frg}
introduces the effective average action, the regulator, the truncation,
and all projections used below.  Section~\ref{sec:flows}
derives the passive, local-transport, and gradient-active flow equations.
Section~\ref{sec:ward} develops the exact current quotient and the
stress-tensor Ward identity and summarizes the explicit two-loop test.
Section~\ref{sec:d3} presents the three-dimensional spectrum.
Section~\ref{sec:fss} derives finite-size scaling and crossover
consequences.  Technical derivations, including the full functional
differentiation of the Wetterich equation and the two-loop calculation, are
collected in the Appendices.

\section{Stochastic model and MSRJD formulation}
\label{sec:model}

\subsection{Conserved stochastic dynamics}

We consider a real scalar density field $\phi(\mathbf x,t)$ and denote its
conserved current by $\mathbf J(\mathbf x,t)$. The continuity equation is
\begin{equation*}
 \partial_t\phi+\grad\cdot\mathbf J=0.
\end{equation*}
The current is decomposed into a deterministic part and conserved Gaussian
noise.  Here $M(\phi)$ is the local mobility, $M_0$ its passive reference
value, $\mu$ is the chemical potential, $\zeta_J$ is the non-gradient current
coupling, $D(\phi)$ is the conserved-noise amplitude, and $\bm\xi$ is a
Gaussian vector noise.  With these definitions,
\begin{equation}
 \mathbf J
 =
 -M(\phi)\grad\mu
 -M_0\zeta_J(\lap\phi)\grad\phi
 -\sqrt{2D(\phi)}\,\bm\xi .
 \label{eq:current}
\end{equation}
The vector noise is white and normalized as follows.  Angle brackets denote
ensemble averages.  Here $\delta_{ij}$ is
the Kronecker delta and $\delta^{(d)}$ and $\delta$ are the spatial and
temporal Dirac delta distributions, respectively:
\begin{align}
 \avg{\xi_i(\mathbf x,t)}&=0,\\
 \avg{\xi_i(\mathbf x,t)\xi_j(\mathbf x',t')}
 &=
 \delta_{ij}\delta^{(d)}(\mathbf x-\mathbf x')\delta(t-t').
 \label{eq:noise}
\end{align}

We write the chemical potential in terms of a local potential $U(\phi)$, a
square-gradient stiffness $K$, and two chemical active couplings $\Lambda$
and $\Xi$.  A prime denotes a derivative with respect to the field:
\begin{equation}
 \mu
 =
 U'(\phi)-K\lap\phi
 +\Lambda(\grad\phi)^2
 +\Xi\phi\lap\phi .
 \label{eq:mu}
\end{equation}
The first two terms derive from the passive static functional
\begin{equation*}
 \mathcal H[\phi]
 =
 \int\dd^dx\left[
 \frac{K}{2}(\grad\phi)^2+U(\phi)
 \right].
\end{equation*}
The gradient terms break equilibrium unless
\begin{equation*}
 \Xi=2\Lambda .
\end{equation*}
Indeed, if
\begin{equation*}
 \mathcal H_{\mathcal K}
 =
 \int\dd^dx\,\frac12 \mathcal K(\phi)(\grad\phi)^2,
\end{equation*}
then
\begin{equation*}
 \frac{\delta\mathcal H_{\mathcal K}}{\delta\phi}
 =
 -\mathcal K(\phi)\lap\phi-\frac12\mathcal K'(\phi)(\grad\phi)^2.
\end{equation*}
Choosing $\mathcal K(\phi)=K-\Xi\phi$ gives the pair
$\Xi\phi\lap\phi+(\Xi/2)(\grad\phi)^2$, hence
$\Lambda=\Xi/2$.

The current proportional to $\zeta_J$ cannot be written as the gradient of
a local scalar chemical potential.  We therefore identify the equilibrium
gradient manifold as
\begin{equation*}
 \Xi=2\Lambda,\qquad \zeta_J=0.
\end{equation*}

The second source of nonequilibrium behavior is the local ratio
\begin{equation}
 \Theta(\phi)=\frac{D(\phi)}{M(\phi)}.
 \label{eq:ThetaDef}
\end{equation}
We denote the equilibrium temperature by $T$.  At the passive reference
point
\begin{equation*}
 M(\phi)=M_0,\qquad
 D(\phi)=M_0T,\qquad
 \Theta(\phi)=T.
\end{equation*}

Two remarks make Eq.~\eqref{eq:ThetaDef} unambiguous.
First, the split of the transport data into $M(\phi)$ and $D(\phi)$, and
hence the statement that constant $\Theta$ is the equilibrium condition,
is prescription dependent when the mobility itself is field dependent.  For
the elementary process
$\dot x=-M(x)U'(x)+\sqrt{2M(x)T}\,\xi$ the It\^o convention gives the
stationary density $P_{\rm s}\propto M^{-1}e^{-U/T}$, whereas the kinetic
(H\"anggi--Klimontovich) convention gives the Gibbs form, even though
$\Theta=T$ is constant in both
cases~\cite{Aron2016,AronBiroliCugliandolo2010}.  Second, in the projected
theory used throughout this work the mobility is held at its passive value
$M_0$, so no noise-induced drift is generated in the It\^o convention and
constant $\Theta(\phi)$ supplies the transport condition for detailed balance,
in addition to $\Xi=2\Lambda$ and $\zeta_J=0$ in the gradient sector.  The
prescription issue is therefore absent from the results below, but it would
have to be revisited in a truncation that also runs $M(\phi)$.

\subsection{It\^o prescription and response functional}

We use the It\^o prescription.  Discretizing the stochastic equation
forward in time makes the functional Jacobian triangular; its determinant
is field independent and is absorbed into the normalization.  Introducing
the response field $\phit$, averaging the Gaussian noise, and integrating
the divergence term by parts gives the MSRJD action
\begin{align}
 S[\phi,\phit]
 =\int\dd t\,\dd^dx\Big\{&
 \phit\big[\partial_t\phi-\grad\cdot\big(M(\phi)\grad\mu\big)\big]
 \nonumber\\
 &-M_0\zeta_J\phit\,
 \grad\cdot\big[(\lap\phi)\grad\phi\big]
 \nonumber\\
 &-D(\phi)(\grad\phit)^2\Big\}.
 \label{eq:MSRJD}
\end{align}
For constant mobility $\grad\cdot(M_0\grad\mu)=M_0\lap\mu$, which is the
form used in the truncation below.
The noise average gives the term $-D(\phi)(\grad\phit)^2$ after
integration by parts.  The retarded response in the time domain is denoted by
$G_R(t,q)$.  The It\^o convention fixes $G_R(t=0,q)=0$;
equal-time response contractions and closed directed response cycles
therefore vanish.  These are the stochastic assumptions needed below.
Equation~\eqref{eq:MSRJD} is the conserved-noise form of the standard
response functional~\cite{MartinSiggiaRose1973,Janssen1976,DeDominicis1976}.

In the passive theory, $\Lambda=\Xi=\zeta_J=0$, evaluated at a uniform
background $\phi=\phi_b$ with $\phit=0$, let $q=|\mathbf q|$ be the
spatial wavenumber and define
\begin{equation*}
 P(q;\phi_b)=U''(\phi_b)+Kq^2,
 \qquad
 \Omega(q;\phi_b)=M_0q^2P(q;\phi_b).
\end{equation*}
The restriction to the passive theory matters because $\Xi$ contributes
linearly to the fluctuation of the chemical potential at $\phi_b\neq0$,
shifting $K\to K-\Xi\phi_b$; in the linearized active analysis of
Sec.~\ref{sec:flows}, inserting this shift into a loop already containing
an active vertex gives a second-order contribution.  The gradient projection
is evaluated at $\phi_b=0$, where the shift itself vanishes.
With Fourier convention $\exp[\ii(\mathbf q\cdot\mathbf x-\omega t)]$, let
$D_0\equiv D(\phi_b)$.  The Gaussian retarded response, advanced response,
and correlation propagators are, respectively,
\begin{align*}
 G_R(\omega,q)&=\frac{1}{-\,\ii\omega+\Omega(q)},\\
 G_A(\omega,q)&=\frac{1}{\ii\omega+\Omega(q)},\\
 C(\omega,q)&=
 \frac{2D_0q^2}{\omega^2+\Omega^2(q)}
 =2D_0q^2G_RG_A .
\end{align*}
The absence of a $\phit\phit$ propagator and the retarded character of
$G_R$ will be central both to the one-loop projections and to the
all-orders triangularity theorem.

\section{Functional RG methodology}
\label{sec:frg}

The calculation must resolve two different departures from equilibrium:
a change in local noise and a change in the deterministic gradient current.
We therefore retain both in one running action, then project the corresponding
vertices separately.  This section specifies the approximations and
normalizations needed to compare their RG eigenvalues.

\subsection{Effective average action and Wetterich equation}

We write frequency--momentum arguments explicitly as pairs
$(\omega,\mathbf q)$ and collect the physical and response fields into
$\Phi_a=(\phi,\phit)$.  Indices $a,b$ run over these two components, and
$\int_{\omega,\mathbf q}\equiv\int\dd\omega/(2\pi)
\int\dd^dq/(2\pi)^d$.  The running infrared scale is denoted by $k$.
We add a quadratic infrared regulator
\begin{equation*}
 \Delta S_k
 =
 \frac12\int_{\omega,\mathbf q}
 \Phi_a(-\omega,-\mathbf q)\,\mathcal R_{k,ab}(q)\,\Phi_b(\omega,\mathbf q).
\end{equation*}
chosen off-diagonal in field space, with scalar profile $R_k(q)$,
\begin{equation}
 \mathcal R_k(q)
 =M_0q^2R_k(q)
 \begin{pmatrix}0&1\\1&0\end{pmatrix}.
 \label{eq:regMatrix}
\end{equation}
The off-diagonal choice is not only causal bookkeeping.  It is the dynamic
implementation of adding the static Hamiltonian regulator
\begin{equation}
 \Delta\mathcal H_k=\frac12\int_{\mathbf q}
 \phi(-\mathbf q)R_k(q)\phi(\mathbf q),
 \label{eq:staticRegulator}
\end{equation}
so the regulated passive dynamics relaxes with respect to
$\mathcal H+\Delta\mathcal H_k$.  No independent $\phit\phit$ regulator is
introduced.  Consequently the equilibrium time-reversal symmetry associated with the
fluctuation--dissipation theorem (FDT), that is the classical
Kubo--Martin--Schwinger (KMS) symmetry, is preserved at every scale $k$, rather than only recovered
at $k=0$~\cite{AronBiroliCugliandolo2010,CrossleyGloriosoLiu2017,
SiebererBuchholdDiehl2016,DuclutDelamotte2017,
HuangFarrellFriedmanZaneGloriosoLucas2023}.
The regulated stationary measure is then exactly
$\propto\exp[-(\mathcal H+\Delta\mathcal H_k)/T]$, a property used in
Appendix~\ref{app:passiveflow} to fix the passive potential flow.
With this convention the regulator modifies the static inverse response to
\begin{equation}
 P_k(q;\phi)=U_k''(\phi)+K_kq^2+R_k(q).
 \label{eq:Pk}
\end{equation}
Here and below, primes on functions of the field denote derivatives with
respect to their field argument.
The effective average action $\Gamma_k$ is the Legendre transform of the
regulated connected generating functional, with $\Delta S_k$ subtracted.
Let $\Gamma_k^{(2)}$ denote its Hessian with
respect to $\Phi_a$.  We use the logarithmic RG time
$\tilde{t}\equiv\ln(k/k_{\rm UV})$, where $k_{\rm UV}$ is the ultraviolet
reference scale, and let $\Tr$ denote the trace over momenta, frequencies,
and field indices.  The effective average action $\Gamma_k$ obeys
\begin{equation}
 \partial_{\tilde{t}}\Gamma_k
 =
 \frac12\Tr\left[
 (\Gamma_k^{(2)}+\mathcal R_k)^{-1}
 \partial_{\tilde{t}}\mathcal R_k
 \right].
 \label{eq:Wetterich}
\end{equation}
This exact flow equation and the derivative expansion used below follow the
standard effective-average-action framework
\cite{Wetterich1993,BergesTetradisWetterich2002,Morris1994,Dupuis2021}.
The anomalous dimension is defined by
$\eta_k=-\partial_{\tilde{t}}\ln K_k$; at a fixed point we write it simply as $\eta$.
We also use the infrared RG time
\begin{equation*}
 \ell=-\tilde{t}=\ln(k_{\rm UV}/k),
 \qquad
 \partial_\ell=-\partial_{\tilde{t}}.
\end{equation*}

\subsection{Truncation}

The truncation used for all three-dimensional results contains
the running local potential $U_k(\phi)$, stiffness $K_k$, active chemical
couplings $\Lambda_k$ and $\Xi_k$, current coupling $\zeta_{J,k}$, and noise
function $D_k(\phi)$.  Explicitly,
\begin{align}
 \Gamma_k
 =\int\dd t\,\dd^dx\Big\{&
 \phit\big[\partial_t\phi-M_0\lap\mu_k\big]
 \nonumber\\
 &-M_0\zeta_{J,k}\phit\,
 \grad\cdot\big[(\lap\phi)\grad\phi\big]
 \nonumber\\
 &-D_k(\phi)(\grad\phit)^2\Big\},
 \label{eq:trunc}
\end{align}
with
\begin{equation}
 \mu_k
 =
 U_k'(\phi)-K_k\lap\phi
 +\Lambda_k(\grad\phi)^2
 +\Xi_k\phi\lap\phi .
 \label{eq:muk}
\end{equation}
Its static sector is the local potential approximation with a running
stiffness, conventionally LPA$'$; because Eq.~\eqref{eq:trunc} in addition
carries a field-dependent noise function and the complete quadratic-gradient
active vertex, we refer to the full truncation as \LPAA{}, where the appended
$A$ denotes the extension by the active sector and is not a standard label,
whenever the
distinction matters.
The mobility is kept at $M_0$ in the projected theory; the local departure
from FDT is then carried by
\begin{equation*}
 D_k(\phi)=M_0\Theta_k(\phi).
\end{equation*}
This is a projection choice, not a statement that a general active theory
cannot generate field-dependent mobility.

Two structural consequences of Eq.~\eqref{eq:trunc} should be recorded.
The coefficient of $\phit\partial_t\phi$ is not renormalized and the
mobility is held fixed, so the regulated relaxation rate obeys
$\Omega_k\sim M_0q^2K_kq^2\sim k^{4-\eta}$ and the truncation carries the
Model-B relation for the dynamic exponent $z$, defined by relaxation
times scaling as length to the power $z$:
\begin{equation}
 z=4-\eta .
 \label{eq:zvalue}
\end{equation}
Moreover, no frequency regulator is introduced.  For purely relaxational
dynamics a momentum regulator is sufficient and avoids the causality and
scheme problems associated with frequency
regulators~\cite{DuclutDelamotte2017,Tauber2014}.

For the passive potential we introduce the dimensionless field $\vphi$ and
the invariant $\rho$ through
\begin{equation}
 \vphi=K_k^{1/2}k^{(2-d)/2}\phi,\qquad
 \rho=\frac12\vphi^2 ,
 \label{eq:dimlessField}
\end{equation}
and
\begin{equation*}
 u_k(\rho)=k^{-d}U_k(\phi).
\end{equation*}
We expand about the running minimum $\rho=\kappa$, with Taylor coefficients
$\lambda_n=u_k^{(n)}(\kappa)$ and highest retained order $N_{\rm poly}$.
Dropping the field-independent constant gives
\begin{equation}
 u_k(\rho)
 =
 \sum_{n=2}^{N_{\rm poly}}
 \frac{\lambda_n}{n!}(\rho-\kappa)^n,
 \qquad N_{\rm poly}=10.
 \label{eq:poly}
\end{equation}

For gradient couplings we distinguish the dimensionful vector
$\mathbf a_k=(\Lambda_k,\Xi_k,\zeta_{J,k})^T$, where $T$ denotes transpose,
from
\begin{equation}
 \bar{\mathbf a}
 =k^{(d-2)/2}K_k^{-3/2}\mathbf a_k
 \equiv(\bar\Lambda,\bar\Xi,\bar\zeta_J)^T,
 \label{eq:dimlessActive}
\end{equation}
where $\Theta_0=1$ fixes the energy unit.  The action uses dimensionful
couplings; stability matrices and quotient coordinates below use the
barred couplings.  Thus their canonical and field-renormalization
contribution is $[(2-d)/2-3\eta_k/2]\bar{\mathbf a}$ in infrared RG time.

\subsection{Smooth regulators and threshold functions}

For the active $p^4$ projection we use only smooth regulators.  The optimized
regulator of Refs.~\cite{Litim2000,Litim2001} has the shape function
$r(y)=(1-y)\theta(1-y)/y$, where $\theta$ is the Heaviside function.  It is continuous but has a discontinuous first
derivative at $y=1$.  Already the threshold function of
Eq.~\eqref{eq:m22explicit} involves $r''$, and the quartic momentum fit of
Eq.~\eqref{eq:p4fit} requires the shape function to be differentiated further
still.  This is the same obstruction that restricts the optimized regulator to
the local potential approximation and to $O(\partial^2)$ of the derivative
expansion, whereas higher orders require well-behaved smooth
profiles~\cite{Balog2019}.  Define the
dimensionless momentum $y=q^2/k^2$.  The two profiles are
\begin{align}
 R_E(q)&=K_kq^2r_E(y),
 &r_E(y)&=\frac{1}{e^y-1},\\
 R_G(q)&=K_kq^2r_G(y),
 &r_G(y)&=\frac{e^{-y}}{y}.
 \label{eq:regs}
\end{align}
Thus $R_G(q)=K_kk^2e^{-y}$, as required.  Both profiles satisfy
$\min_y y[1+r(y)]=1$, so that the dimensionless inverse propagator obeys
$P_k(q;0)/(K_kk^2)\ge1+U_k''(0)/(K_kk^2)$.
At the fixed points found below this lower bound is positive despite the
negative curvature at the symmetric point.  For a generic smooth shape
function $r(y)$, a prime denotes differentiation with respect to $y$.
A dimensionless mass argument $w$ is defined by $U_k''=K_kk^2w$.
A common threshold function for the potential sector is
\begin{equation}
 l_0^d(w;\eta)
 =
 -\frac12\int_0^\infty\dd y\,
 y^{d/2}
 \frac{\eta r(y)+2yr'(y)}
 {y[1+r(y)]+w}.
 \label{eq:l0}
\end{equation}
Let $\Gamma(z)$ denote the Euler gamma function.  The geometrical factor is
\begin{equation*}
 v_d^{-1}=2^{d+1}\pi^{d/2}\Gamma(d/2).
\end{equation*}

\subsection{Projections}

We denote by $\Gamma_k^{(n,m)}$ the one-particle-irreducible (1PI) vertex
with $n$ response-field legs and $m$ physical-field legs.  The local noise
function is defined by
\begin{equation}
 D_k(\phi)
 =
 -\frac12
 \lim_{q\to0}
 \frac{\Gamma_k^{(2,0)}(q,-q;\phi)}{q^2}.
 \label{eq:DprojMain}
\end{equation}
The running square-gradient stiffness \(K_k\) is extracted from the small-momentum expansion of the two-point vertex.  For the active three-leg projection we set all external frequencies to zero.  The symbols $Q,p_1,p_2$ below then
denote spatial momenta (boldface is suppressed), all taken as incoming.
Frequency--momentum arguments elsewhere are written explicitly as pairs:
\begin{equation*}
 Q+p_1+p_2=0.
\end{equation*}
Here $Q^2=|Q|^2$, and products such as $Q\cdot p_i$ are Euclidean spatial
products.
The three basis tensors are
\begin{align}
 V_\Lambda&=-2Q^2(p_1\cdot p_2),\\
 V_\Xi&=-Q^2(p_1^2+p_2^2),\\
 V_J&=(Q\cdot p_1)p_2^2+(Q\cdot p_2)p_1^2.
 \label{eq:activeVertices}
\end{align}
Writing
\begin{equation}
 \kinS=p_1^2+p_2^2,\qquad
 \kinT=p_1\cdot p_2,\qquad
 \kinV=p_1^2p_2^2,
 \label{eq:invariants}
\end{equation}
we reserve these typewriter letters for scalar momentum invariants, not vectors.
The fourth-order part of the flow of $\Gamma_k^{(1,2)}$ is denoted by
$X_{p^4}$.  We denote the coefficients of the four invariant monomials by
$c_{\kinS^2},c_{\kinS\kinT},c_{\kinT^2}$, and $c_{\kinV}$;
$c_\alpha$ denotes any one of them, with $\alpha$ labeling its monomial.
The vertex polynomial is reconstructed
as
\begin{equation}
 \frac{X_{p^4}}{M_0}
 =
 c_{\kinS^2}\kinS^2+c_{\kinS\kinT}\kinS\kinT+c_{\kinT^2}\kinT^2+c_{\kinV}\kinV.
 \label{eq:p4fit}
\end{equation}
At this stage $X_{p^4}$ is the dimensionful fluctuation contribution
obtained from the Wetterich trace, before rescaling fields and momenta.
We denote the corresponding coupling flows by
$\beta_{\Lambda,\rm loop}\equiv\partial_\ell\Lambda_k$,
$\beta_{\Xi,\rm loop}\equiv\partial_\ell\Xi_k$, and
$\beta_{J,\rm loop}\equiv\partial_\ell\zeta_{J,k}$.
The subscript ``loop'' denotes the flow of the dimensionful couplings, before rescaling fields
and momenta. Canonical scaling and anomalous-dimension terms enter after conversion to dimensionless couplings, as made explicit in Eq.~\eqref{eq:lambdaBetaExplicit}.
Matching the local polynomial to
\begin{equation*}
 \frac{X_{p^4}}{M_0}
 =\beta_{\Lambda,\rm loop}V_\Lambda
 +\beta_{\Xi,\rm loop}V_\Xi
 +\beta_{J,\rm loop}V_J
\end{equation*}
gives the projections below, since
$V_\Lambda=-2\kinS\kinT-4\kinT^2$,
$V_\Xi=-\kinS^2-2\kinS\kinT$, and $V_J=-\kinS\kinT-2\kinV$.
The coefficients $c_\alpha$ thus encode the dependence of the vertex
flow on the running couplings; they are not independent parameters.
The dimensionful coupling projections are
\begin{equation}
 \partial_\ell\Lambda_k=-\frac14c_{\kinT^2},\quad
 \partial_\ell\Xi_k=-c_{\kinS^2},\quad
 \partial_\ell\zeta_{J,k}=-\frac12c_{\kinV} .
 \label{eq:activeProj}
\end{equation}
The four coefficients are not independent.  For conserved dynamics the MSRJD
action is invariant under a constant shift of the response field,
$\phit\to\phit+c$, with constant $c$: the deterministic part changes by
$c\int\dd t\,\dd^dx\,[\partial_t\phi+\grad\cdot\mathbf J_{\rm det}]$,
where $\mathbf J_{\rm det}$ is the deterministic current of Eq.~\eqref{eq:current}.
This is a boundary term, the noise term depends only on $\grad\phit$, and the regulator
of Eq.~\eqref{eq:regMatrix} is invariant because $q^2R_k(q)\to0$ as $q\to0$.
The associated Ward identity states that every vertex with a vanishing
external response momentum vanishes,
$\Gamma_k^{(1,n)}(Q=0;p_1,\ldots,p_n)=0$.  Setting $Q=0$, hence $p_2=-p_1$,
$\kinS=2p_1^2$, $\kinT=-p_1^2$ and $\kinV=p_1^4$, in Eq.~\eqref{eq:p4fit} gives
$p_1^4[4c_{\kinS^2}-2c_{\kinS\kinT}+c_{\kinT^2}+c_{\kinV}]$, so that
\begin{equation}
 c_{\kinS\kinT}
 =
 \frac12c_{\kinT^2}+2c_{\kinS^2}+\frac12c_{\kinV} .
 \label{eq:closure}
\end{equation}
Equation~\eqref{eq:closure} is therefore an exact consequence of the
conservation law and not a test of the assumed operator basis; numerically it
measures the quadrature and fitting error alone.  It also settles the
completeness of the gradient basis at this order.  The space of symmetric
quartic momentum polynomials is four dimensional, conservation removes one
dimension, and $V_\Lambda$, $V_\Xi$, $V_J$ are three linearly independent
elements of the remaining three-dimensional subspace.  They therefore span it,
and no further local active operator exists at order $p^4$.  For comparison,
the structure $\phit\phi\nabla^4\phi$, whose vertex is $\kinS^2-2\kinV$,
violates Eq.~\eqref{eq:closure} and is correctly absent.

\section{Flow equations around the passive fixed point}
\label{sec:flows}

We first determine the passive fixed point that supplies the propagators
and vertices.  We then solve two linear stability problems on that
background: one for the field-dependent noise ratio and one for the three
gradient couplings.  Their eigenvalues are combined in Sec.~\ref{sec:d3}.

\subsection{Passive flow}

Because the regulator of Eq.~\eqref{eq:regMatrix} deforms only the static
Hamiltonian, the equal-time sector of the regulated theory is the static
theory regulated by $R_k$, and the potential flow is the static Wetterich
flow.  In this section we suppress the subscript $k$ on dimensionless
running quantities, and a prime on $u(\rho)$ denotes a derivative with
respect to $\rho$.  The field-dependent dimensionless longitudinal mass is
\begin{equation*}
 w(\rho)=u'(\rho)+2\rho u''(\rho).
\end{equation*}
The one-loop kernel carries an explicit factor of the local
noise-to-mobility ratio, because the equal-time correlation function obeys
$\int\dd\omega\,C/2\pi=\Theta_k(\phi)/P_k$.  With this factor retained,
\begin{equation}
 \partial_{\tilde{t}}u
 =
 -du+(d-2+\eta)\rho u'
 +2v_d\,\Theta_k\,l_0^d(w;\eta),
 \label{eq:potFlow}
\end{equation}
which reduces to the familiar equilibrium form at $\Theta_k=\Theta_0=1$.
Retaining it matters only for the coupled stability analysis of
Sec.~\ref{sec:localFDT}; the passive fixed point itself is unchanged.
Appendix~\ref{app:passiveflow} derives Eq.~\eqref{eq:potFlow} and explains
why the naive evaluation of the dynamic trace at $\phit=0$ cannot be used
for this purpose.

Let $F(\rho)$ denote the right-hand side of Eq.~\eqref{eq:potFlow} at
$\Theta_k=1$, and define
$\beta_\kappa\equiv\partial_{\tilde{t}}\kappa$ and
$\beta_{\lambda_n}\equiv\partial_{\tilde{t}}\lambda_n$.  The running minimum and
Taylor couplings satisfy
\begin{align}
 \beta_\kappa
 &=
 -\frac{F'(\kappa)}{\lambda_2},
 \label{eq:betaKappa}\\
 \beta_{\lambda_n}
 &=
 F^{(n)}(\kappa)+\lambda_{n+1}\beta_\kappa,
 \qquad 2\le n<N_{\rm poly},
 \label{eq:betaLambdaN}\\
 \beta_{\lambda_{N_{\rm poly}}}
 &=
 F^{(N_{\rm poly})}(\kappa).
 \label{eq:betaLambdaLast}
\end{align}
These equations follow by differentiating the identity
$u'(\kappa)=0$ and the definitions
$\lambda_n=u^{(n)}(\kappa)$, with
$\lambda_{N_{\rm poly}+1}=0$ at the truncation boundary.

The anomalous dimension is obtained self-consistently from the standard
transverse projection
\begin{equation}
 \eta
 =
 \frac{16v_d}{d}\,
 \kappa\lambda_2^2
 m_{22}^d(w_\kappa,\eta),
 \qquad
 w_\kappa=2\kappa\lambda_2 .
 \label{eq:etaFlow}
\end{equation}
Here $m_{22}^d$ is the dimensionless threshold integral for the stiffness
projection.  Its compact definition and full smooth-regulator expression are given in Appendix~\ref{app:passiveflow}.

\subsection{Local fluctuation--dissipation violation}
\label{sec:localFDT}

We use the spatial measure
$\int_{\mathbf q}\equiv\int\dd^dq/(2\pi)^d$.  Write
\begin{equation}
 \Theta_k(\phi)=\Theta_0[1+\vartheta_k(\vphi)]
 \label{eq:varthetaDef}
\end{equation}
near the passive fixed point.  The local noise projection of the Wetterich
equation yields, before linearization,
\begin{align}
 \partial_\ell D_k(\phi)
 &=-\frac12D_k''(\phi)\Theta_k(\phi)
 \nonumber\\[-2pt]
 &\quad\times
 \int_{\mathbf q}
 \frac{\partial_\ell R_k(q)}
 {[U_k''(\phi)+K_kq^2+R_k(q)]^2}.
 \label{eq:Dflow}
\end{align}
Only the tadpole with the four-leg noise vertex contributes at this order,
because it is the only vertex in Eq.~\eqref{eq:trunc} carrying two response
legs; the trace therefore closes on the correlation propagator.  The
diagrammatic derivation, the classification of the remaining one-loop
classes, and the proof that the
$(\Lambda,\Xi,\zeta_J)$ sector generates only higher-gradient noise in this
projection are given in Appendix~\ref{app:thetaflow}.

To avoid mixing derivatives with respect to $\rho$ and $\vphi$, define
\begin{equation*}
 \bar u_*(\vphi)\equiv u_*(\rho=\vphi^2/2),\qquad
 w_*(\vphi)\equiv\partial_{\vphi}^2\bar u_*(\vphi).
\end{equation*}
A prime on $\vartheta$ denotes $\partial_{\vphi}$, while
$l_0^{d\,\prime}(w;\eta)\equiv\partial l_0^d(w;\eta)/\partial w$.
Linearizing Eq.~\eqref{eq:Dflow} at
$\Theta_k=\Theta_0$, $u=u_*$ gives
\begin{align}
 \partial_\ell\vartheta
 &=
 \mathcal{L}_\Theta\,\vartheta,
 \nonumber\\
 \mathcal{L}_\Theta
 &=-\Dphi\,\vphi\,\partial_\vphi
 -2v_d\,l_0^{d\,\prime}(w_*(\vphi);\eta_*)\,\partial_\vphi^2 ,
 \label{eq:thetaFlow}
\end{align}
where
\begin{equation*}
 \Dphi=\frac{d-2+\eta_*}{2}.
\end{equation*}
Although the reference potential is even, its perturbations need not be:
we write $\bar u_k(\vphi)=u_*(\vphi^2/2)+\delta\bar u_k(\vphi)$.
The Taylor expansion in $\rho$ determines the even reference solution;
the stability problem also admits odd functions $\delta\bar u(\vphi)$.
For comparison, the fixed-$\eta$ linearized potential operator is
\begin{equation}
 \mathcal{L}_U^{(0)}
 =
 d-\Dphi\vphi\partial_\vphi
 -2v_dl_0^{d\,\prime}(w_*(\vphi);\eta_*)\partial_\vphi^2 .
 \label{eq:LU0}
\end{equation}
Hence
\begin{equation}
 \mathcal{L}_\Theta=\mathcal{L}_U^{(0)}-d.
 \label{eq:operatorIdentity}
\end{equation}
In particular $\vartheta_1(\vphi)\propto\vphi$ has
$\vartheta_1''=0$, so within the truncation
\begin{equation}
 \yth=-\Dphi.
 \label{eq:ytheta}
\end{equation}

\begin{figure}[tb]
\centering
\includegraphics[width=\columnwidth]{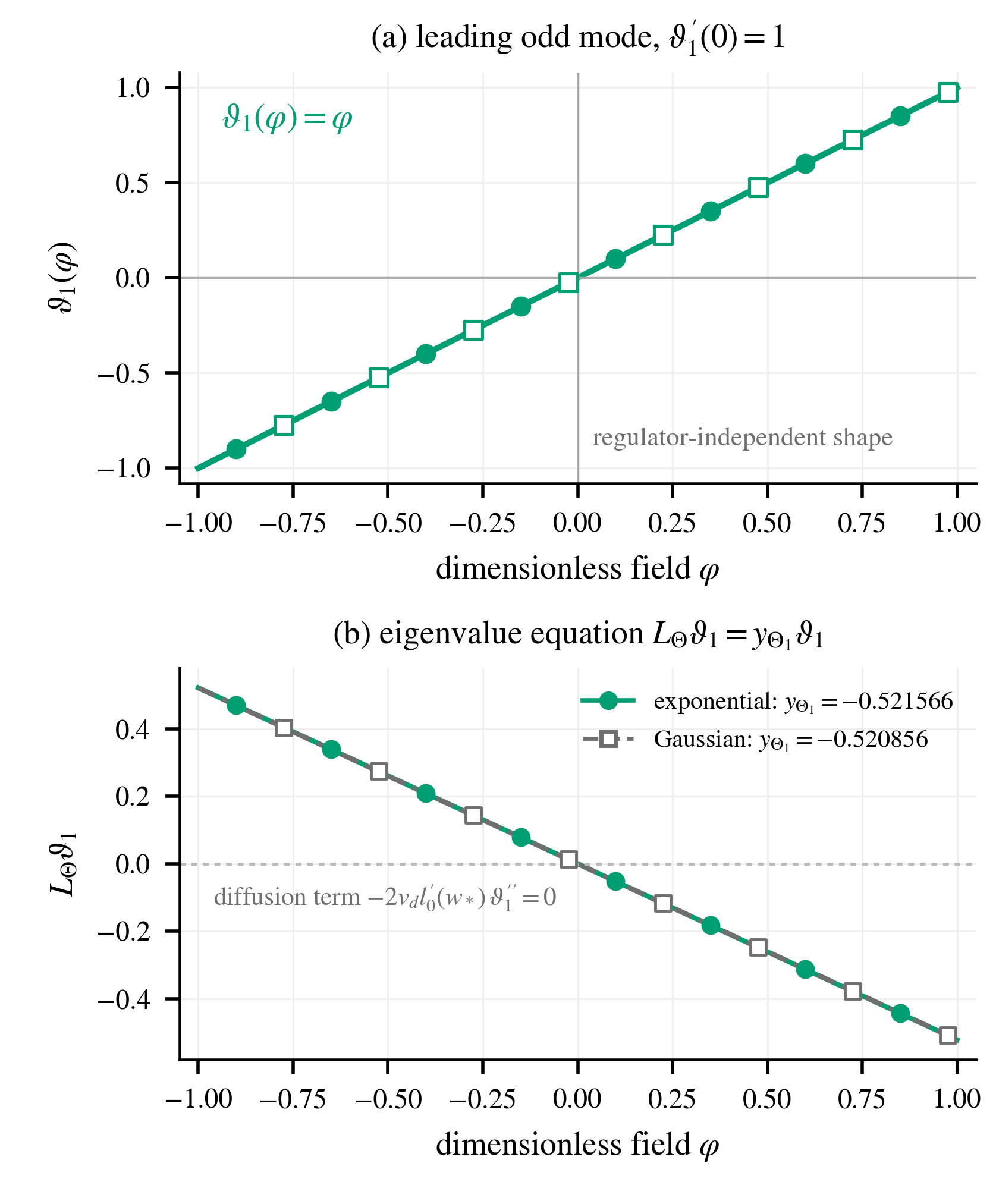}
\caption{Leading odd local-transport eigenfield at the passive fixed point.
(a) With $\vartheta_1'(0)=1$, the eigenfunction is
$\vartheta_1(\vphi)=\vphi$ and has the same shape for the exponential
(filled circles) and Gaussian (open squares) regulators.
(b) Since $\vartheta_1''=0$, the field-dependent diffusion term in
$\mathcal{L}_\Theta$ vanishes identically.  The eigenvalue equation is saturated by
the scaling drift and gives $y_{\Theta_1}=-\Dphi$.}
\label{fig:localEigenfield}
\end{figure}

Figure~\ref{fig:localEigenfield} makes the simplicity of this
mode explicit.  With the normalization $\vartheta_1'(0)=1$, both regulators
give the same straight eigenfunction $\vartheta_1(\vphi)=\vphi$.  The two
curves in panel (a) therefore coincide exactly; the markers distinguish the
regulator calculations rather than different functional shapes.  Panel (b)
separates the two terms in $\mathcal{L}_\Theta$: because
$\vartheta_1''(\vphi)=0$, the field-dependent diffusion term vanishes point
by point, and the scaling drift alone gives
$\mathcal{L}_\Theta\vartheta_1=-\Dphi\vartheta_1$.  The figure displays the closed
transport component of the eigenvector.  The induced odd potential
admixture discussed below is not needed to determine the eigenvalue and is
not included in the plot.

Three properties make this result more robust than a generic truncation
estimate.  First, the diffusion coefficient in Eq.~\eqref{eq:thetaFlow}
depends on the field through $w_*(\vphi)$, so monomials are in general not
eigenfunctions; exactly two of them are, namely $\vartheta=\rm const$ with
eigenvalue $0$ and $\vartheta_1\propto\vphi$ with eigenvalue $-\Dphi$,
precisely because both have vanishing second derivative.  Equation
\eqref{eq:ytheta} has a form independent of the threshold function and of
the regulator; its numerical value still depends on the estimated $\eta_*$.
The higher local modes depend on the threshold function and are not quoted
here.  Second, $\vartheta$ is odd and $\eta$ is even under
$\phi\to-\phi$, so the two do not mix at linear order and the
fixed-$\eta$ restriction in Eq.~\eqref{eq:LU0} does not affect
Eq.~\eqref{eq:ytheta}.  Third, the coupled system is triangular.  Because
the flow of $\vartheta$ is proportional to $\vartheta$, it receives no
feedback from $\delta\bar u$ at the passive fixed point, whereas $\vartheta$
does feed the potential through the explicit factor in
Eq.~\eqref{eq:potFlow}:
\begin{align}
 \partial_\ell
 \begin{pmatrix}\delta\bar u\\ \vartheta\end{pmatrix}
 &=
 \begin{pmatrix}\mathcal{L}_U&\mathcal B_{U\Theta}\\ 0&\mathcal{L}_\Theta\end{pmatrix}
 \begin{pmatrix}\delta\bar u\\ \vartheta\end{pmatrix},
 \nonumber\\
 \mathcal B_{U\Theta}\vartheta&=-2v_d\,\vartheta\,l_0^d(w_*;\eta_*).
 \label{eq:triangularTheta}
\end{align}
Here $\mathcal{L}_U$ is the potential stability operator and $\mathcal B_{U\Theta}$
maps a transport perturbation into the potential flow.  In the odd sector
considered here $\delta\eta=0$, so $\mathcal{L}_U=\mathcal{L}_U^{(0)}$.
The spectrum is the union of those of $\mathcal{L}_U$ and $\mathcal{L}_\Theta$, so
Eq.~\eqref{eq:ytheta} is unaffected; the corresponding scaling field
however acquires a potential admixture, which is odd and therefore
contributes to the ordering-field direction used in
Sec.~\ref{sec:fss}.  Appendix~\ref{app:thetaflow} records an exact zero
mode of Eq.~\eqref{eq:triangularTheta} that tests this structure.

This result has a useful operator interpretation that does not rely on the
detailed threshold function.  At the passive Model-B fixed point the
response-field dimension and dynamic exponent are
\begin{equation*}
 \Delta_{\phit}=d-\Dphi,\qquad z=4-\eta,
\end{equation*}
the latter being the exact relaxational
identity~\cite{HalperinHohenbergMa1976,Tauber2014}.
The local-transport insertion is represented by
$\mathcal O_\Theta=\phi(\grad\phit)^2$.  Its dimension-counting eigenvalue is
therefore
\begin{align}
 y_\Theta^{(0)}
 &=d+z-\big(\Dphi+2\Delta_{\phit}+2\big)
 \nonumber\\
 &=z-d+\Dphi-2=-\Dphi.
 \label{eq:yThetaCounting}
\end{align}
Thus Eq.~\eqref{eq:ytheta} is equivalently the statement that the present
projection generates no additional contact anomaly for
$\mathcal O_\Theta$.  If such a composite-operator contribution is denoted
by $\gamma_{\Theta,\rm ct}$, the truncation gives
$\gamma_{\Theta,\rm ct}=0$.  Unlike the current sector below, no all-orders
Ward identity based proof of this vanishing is claimed.

\subsection{Gradient-active flow}

The two projections use different backgrounds for a controlled reason.  The
local-transport eigenproblem retains the full fixed function
$w_*(\vphi)$ because it resolves arbitrary field dependence of
$\Theta(\phi)$.  The gradient-active sector instead probes the lowest-field
three-leg vertex $\Gamma^{(1,2)}$ at the symmetric background $\phi_b=0$;
its inputs $w_{0,*}=u'_*(0)$ and $g_{4,*}=3u''_*(0)$ are derivatives at the origin of the same
global fixed function $u_*$.  This choice isolates the closed
$(\Lambda,\Xi,\zeta_J)$ basis without importing higher-field active
operators.  Moreover the combined Ising transformation
$(\phi_b,\phi,\phit,\bar{\mathbf a})\mapsto(-\phi_b,-\phi,-\phit,-\bar{\mathbf a})$
implies for the linearized
active matrix
\begin{align}
 \mathsf M_{\rm act}(\phi_b)&=\mathsf M_{\rm act}(-\phi_b),\nonumber\\
 \mathsf M_{\rm act}(\phi_b)
 &=\mathsf M_{\rm act}(0)+\Order(\phi_b^2).
 \label{eq:backgroundSensitivity}
\end{align}
The three active terms are even in $\phi$, so the reversal of $\phit$ is what
forces $\bar{\mathbf a}\to-\bar{\mathbf a}$; the passive fixed-point potential is even and
therefore invariant.  The absent linear term is an internal sensitivity check.  The quadratic
background dependence mixes higher-field active operators and is outside
the retained basis; quantifying it requires an enlarged truncation and is a
limitation of the present numerical spectrum.

We evaluate the active stability matrix to first order in the coupling vector $\bar{\mathbf a}$:
\begin{equation*}
 \bar{\mathbf a}=(\bar\Lambda,\bar\Xi,\bar\zeta_J)^T.
\end{equation*}
The common canonical and field-renormalization contribution is denoted by
$y_{\rm grad}^{(0)}$ and equals
\begin{equation}
 y_{\rm grad}^{(0)}=\frac{2-d}{2}-\frac32\eta_* .
 \label{eq:ca}
\end{equation}
The loop part follows from the third functional derivative of the Wetterich
equation.  At the symmetric passive background only two bubble topologies
survive at linear order in $\bar{\mathbf a}$.  Before writing them, we introduce
the notation used in the formulas.  The index
$j\in\{\Lambda,\Xi,J\}$ labels the active vertex $V_j$ in
Eq.~\eqref{eq:activeVertices}; $g_{4,*}$ is the dimensionless passive quartic
vertex at the Wilson--Fisher fixed point.  After the frequency integration,
we use $\int_q\equiv\int\dd^dq/(2\pi)^d$ for the remaining spatial loop
integral.  For the two external physical momenta $p_1,p_2$ and the external
response momentum $Q=-p_1-p_2$, define
\begin{equation}
 q_i=|q+p_i|,\qquad q_Q=|q+Q|,\qquad i=1,2,
 \label{eq:shiftedMomenta}
\end{equation}
and write $\bar{\imath}=3-i$ for the other external physical leg in the first
topology.  The symbols $q_i$ and $q_Q$ denote magnitudes; $q$ in vertex arguments
denotes the corresponding loop vector.

For compactness, all loop momenta in the following kernel are measured in
units of $k$.  We use the equilibrium normalization $M_0=\Theta_0=1$ in the
dimensionless loop kernel; the corresponding factors are restored by the
coupling definitions.  Let $w_{0,*}$ be the dimensionless fixed-point mass and
$R_*(q)=q^2r(q^2)$ the dimensionless regulator contribution.  Define
\begin{equation}
 P_*(q)=w_{0,*}+q^2+R_*(q),\qquad E_q=q^2P_*(q).
 \label{eq:dimensionlessEnergy}
\end{equation}
The exact frequency integral of one correlation line of momentum $q$ and one
response line of momentum $q'$ is
\begin{align}
 \mathcal F(q,q')
 &=\int\frac{\dd\omega}{2\pi}C(\omega,q)G_R(\omega,q')\nonumber\\
 &=\frac{q^2}{E_q(E_q+E_{q'})} .
 \label{eq:Fkernel}
\end{align}
We denote its single-scale derivative by
$\dot{\mathcal F}\equiv\widetilde\partial_\ell\mathcal F$, where
$\widetilde\partial_\ell$ acts only on the regulator.  Explicitly,
\begin{align}
 \dot{\mathcal F}(q,q')
 ={}&-q^2\frac{2E_q+E_{q'}}{E_q^2(E_q+E_{q'})^2}
 \widetilde\partial_\ell E_q
 \nonumber\\
 &-q^2\frac{1}{E_q(E_q+E_{q'})^2}
 \widetilde\partial_\ell E_{q'} .
 \label{eq:Fdot}
\end{align}

With these definitions, the momentum-dependent loop contributions whose
fourth-order parts enter the active projection are
\begin{align}
 X_{T1}^{(j)}
 &=
 -g_{4,*}Q^2
 \sum_{i=1}^2
 \int_q
 V_j(-q-p_i;p_i,q)\,
 \dot{\mathcal F}(q,q_i),
 \label{eq:T1}\\
 X_{T2}^{(j)}
 &=
 -g_{4,*}
 \int_q
 q_Q^2
 V_j(Q;q,-Q-q)\,
 \dot{\mathcal F}(q,q_Q).
 \label{eq:T2}
\end{align}
Their sum $X^{(j)}=X_{T1}^{(j)}+X_{T2}^{(j)}$ is the derivative of the
vertex flow with respect to $\bar a_j$.  Its fourth-order projection gives
the $j$-th column of the loop matrix $\Delta\mathsf M_{\rm act}$.
The equilibrium normalization
$M_0=\Theta_0=1$ is used inside the kernel and the correction to it is of
order $\bar{\mathbf a}\vartheta$.
The corresponding MSRJD contractions are drawn below.  In $T1$ the active
vertex carries $p_i$ and its response leg is internal.  The response line
therefore points from the active vertex to the passive one.  In $T2$ the
active vertex carries $Q$ and no external physical leg.  The passive response
leg is internal, so this response line points from the passive vertex to the
active one.  In both cases the second internal line is a correlation.

\begin{center}
\centering
\begin{tikzpicture}[x=1.05cm,y=0.78cm,font=\scriptsize]
  \node[activevertex] (a1) at (0,1.4) {$j$};
  \node[passivevertex] (b1) at (1.7,1.4) {};
  \node[anchor=east] at (-1.18,1.4) {$T1:$};
  \draw[diagline] (-0.85,1.4) -- (a1) node[midway,above] {$p_1$};
  \draw[response] (b1) -- (2.55,1.85) node[pos=1,right] {$Q$};
  \draw[diagline] (b1) -- (2.55,0.95) node[pos=1,right] {$p_2$};
  \draw[response] (a1) to[out=35,in=145] (b1);
  \draw[correlation] (a1) to[out=-35,in=-145] (b1);

  \node[activevertex] (a2) at (0,-0.2) {$j$};
  \node[passivevertex] (b2) at (1.7,-0.2) {};
  \node[anchor=east] at (-1.18,-0.2) {$T2:$};
  \draw[response] (-0.85,-0.2) -- (a2) node[midway,above] {$Q$};
  \draw[diagline] (b2) -- (2.55,0.25) node[pos=1,right] {$p_1$};
  \draw[diagline] (b2) -- (2.55,-0.65) node[pos=1,right] {$p_2$};
  \draw[response] (b2) to[out=145,in=35] (a2);
  \draw[correlation] (a2) to[out=-35,in=-145] (b2);
\end{tikzpicture}
\end{center}
The $T1$ drawing shows the $i=1$ representative; the $i=2$ term in
Eq.~\eqref{eq:T1} is obtained by exchanging $p_1$ and $p_2$.  The two
drawings show the one-loop active topologies.  Plain external single
lines are amputated physical-field legs, arrowed external single lines are
amputated response-field legs, directed internal single lines are retarded
response propagators, and double lines are internal correlations.  The
opposite internal-response orientations follow from which vertex supplies
the internal response leg.
Equations~\eqref{eq:Fdot}--\eqref{eq:T2}, followed by the invariant
projection Eq.~\eqref{eq:activeProj}, define the loop correction matrix
$\Delta\mathsf M_{\rm act}$ without any additional approximation
in the internal shifted momentum.  Their complete
derivation from repeated differentiation of
$(\Gamma_k^{(2)}+\mathcal R_k)^{-1}$ is given in
Appendix~\ref{app:activeflow}, including the momentum routing.
The graphical conventions are collected in Appendix~\ref{app:diagrammatics},
and the numerical reconstruction is described in Appendix~\ref{app:numerics}.

After rescaling the loop flow to dimensionless variables, we reuse the
symbols $c_\alpha$ for its coefficients in the same monomial basis.  The
linear response to the active couplings is defined, with $\alpha\in\{\kinS^2,\kinS\kinT,\kinT^2,\kinV\}$, by
\begin{align}
 c_\alpha(\bar{\mathbf a})
 &=\sum_{j=\Lambda,\Xi,J}c_{\alpha j}\bar a_j
   +\Order(\bar{\mathbf a}^{2}),\nonumber\\
 c_{\alpha j}
 &\equiv\left.\frac{\partial c_\alpha}{\partial\bar a_j}
       \right|_{u_*,\eta_*,\bar{\mathbf a}=0}.
 \label{eq:coefficientLinearization}
\end{align}
Equivalently, $X^{(j)}=\partial X/\partial\bar a_j|_*$ is a single
infinitesimal active insertion with passive propagators and vertices, and
$X=\sum_j\bar a_jX^{(j)}+\Order(\bar{\mathbf a}^{2})$.
The numbers $c_{\alpha j}$ depend on the passive fixed-point data and
regulator; the running couplings multiply them.  For example,
\begin{equation}
 \partial_\ell\bar\Lambda
 =y_{\rm grad}^{(0)}\bar\Lambda
 -\frac14\sum_{j=\Lambda,\Xi,J}c_{\kinT^2 j}\bar a_j
 +\Order(\bar{\mathbf a}^{2}).
 \label{eq:lambdaBetaExplicit}
\end{equation}
The first term comes from rescaling; the sum is the loop contribution.
Let $\Id$ denote the $3\times3$ identity matrix.  At linear order,
\begin{align}
 \partial_\ell\bar{\mathbf a}
 &=\mathsf M_{\rm act} \bar{\mathbf a}=
 \left[y_{\rm grad}^{(0)}\Id+\Delta\mathsf M_{\rm act}\right]\bar{\mathbf a},\\
 \mathsf M_{\rm act}&=\begin{pmatrix}
     y_{\rm grad}^{(0)}-\frac{1}{4}c_{\kinT^2\,\Lambda} & -\frac{1}{4}c_{\kinT^2\,\Xi} & -\frac{1}{4}c_{\kinT^2\,J}\\
     -c_{\kinS^2\,\Lambda} & y_{\rm grad}^{(0)}-c_{\kinS^2\,\Xi} &-c_{\kinS^2\,J}\\
     0 & 0 & y_{\rm grad}^{(0)}
 \end{pmatrix}. \nonumber
 \label{eq:activeFlow}
\end{align}
Here the second index of $c_{\alpha j}$ labels the differentiated active coupling and hence the column of the stability matrix.
We finally pass to
\begin{equation}
 \begin{pmatrix}\bar\Lambda\\\bar\Xi\\\bar\zeta_J\end{pmatrix}
 =
 \begin{pmatrix}
 1&0&0\\2&1&0\\0&0&1
 \end{pmatrix}
 \begin{pmatrix}\aeq\\\ac\\\aJ\end{pmatrix}.
 \label{eq:Tmatrix}
\end{equation}
Let $\mathcal V_{\rm grad}$ be the three-dimensional space of gradient
couplings.  The equilibrium tangent is the $\aeq$ axis.  The lower
$2\times2$ block in $(\ac,\aJ)$, denoted by $\mathsf M_{\rm NE}$, acts on
$\mathcal V_{\rm grad}/\mathcal V_{\rm eq}$, where
$\mathcal V_{\rm eq}=\Span\{(1,2,0)\}$.  It retains both nonequilibrium
directions.  The one-dimensional current quotient used in the Ward proof
instead removes the entire chemical plane
$\mathcal V_{\rm chem}=\Span\{(1,0,0),(0,1,0)\}$.
The current projector $\Pi_J$, defined in Eq.~\eqref{eq:PiJ}, therefore
also removes the nonequilibrium chemical direction retained by
$\mathsf M_{\rm NE}$.  The spectrum needs the two-dimensional quotient;
the Ward argument needs only its current component.

\section{Ward structure and two-loop current renormalization}
\label{sec:ward}

The flow calculation gives a simple current eigenvalue even though the corresponding eigenvector mixes strongly with the chemical direction.  We now ask which part of that simplicity is
fixed by symmetry and which part requires an explicit calculation.
The three-dimensional calculation above and the Ward analysis below are two
complementary descriptions of the same renormalized theory.  In
Secs.~\ref{sec:frg}--\ref{sec:flows} we followed the scale-dependent effective
average action $\Gamma_k$ and used the Wetterich equation to estimate the
fixed-point spectrum in $d=3$.  In this section we instead consider the
regulator-free 1PI action
\begin{equation}
 \Gamma=\lim_{k\to0}\Gamma_k .
 \label{eq:EAAto1PI}
\end{equation}
The translation Ward identity constrains this full 1PI action independently
of how the RG flow is parametrized.  To test whether the current operator
requires an additional local contact counterterm, we subsequently expand its
renormalized 1PI vertices near $d=4$ in powers of the quartic coupling and use
dimensional regularization with minimal subtraction.  This perturbative test
does not supply or modify the beta functions used in the numerical Wetterich
flow; it checks the operator statement by an independent calculation.

At nonzero $k$, a local-translation identity for $\Gamma_k$ generally
contains an explicit regulator insertion.  The regulator used here preserves
global translations and equilibrium FDT, and the additional insertion
vanishes together with $R_k$ as $k\to0$.  The Ward identity used below is
therefore the regulator-free identity for Eq.~\eqref{eq:EAAto1PI}, not an
unmodified identity imposed on the truncated $\Gamma_k$ at every scale.

\subsection{Translation Ward identity}

Let $\Gamma_{\rm stat}[\phi]$ denote the static sector of the regulator-free
renormalized 1PI action in Eq.~\eqref{eq:EAAto1PI}.  In the derivative sector
retained in the present calculation,
define the equation-of-motion operator by
\begin{equation}
 \EOM(x)=\frac{\delta\Gamma_{\rm stat}}{\delta\phi(x)}
 =-K\lap\phi+U'(\phi).
 \label{eq:EOMmain}
\end{equation}
Coupling the renormalized static theory to a background metric and varying
under an infinitesimal local translation gives
\begin{equation}
 \partial_jT_{ij}=-\EOM\,\partial_i\phi .
 \label{eq:WardMain}
\end{equation}
Using Eq.~\eqref{eq:EOMmain},
\begin{align}
 K(\lap\phi)\partial_i\phi
 &=
 U'(\phi)\partial_i\phi-\EOM\partial_i\phi
 \nonumber\\
 &=
 \partial_iU+\partial_jT_{ij}.
 \label{eq:currentDecomp}
\end{align}
We write $[\mathcal O]_{\rm quot}$ for the equivalence class of an operator
$\mathcal O$ modulo local chemical-gradient currents.  Therefore
\begin{equation}
 [(\lap\phi)\partial_i\phi]_{\rm quot}
 =
 K^{-1}[\partial_jT_{ij}]_{\rm quot} .
 \label{eq:stressRep}
\end{equation}

The corresponding MSRJD perturbation is
\begin{equation*}
 S_J
 =
 M_0\zeta_J
 \int\dd t\,\dd^dx\,
 (\partial_i\phit)(\lap\phi)\partial_i\phi .
\end{equation*}
Using Eq.~\eqref{eq:stressRep} and integrating by parts,
\begin{equation}
 [S_J]_{\rm quot}
 =
 -\frac{M_0\zeta_J}{K}
 \int\dd t\,\dd^dx\,
 (\partial_i\partial_j\phit)T_{ij}.
 \label{eq:SJstress}
\end{equation}

\subsection{Exact quotient and triangular mixing}

For the symmetric quartic momentum polynomial
\begin{equation*}
 P_4=c_{\kinS^2}\kinS^2+c_{\kinS\kinT}\kinS\kinT+c_{\kinT^2}\kinT^2
 +c_{\kinV}\kinV,
\end{equation*}
the symbol $\Pi_J$ denotes a linear coefficient-extraction functional.  We normalize it by
\begin{equation}
 \Pi_J[P_4]=-\frac12c_{\kinV}.
 \label{eq:PiJ}
\end{equation}
Restricted to the conserving quartic vertex space, it extracts the coefficient
of the current quotient representative $V_J$ modulo the chemical subspace:
$\Pi_J[V_J]=1$, while $\Pi_J[V_\Lambda]=\Pi_J[V_\Xi]=0$.  Equivalently,
$\Pi_J[Q^2\kinS]=\Pi_J[Q^2\kinT]=0$.  Thus applying $\Pi_J$ to a vertex flow returns
only its current-like quotient component.
Every chemical response vertex carries an explicit $Q^2$.  Locality
prevents a UV counterterm from canceling this factor.  The only alternative
attachment of the external response field is to the additive-noise vertex;
the remaining response line then generates a closed directed retarded
cycle, which vanishes in the It\^o prescription.  Let $g_{\rm chem}$ denote
any local chemical coupling and let
$\beta_J\equiv\partial_\ell\bar\zeta_J$ be the current beta function.  Hence, to
all loop orders at the passive fixed point,
\begin{equation}
 \left.
 \frac{\partial\beta_J}{\partial g_{\rm chem}}
 \right|_{\zeta_J=0}=0 .
 \label{eq:triangularExact}
\end{equation}
The proof is given in Appendix~\ref{app:wardproof}.  Equation~\eqref{eq:triangularExact} states only that chemical couplings do not feed into $\beta_J$. The reverse mixing is allowed: a bare current perturbation does generate a chemical component. In the ordered basis $(a_{\rm ch},a_J)$ the quotient stability matrix is therefore upper triangular, with a vanishing $(2,1)$ entry.

\subsection{Two-loop test of the current contact anomaly}

Once the renormalized two-point vertex is known, the stress representation fixes the fully contracted current/stress insertion.  The insertion then cancels against the ordinary field counterterm.  This cancellation is kinematics forced by the Ward identity, not by itself a new dynamical result.  A coincident
dynamic product can nevertheless possess an additional local counterterm.
For the $4-\eps$ expansion set $\eps\equiv4-d$ and use the dimensionless
renormalized quartic coupling $\widetilde g_4$ in the minimal-subtraction
scheme, defined by $\widetilde{\lambda}=8\pi^2\widetilde g_4$, where
$\widetilde{\lambda}$ is the conventional static $\phi^4$ coupling.
The tilde distinguishes this perturbative normalization from the FRG
vertex $g_{4,*}=3u_*''(0)$ used in the active loop integrals; it does not
assert a direct numerical conversion between the two schemes.
Its fixed-point value is denoted by $\widetilde g_{4,*}$.
We denote the
dimensional-regularization scale by $\mu_R$ and by $b_J$ the coefficient of
the possible two-loop simple pole in the additional current factor:
\begin{equation*}
 Z_{J,\rm ct}
 =
 1+\frac{b_J\widetilde g_4^2}{\eps}+\Order(\widetilde g_4^3).
\end{equation*}
We define the infrared-RG contact contribution by
$\gamma_{J,\rm ct}\equiv-\mu_R\dd\ln Z_{J,\rm ct}/\dd\mu_R$ at the
Wilson--Fisher fixed point and define $c_{J,\rm ct}$ as its order-$\eps^2$
coefficient:
\begin{equation*}
 \gamma_{J,\rm ct}=c_{J,\rm ct}\eps^2+\Order(\eps^3).
\end{equation*}
Appendix~\ref{app:twoloop} evaluates all two-loop 1PI graphs linear in
$\zeta_J$ and quadratic in the Wilson--Fisher coupling.  The dynamical
time-ordering sums reduce topology by topology to the corresponding static
stress-tensor graphs.  The only nontrivial momentum-dependent static contribution at two-loop
order is given by a three-propagator integral.  We write
$\int_q\equiv\int\dd^dq/(2\pi)^d$, and analogously for $\int_r$; $p$ is the
external spatial momentum.  Then
\begin{equation*}
 I(p)
 =
 \mu_R^{2\eps}\int_q\int_r
 \frac{1}{q^2r^2(p-q-r)^2},
\end{equation*}
whose pole is
\begin{equation*}
 I(p)\big|_{1/\eps}
 =
 -\frac{p^2}{2(16\pi^2)^2}\frac1\eps .
\end{equation*}
With $\widetilde{\lambda}=8\pi^2\widetilde g_4$, the corresponding 1PI self-energy pole is
\begin{equation*}
 \Sigma^{(2)}(p)\big|_{1/\eps}
 =
 \frac{\widetilde g_4^2}{48\eps}p^2.
\end{equation*}
The translation Ward identity converts this into a current/stress insertion
pole $+(\widetilde g_4^2/48\eps)V_J$.  In the perturbative normalization used here,
the stiffness factor $K^{-1}$ in Eq.~\eqref{eq:stressRep} becomes the
ordinary inverse field renormalization $Z_\phi^{-1}$ after canonical
normalization, with $\phi=Z_\phi^{1/2}\phi_0$ and $\phi_0$ the bare field, and supplies the opposite counterterm.  The nontrivial
dynamical statement is instead that the causal frequency/time-ordering sums
reduce both allowed two-loop topologies to their static partners with unit
weights, $W_{\rm serial}=W_{\rm triangle}=1$, and that no further local
quotient counterterm survives.  After ordinary subdivergence subtraction,
\begin{equation}
 b_J=0,\qquad c_{J,\rm ct}=0\quad\text{at two loops}.
 \label{eq:cJzero}
\end{equation}

Thus the current exponent through the first nontrivial anomalous order is
\begin{equation}
 y_J
 =
 \frac{2-d}{2}-\frac{3\eta}{2}
 +\Order(\eps^3).
 \label{eq:yJ2loop}
\end{equation}
This is the current counterpart of Eq.~\eqref{eq:yThetaCounting}.  The
insertion
$\mathcal O_J=(\partial_i\phit)(\lap\phi)\partial_i\phi$ has the
dimension-counting eigenvalue
\begin{align}
 y_J^{(0)}
 &=d+z-\big(\Delta_{\phit}+2\Dphi+4\big)
 \nonumber\\
 &=-\Dphi-\eta
 =\frac{2-d}{2}-\frac{3\eta}{2}.
 \label{eq:yJCounting}
\end{align}
The two results therefore have the same structural meaning: no additional
contact anomaly beyond equilibrium field scaling.  Their difference
$y_J^{(0)}-y_\Theta^{(0)}=-\eta$ follows from dimension counting alone.
For $\mathcal O_J$ the two-loop calculation tests this statement explicitly;
for $\mathcal O_\Theta$ it remains a truncation result.
For the one-component Ising theory,
\begin{equation*}
 \eta_*=\frac{\eps^2}{54}+\Order(\eps^3),
\end{equation*}
so
\begin{equation*}
 y_J
 =
 -1+\frac{\eps}{2}
 -\frac{\eps^2}{36}
 +\Order(\eps^3).
\end{equation*}

\section{Three-dimensional nonequilibrium spectrum}
\label{sec:d3}

We now combine the local and gradient sectors to obtain the three retained
odd correction exponents in $d=3$.  We first report the internally
consistent FRG result, then use the operator relations to separate numerical
precision from the physical assumptions entering crossover predictions.
We solve the passive fixed point of the \LPAA{} truncation self-consistently at $N_{\rm poly}=10$ for both smooth regulators.  We denote the correlation-length exponent by $\nu$
and the leading equilibrium correction-to-scaling exponent by $\omega_{\rm eq}$.  For
the fixed-point potential $\bar u_*(\vphi)=u_*(\vphi^2/2)$ introduced above,
we record
the two passive quantities that enter the active loop integrals, namely the
curvature at the symmetric point and the corresponding four-point vertex,
\begin{align}
 w_{0,*}&\equiv\partial_\vphi^2\bar u_*\big|_{\vphi=0}=u_*'(0),\nonumber\\
 g_{4,*}&\equiv\partial_\vphi^4\bar u_*\big|_{\vphi=0}=3u_*''(0),
 \label{eq:mstaru4star}
\end{align}
where primes on $u_*(\rho)$ denote derivatives with respect to $\rho$.
The quantities entering the active loop projection are
listed in Table~\ref{tab:passive}.

\begin{table}[b]
\caption{Passive fixed-point input in $d=3$.  The digits are quoted for
reproducibility of the flow, not as an estimate of physical accuracy.  The
quoted order lies on the plateau of the polynomial sequence, which is scanned
to $N_{\rm poly}=12$ in Appendix~\ref{app:numerics}.}
\label{tab:passive}
\begin{ruledtabular}
\begin{tabular}{lcc}
 & exponential & Gaussian\\
\colrule
$\eta_*$ & 0.04313150 & 0.04171172\\
$\nu$ & 0.64360760 & 0.64122385\\
$\omega_{\rm eq}$ & 0.70339711 & 0.70736409\\
$w_{0,*}$ & -0.30210395 & -0.23755767\\
$g_{4,*}$ & 17.77014338 & 15.93215480
\end{tabular}
\end{ruledtabular}
\end{table}

Superscripts $E$ and $G$ denote the exponential and Gaussian regulators,
respectively.  The estimate $\eta_*^{E}=0.04313$ is about $19\%$ above the
conformal-bootstrap reference value
$\eta_{\rm ref}=0.0362978(20)$~\cite{Kos2016}.
Independent Monte Carlo results support this reference
estimate~\cite{Hasenbusch2010}.  The offset is a truncation effect.  The passive ansatz keeps a field-independent
stiffness and no momentum dependence beyond $q^2$ in the two-point
function.  In addition, the transverse $O(N)$ projection is analytically
continued to $N=1$, where there is no physical transverse mode.
Both choices can affect $\eta$; the present calculation does not quantify
their separate contributions.  Higher derivative expansions improve
Ising estimates~\cite{Balog2019}, but regulator variation within the
present ansatz cannot determine its full systematic error.

The consequence for the present work is direct.  The splitting of
the two slowest modes \emph{is} $\eta$, by Eq.~\eqref{eq:yJCounting}, so the
truncation's own prediction for that splitting inherits the same $19\%$ error.
Improving the passive estimate requires an enlarged derivative expansion
and a reassessment of the stiffness projection.  Such a calculation would
leave the exact conservation and translation identities intact, while
also allowing the truncation-dependent operator statements to be tested.  The role of the
truncation here is to establish the structure of the active stability matrix
and to supply the one exponent that is not fixed by that structure; quantitative crossover estimates can instead use
$\eta_{\rm ref}$ from Ref.~\cite{Kos2016}.
The tabled active matrices retain the self-consistent FRG fixed point.
Separately, evaluating our operator relations with $\eta_{\rm ref}$
gives the conditional reference-informed estimates
\begin{align}
 y_{\Theta_1,\rm ref}
 &=-\frac{1+\eta_{\rm ref}}{2}
 \simeq-0.518149,
 \nonumber\\
 y_{J,\rm ref}
 &=-\frac{1+3\eta_{\rm ref}}{2}
 \simeq-0.554447.
 \label{eq:referenceImproved}
\end{align}
The corresponding inverse splitting is
$\eta_{\rm ref}^{-1}\simeq27.5$.
These are conditional on the absence of a further contact anomaly in either
sector, in the sense made precise in Sec.~\ref{sec:discussion}; the six digits
propagate the precision of $\eta_{\rm ref}$ and are not a claim about the
operator calculation itself.  Granting the condition, they are the appropriate
numbers for the physical crossover estimate;
the truncation pair remains useful for assessing internal regulator and
truncation consistency.

For the complete \(T1+T2\) projection we evaluate the loop integrals by tensor-product Gauss--Legendre quadrature, with radial order \(n_r=160\) and angular order \(q_{\rm ang}=28\).  The passive fixed-point potential is truncated at \(N_{\rm poly}=10\).
The external momenta are parameterized as
$$ p_1=\delta_p\,v_1,\qquad p_2=\delta_p\,v_2, $$
where \(\delta_p\) controls their common scale, while the fixed dimensionless vectors \(v_1\) and \(v_2\) specify the momentum configuration. Several generic choices of \((v_1,v_2)\) are used to reconstruct the quartic momentum polynomial (see Appendix~\ref{app:numerics}).
For each configuration, the loop result is fitted over \(0<\delta_p\le \delta_{p,\max}\). The \(\delta_p\to0\) limit is estimated by averaging the stable plateau obtained for
\( \delta_{p,\max}=0.010,\;0.012,\;0.014, \)
as described in Appendix~\ref{app:numerics}.  The raw active matrices are
\begin{equation}
 \mathsf M_{\rm act}^{E}
 =
 \begin{pmatrix}
 -1.381127&-0.106297&-0.081643\\
 -0.970217&-1.108613&-1.388838\\
 0&0&-0.564697
 \end{pmatrix},
 \label{eq:MEraw}
\end{equation}
and
\begin{equation}
 \mathsf M_{\rm act}^{G}
 =
 \begin{pmatrix}
 -1.368539&-0.065109&-0.080597\\
 -0.951150&-1.023181&-1.314625\\
 0&0&-0.562568
 \end{pmatrix}.
 \label{eq:MGraw}
\end{equation}
The exact triangularity theorem of Sec.~\ref{sec:ward} requires only the two
off-diagonal statements
$(\mathsf M_{\rm act})_{31}=(\mathsf M_{\rm act})_{32}=0$: chemical
operators do not generate the current quotient.  The vanishing of the whole
loop correction to the third row, including its diagonal element, is
separate and is also exact at one loop.  It follows from the two
topologies themselves.  In $T1$ the passive vertex carries the external
response leg, so Eq.~\eqref{eq:T1} has an explicit factor $Q^2$ and its
quartic part lies in $\Span\{Q^2\kinS,Q^2\kinT\}$, the chemical span
annihilated by $\Pi_J$.  In $T2$ the integrand of Eq.~\eqref{eq:T2} depends on the external
momenta only through $Q$, so its quartic part can only be proportional to
$Q^4=\kinS^2+4\kinS\kinT+4\kinT^2$, for which $c_{\kinV}=0$.  Hence
$\Delta M_{33}=0$ analytically, which is the one-loop statement
$\gamma_{J,\rm ct}=0$ for the current contact anomaly; the residual quoted in
Appendix~\ref{app:numerics} measures the momentum fit alone.  Accordingly,
the displayed third diagonal entry is only the common canonical/anomalous
contribution $y_{\rm grad}^{(0)}$; it is not implied by triangularity alone.
A further internal check is that the equilibrium tangent
$\mathbf e_{\rm eq}=(1,2,0)$ is left invariant by
Eqs.~\eqref{eq:MEraw}--\eqref{eq:MGraw} to the quoted precision, which is
what makes the quotient construction meaningful.

In quotient coordinates Eq.~\eqref{eq:introQuot}, the genuine active
blocks are
\begin{align}
 \mathsf M_{\rm NE}^E&=
 \begin{pmatrix}
 -0.896019&-1.225552\\
 0&-0.564697
 \end{pmatrix},
 \label{eq:MQE}\\
 \mathsf M_{\rm NE}^G&=
 \begin{pmatrix}
 -0.892963&-1.153431\\
 0&-0.562568
 \end{pmatrix}.
 \label{eq:MQG}
\end{align}
The chemical eigenvector is $(1,0)$ in $(a_{\rm ch},a_J)$ coordinates.  The
current-like eigenvector is
\begin{equation}
 v_J^{\rm NE}=(\chi,1),\qquad
 \chi=-\frac{(\mathsf M_{\rm NE})_{12}}
 {(\mathsf M_{\rm NE})_{11}-(\mathsf M_{\rm NE})_{22}},
 \label{eq:chi}
\end{equation}
giving
\begin{equation*}
 v_J^{{\rm NE},E}\propto(-3.699,1),\qquad
 v_J^{{\rm NE},G}\propto(-3.491,1).
\end{equation*}
With unit normalization of the current coordinate, the gradient scaling
fields are explicitly
\begin{equation}
 u_J=\aJ,\qquad u_{\rm ch}=\ac-\chi\aJ,
 \qquad\partial_\ell u_i=y_i u_i.
 \label{eq:gradientEigenfields}
\end{equation}
Thus $\ac=u_{\rm ch}+\chi u_J$ is a coordinate, rather than an
eigenfield by itself.  A bare current perturbation develops a substantial chemical component
under coarse graining even though its eigenvalue is controlled by the
current diagonal entry.

\begin{figure}[tb]
\centering
\includegraphics[width=\columnwidth]{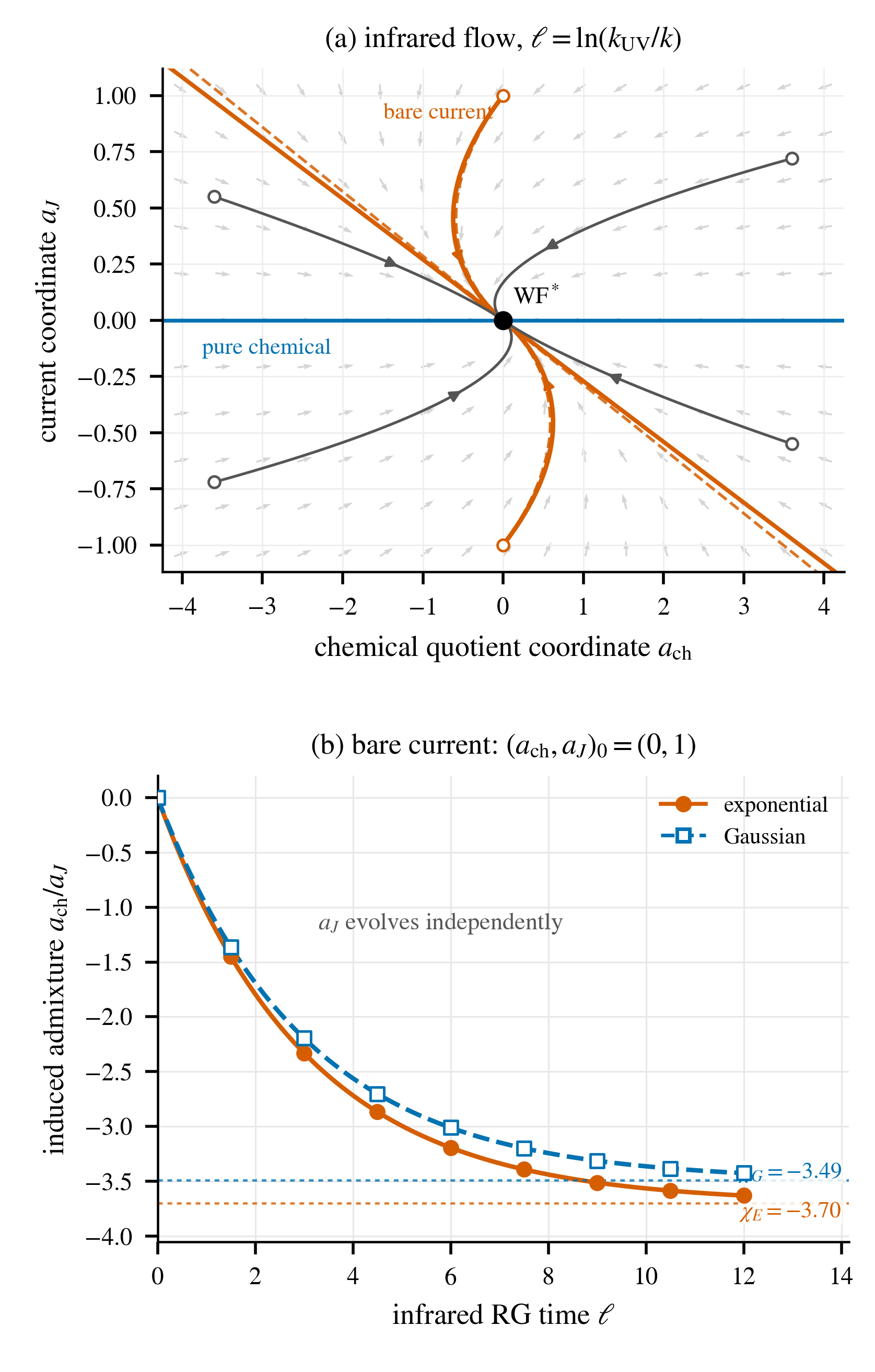}
\caption{Linearized infrared flow in the quotient plane.
(a) The chemical axis $a_J=0$ is invariant, whereas a current perturbation
induces a chemical component and approaches the slow current eigenline
before reaching WF$^*$.  Arrows point toward increasing infrared RG time
$\ell$.  Solid trajectories use the exponential regulator; dashed curves
show the Gaussian bare-current trajectory and eigenline.
(b) Chemical admixture generated from $(a_{\rm ch},a_J)_{\ell=0}=(0,1)$, approaching
$\chi_E=-3.70$ and $\chi_G=-3.49$.  The one-way generation directly
visualizes the triangular quotient flow.}
\label{fig:quotientFlow}
\end{figure}

The resulting flow in the quotient plane is shown in
Fig.~\ref{fig:quotientFlow}.  The triangular matrix makes the chemical axis
$a_J=0$ invariant, because a chemical perturbation cannot generate the
current coordinate.  The opposite generation, current into chemical, is allowed.  For a unit bare
current perturbation, $(a_{\rm ch},a_J)_{\ell=0}=(0,1)$, the linear solution is
\begin{equation}
 a_J(\ell)=e^{y_J\ell},
 \qquad
 \frac{a_{\rm ch}(\ell)}{a_J(\ell)}
 =\chi\left[1-e^{(y_{\rm ch}-y_J)\ell}\right].
 \label{eq:quotientTrajectory}
\end{equation}
Because $y_J>y_{\rm ch}$, the ratio approaches $\chi$ while both coordinates
decay to zero.  Thus a trajectory starting on the bare $a_J$ axis bends
toward the current eigenline $a_{\rm ch}=\chi a_J$ before reaching the passive fixed
point.  The close exponential and Gaussian trajectories show that this
geometric interpretation is insensitive to the regulator choice.

For completeness, the equilibrium tangent itself is an exact eigenvector of
the raw matrix,
\begin{equation}
 y_{\rm eq}^{E}=-1.593722,\qquad
 y_{\rm eq}^{G}=-1.498756.
 \label{eq:yeqDiagnostic}
\end{equation}
It is removed by the quotient and is therefore not a physical active
scaling field.  Its regulator spread is nevertheless a useful truncation
diagnostic: the relative exponential--Gaussian difference is $6.14\%$,
more than an order of magnitude larger than for the quotient eigenvalues.

\begin{figure}[!htbp]
\centering
\includegraphics[width=\columnwidth]{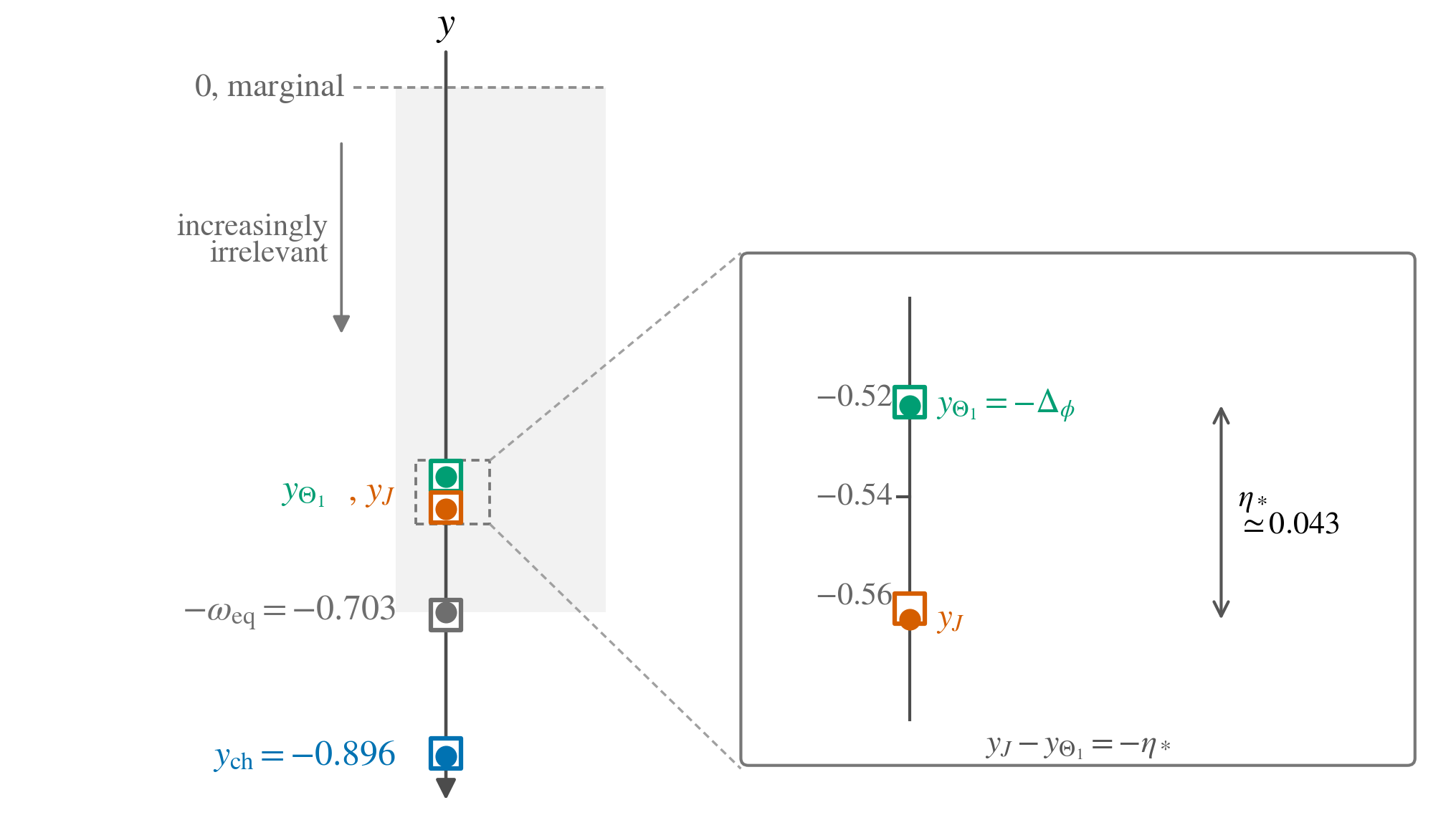}
\caption{Three-dimensional correction-to-scaling spectrum for the
exponential (filled circles) and Gaussian (open squares) regulators.  The
shaded band $-\omega_{\rm eq}^{E}<y<0$ contains irrelevant
perturbations that decay more slowly than the leading equilibrium
correction for the exponential regulator, using Table~\ref{tab:passive}.  The inset resolves the
near-degenerate pair $y_{\Theta_1}$ and $y_J$, whose separation is
$y_J-y_{\Theta_1}=-\eta_*$.}
\label{fig:spectrum}
\end{figure}

Table~\ref{tab:spectrum} collects the three odd nonequilibrium modes retained
in this analysis. Both regulators give the same ordering, 
\begin{equation}
\yth>\yJ>\ych,\qquad \yth,\yJ,\ych<0, 
\end{equation} 
so that all three perturbations are irrelevant.
\begin{table}[t]
\caption{Leading odd nonequilibrium exponents in $d=3$.}
\label{tab:spectrum}
\begin{ruledtabular}
\begin{tabular}{lcc}
 & exponential & Gaussian\\
\colrule
$y_{\Theta_1}$ & $-0.521566$ & $-0.520856$\\
$y_J$ & $-0.564697$ & $-0.562568$\\
$y_{\rm ch}$ & $-0.896019$ & $-0.892963$
\end{tabular}
\end{ruledtabular}
\end{table}
The spectrum is displayed in Fig.~\ref{fig:spectrum}; the two slowest modes
lie above the leading equilibrium correction $-\omega_{\rm eq}$ and are separated only
by $\eta_*$.
Within the common \LPAA{} truncation,
\begin{equation}
 \yJ-\yth=-\eta_* .
 \label{eq:splitD3}
\end{equation}
At fixed regulator and $N_{\rm poly}=10$, the entries of
Table~\ref{tab:spectrum} have numerical variations of order a few $10^{-7}$;
this is not their physical accuracy.  Among the checks performed at fixed ansatz, the regulator spread dominates; it exceeds the numerical errors controlled in Appendix~\ref{app:numerics}.  For $y_{\rm ch}$ the range over $N_{\rm poly}=10,11,12$ is below $2.4\times10^{-6}$, the momentum window contributes $6\times10^{-8}$, and the radial quadrature $1.7\times10^{-7}$.  The regulator spread is $3.1\times10^{-3}$.  Quoting the mean of the two regulators with half their
difference gives the two-regulator spread estimates
\begin{align}
 y_{\Theta_1}&=-0.5212(4), &
 y_J&=-0.5636(11),
 \nonumber\\
 y_{\rm ch}&=-0.8945(15). &&
 \label{eq:physicalValues}
\end{align}
For the first two this merely restates the spread of $\eta_*$, because they
follow from it through the operator relations Eqs.~\eqref{eq:ytheta} and
\eqref{eq:yJCounting}.  Those relations are exact dimension counts, but each presupposes that no further contact anomaly is present.  For $\mathcal O_\Theta$ this absence is a result of the present truncation; for $\mathcal O_J$ it is verified only through two loops.  Only $y_{\rm ch}$ is an independent numerical output.  Equation~\eqref{eq:yeqDiagnostic} shows why the
equilibrium tangent must not be used as an equally precise physical output.
The stability of the eigenvalues under increasing polynomial order is
quantified in Appendix~\ref{app:numerics} and
Fig.~\ref{fig:numericalConvergence}(a).  The parentheses in
Eq.~\eqref{eq:physicalValues} are a two-regulator spread estimate and not a
complete systematic error: they measure the sensitivity to the regulator
profile at fixed truncation, and do not bound the error of the
\LPAA{} ansatz itself, which is not accessible from within it.  This caveat
matters most for $y_{\rm ch}$, the only entry that is not tied to $\eta_*$ by
an operator relation.

\section{Finite-size scaling and long crossovers}
\label{sec:fss}

The spectrum tells us which perturbations decay slowly.  What a simulation measures depends in addition on which eigenfields can contribute to the chosen observable.  Ising parity supplies that selection rule, and it gives different finite-size signatures in odd and in even observables.

Let $\tau$ denote the thermal scaling field, $h$ the ordering-field scaling
field, and $u_\omega$ the leading equilibrium irrelevant scaling field in
the sense of corrections to scaling~\cite{Wegner1972}.  We
denote their RG eigenvalues by
\begin{equation*}
 y_t=\frac1\nu,\qquad
 y_h=\frac{d+2-\eta}{2},\qquad
 y_\omega=-\omega_{\rm eq}.
\end{equation*}
These are generic scaling relations, without a specified numerical input. The subscript $\rm ref$ identifies literature inputs and estimates derived from them: $\eta_{\rm ref}$ is taken from Ref.~\cite{Kos2016}, and
$\omega_{\rm eq,ref}=0.8303(18)$ from Ref.~\cite{ElShowk2014}.
The active estimates $y_{\Theta_1,\rm ref}$ and $y_{J,\rm ref}$ are defined in Eq.~\eqref{eq:referenceImproved}.
Superscripts $E$ and $G$ instead identify our regulator-specific FRG results.
The three odd nonequilibrium scaling fields are
\begin{equation*}
 u_\Theta,\qquad u_J,\qquad u_{\rm ch}.
\end{equation*}
Here $u_\Theta$ is the amplitude of the normalized local mode
$\vartheta_1(\vphi)=\vphi$, including its induced potential component;
$u_J,u_{\rm ch}$ are the eigenfields of Eq.~\eqref{eq:gradientEigenfields}.
All amplitudes in this section are specified at the matching scale.
Their eigenvalues are $y_\Theta\equiv y_{\Theta_1}$, $y_J$, and
$y_{\rm ch}$, respectively.  Let $L$ denote the linear system size measured in units of a fixed
microscopic matching length $L_0$, and let $R$ be a dimensionless observable.
The finite-size scaling function $\mathcal S_R$ of $R$ is defined by
\begin{align}
 R={}&
 \mathcal S_R\Big(
 \tau L^{y_t},hL^{y_h},u_\omega L^{-\omega_{\rm eq}},
 \nonumber\\
 &\hspace{1.8cm}
 u_\Theta L^{y_\Theta},
 u_JL^{y_J},
 u_{\rm ch}L^{y_{\rm ch}},
 \ldots
 \Big).
 \label{eq:FSSmaster}
\end{align}
The emergent Ising transformation, $\phi\to -\phi$, makes
$h,u_\Theta,u_J,u_{\rm ch}$ odd and $\tau,u_\omega$ even.  Because
$u_\Theta$ is odd, the potential admixture generated by the triangular
structure of Eq.~\eqref{eq:triangularTheta} is itself odd and shifts the
tuning of $h$; the expressions below refer, as usual, to the tuned critical
point at which the relevant odd direction has been set to zero.

Let $R_{\rm odd}$ be an observable odd under the emergent Ising symmetry.
At $\tau=h=0$, denote the corresponding nonuniversal linear amplitudes by
$A_\Theta,A_J$, and $A_{\rm ch}$.  Its leading expansion is
\begin{equation}
 R_{\rm odd}
 =
 A_\Theta u_\Theta L^{y_\Theta}
 +A_Ju_JL^{y_J}
 +A_{\rm ch}u_{\rm ch}L^{y_{\rm ch}}
 +\cdots .
 \label{eq:FSSodd}
\end{equation}
For an even dimensionless observable, let $R^*$ be its fixed-point value,
$c_\omega$ the leading equilibrium correction amplitude, and $B_{ij}$ the
quadratic nonequilibrium amplitudes, taken symmetric in $i,j$.  Then
\begin{align}
 R_{\rm even}&-R^*
 =
 c_\omega u_\omega L^{-\omega_{\rm eq}}
 +B_{\Theta\Theta}u_\Theta^2L^{2y_\Theta} \nonumber
 \\
 &+
 2B_{\Theta J}u_\Theta u_JL^{y_\Theta+y_J}
 +B_{JJ}u_J^2L^{2y_J}
 +\cdots .
 \label{eq:FSSeven}
\end{align}
Thus an odd nonequilibrium perturbation contributes only at second order to
an even dimensionless observable.  Generically its leading active term is
$B_{\Theta\Theta}u_\Theta^2L^{2y_\Theta}$, provided this amplitude is
nonzero.  The mixed $\Theta J$ term becomes leading only if the slower
contribution is absent or suppressed.  Using Eq.~\eqref{eq:referenceImproved}, based on
$\eta_{\rm ref}$ from Ref.~\cite{Kos2016}, the three powers are
$2y_{\Theta_1,\rm ref}\simeq-1.036298$,
$y_{\Theta_1,\rm ref}+y_{J,\rm ref}\simeq-1.072596$, and
$2y_{J,\rm ref}\simeq-1.108894$.

In an odd observable at $h=0$, a pure term
$u_\omega L^{-\omega_{\rm eq}}$ is forbidden by parity.  The leading
equilibrium irrelevant field instead dresses the odd terms:
\begin{equation}
 \delta R_{\rm odd}
 =\sum_{i=\Theta,J,\rm ch} C_i u_i u_\omega
 L^{y_i-\omega_{\rm eq}}+\cdots .
 \label{eq:oddWegnerDressing}
\end{equation}
Here $C_i$ are observable-dependent amplitudes.  Comparing an odd active
power directly with $L^{-\omega_{\rm eq}}$ is therefore a comparison
between channels, not between two allowed terms of one odd observable.

For fixed-density MIPS simulations the spatially uniform density mode is
constrained, so Eq.~\eqref{eq:FSSodd} should be tested with a block or
subvolume order parameter rather than the global density.  Let $m_b$ denote
the centered, field-mixed density in a block of linear size $L_b$ (the
mixing removes the leading liquid--gas asymmetry in the passive reference
distribution~\cite{BruceWilding1992}).  Let $P_{L_b}(m_b)$ be the probability density of the block variable, $\sigma_b=\sqrt{\avg{m_b^2}}$ its standard deviation, $x=m_b/\sigma_b$, and $\widehat P_b(x)=\sigma_b P_{L_b}(\sigma_b x)$ the standardized density.  Three directly measurable odd observables are the block skewness:
\begin{equation}
 S_3(L_b)=
 \frac{\avg{m_b^3}}{\avg{m_b^2}^{3/2}},
 \label{eq:blockSkewness}    
\end{equation}
odd part of the standardized block-density histogram:
 \begin{equation}
      \widehat P_{\rm odd}(x;L_b)=
 \frac{\widehat P_b(x)-\widehat P_b(-x)}{2},
 \label{eq:oddHistogram}
 \end{equation}
 and the normalized fifth moment:
\begin{equation}
    R_5(L_b)=
 \frac{\avg{m_b^5}}{\avg{m_b^2}^{5/2}}.
 \label{eq:fifthMoment} 
\end{equation}
For a block inside a box of size $L_{\rm box}$, the scaling function
also depends on $L_b/L_{\rm box}$.  One should either work in the
mesoscopic regime $1\ll L_b\ll L_{\rm box}$ or compare a family of
systems at fixed block-to-box ratio.  Standardizing the histogram makes
its odd part dimensionless and suitable for the same parity expansion.
Several block sizes provide sensitivity to the powers in
Eq.~\eqref{eq:FSSodd} while respecting global conservation, as in
block-based protocols for critical active
particles~\cite{Siebert2018,PartridgeLee2019,Maggi2021}.
The small separation of the two slow exponents nevertheless makes their
individual extraction ill-conditioned: joint fits to several observables
or independent constraints on amplitudes are needed to distinguish them.

The near degeneracy of $u_\Theta$ and $u_J$ controls what a finite-size fit can resolve.  For
\begin{equation*}
 R_{\rm odd}(L)=
 A_\Theta u_\Theta L^{y_\Theta}
 +A_Ju_JL^{y_J},
\end{equation*}
equal magnitudes occur at
\begin{equation}
 L_\times
 =
 \left|
 \frac{A_Ju_J}{A_\Theta u_\Theta}
 \right|^{1/(y_\Theta-y_J)}.
 \label{eq:Lcross}
\end{equation}
Here $L_\times$ is dimensionless; the physical crossover length is
$L_0L_\times$.  Within the truncation relation
$y_\Theta-y_J=\eta_*$,
\begin{equation}
 L_\times
 =
 \left|
 \frac{A_Ju_J}{A_\Theta u_\Theta}
 \right|^{1/\eta_*}.
 \label{eq:LcrossEta}
\end{equation}
For the conditional estimates in Eq.~\eqref{eq:referenceImproved},
the splitting is instead
$y_{\Theta_1,\rm ref}-y_{J,\rm ref}=\eta_{\rm ref}$.
Using $\eta_{\rm ref}=0.0362978(20)$~\cite{Kos2016}
gives $1/\eta_{\rm ref}\simeq27.5$.
Define the signed matching-scale ratio
$\mathcal A_0=A_Ju_J/(A_\Theta u_\Theta)$ and its magnitude
$r_A=|\mathcal A_0|$.  A crossover above the matching scale requires
\begin{equation}
 r_A>1.
 \label{eq:crossoverCondition}
\end{equation}
This condition alone does not make the crossover inaccessible.
Using the conditional splitting $\eta_{\rm ref}$~\cite{Kos2016},
$r_A=1.1$, $1.2$, and $2$ give
$L_\times\simeq14$, $152$, and $2.0\times10^8$, respectively.
With the same reference input, the crossover exceeds the largest
available size $L_{\max}$ only if
$r_A>L_{\max}^{\eta_{\rm ref}}$.
If $r_A\le1$, the formal crossing lies at or below the matching scale.
Opposite signs produce a zero of the two-term observable at equal
magnitudes; its logarithmic derivative has a pole there, not a new
asymptotic power law.

We characterize a local apparent power law by the effective exponent
\begin{equation*}
 y_{\rm eff}(L)
 =
 \frac{\dd\ln|R_{\rm odd}|}{\dd\ln L}.
\end{equation*}
Away from a zero of $R_{\rm odd}$,
\begin{equation}
 y_{\rm eff}
 =
 \frac{y_\Theta+y_J\mathcal A(L)}{1+\mathcal A(L)},
 \qquad
 \mathcal A(L)=\mathcal A_0 L^{y_J-y_\Theta}.
 \label{eq:yeff}
\end{equation}
A broad apparent single-power regime therefore need not identify the true asymptotic exponent: the fitted slope is an amplitude-weighted blend of $y_\Theta$ and $y_J$.

\begin{figure}[tb]
\centering
\includegraphics[width=\columnwidth]{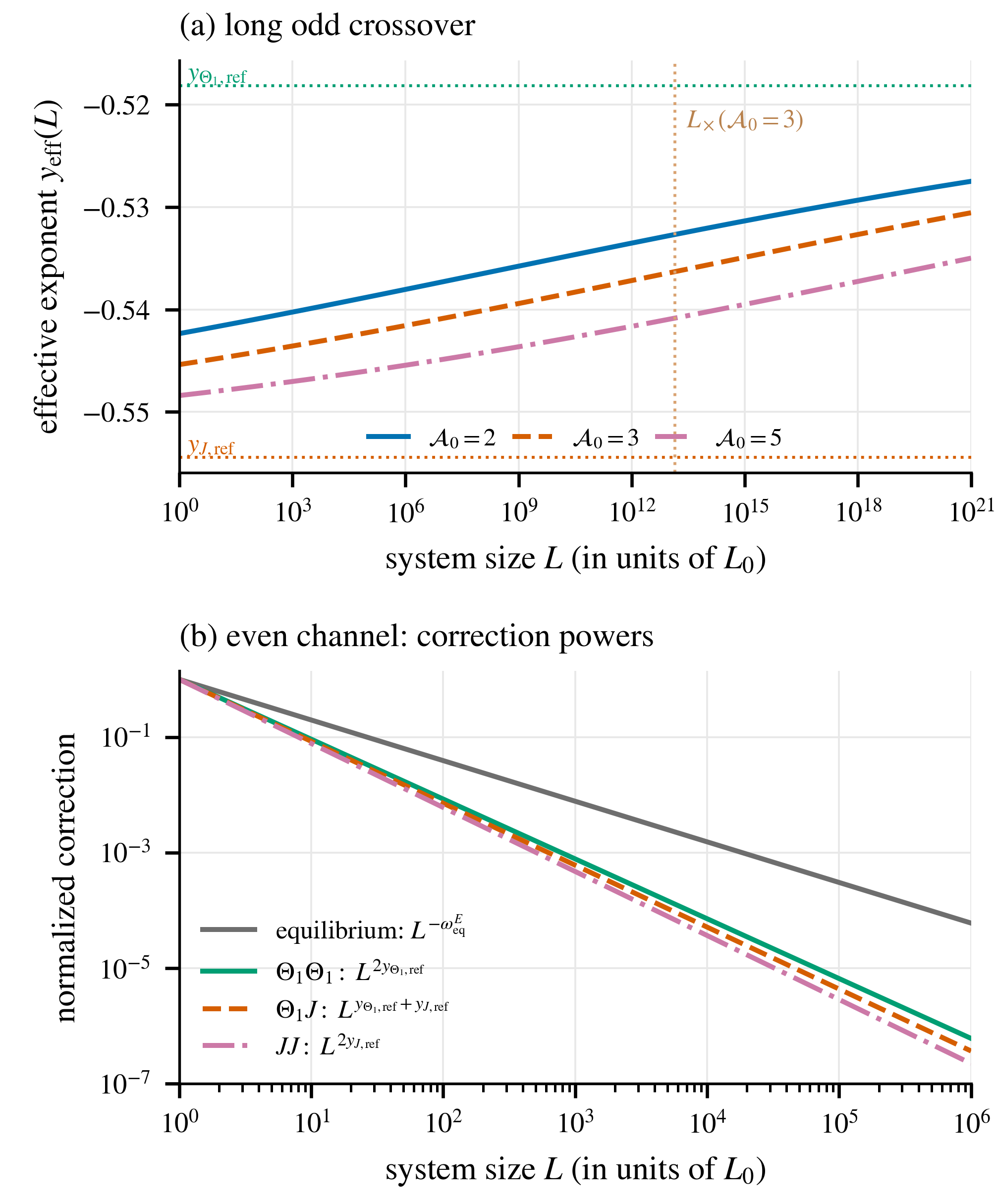}
\caption{Long crossover caused by the near-degenerate local and
current-like odd scaling fields.
The active powers use the conditional estimates
$y_{\Theta_1,\rm ref}$ and $y_{J,\rm ref}$ of
Eq.~\eqref{eq:referenceImproved}, obtained from
$\eta_{\rm ref}$~\cite{Kos2016}.
(a) Effective exponent of Eq.~\eqref{eq:yeff} for three initial amplitude
ratios $\mathcal A_0=A_Ju_J/(A_\Theta u_\Theta)$.  For $\mathcal A_0=3$ the two terms become
equal only near $L_\times\simeq1.4\times10^{13}$, so the measured slope is
an amplitude-dependent interpolation between
$y_{J,\rm ref}$ and $y_{\Theta_1,\rm ref}$ over
the entire accessible range.  Ratios $|\mathcal A_0|\le1$ do not produce a
crossover above the matching scale.
(b) For an even observable the nonequilibrium corrections enter
quadratically, Eq.~\eqref{eq:FSSeven}, and all three decay faster than the
leading equilibrium correction $L^{-\omega_{\rm eq}^{E}}$.
Here $\omega_{\rm eq}^{E}=0.70339711$ is our
exponential-regulator FRG result from Table~\ref{tab:passive},
not the literature reference value $\omega_{\rm eq,ref}$.
The full prefactor of each plotted term, including the factor two
in the mixed contribution, is set to unity.}
\label{fig:crossover}
\end{figure}

Figure~\ref{fig:crossover} illustrates these consequences using $y_{\Theta_1,\rm ref}$ and $y_{J,\rm ref}$ from
Eq.~\eqref{eq:referenceImproved}, based on
$\eta_{\rm ref}$~\cite{Kos2016}.
Panel (a) shows the long, amplitude-dependent interpolation in an odd observable; no crossover occurs above the matching scale when $|\mathcal A_0|\le1$.
Panel (b) compares the three quadratic active corrections with the equilibrium FRG power $L^{-\omega_{\rm eq}^{E}}$
from Table~\ref{tab:passive}.

\subsection{What is accessible at simulation sizes}

\begin{figure}[tb]
\centering
\includegraphics[width=\columnwidth]{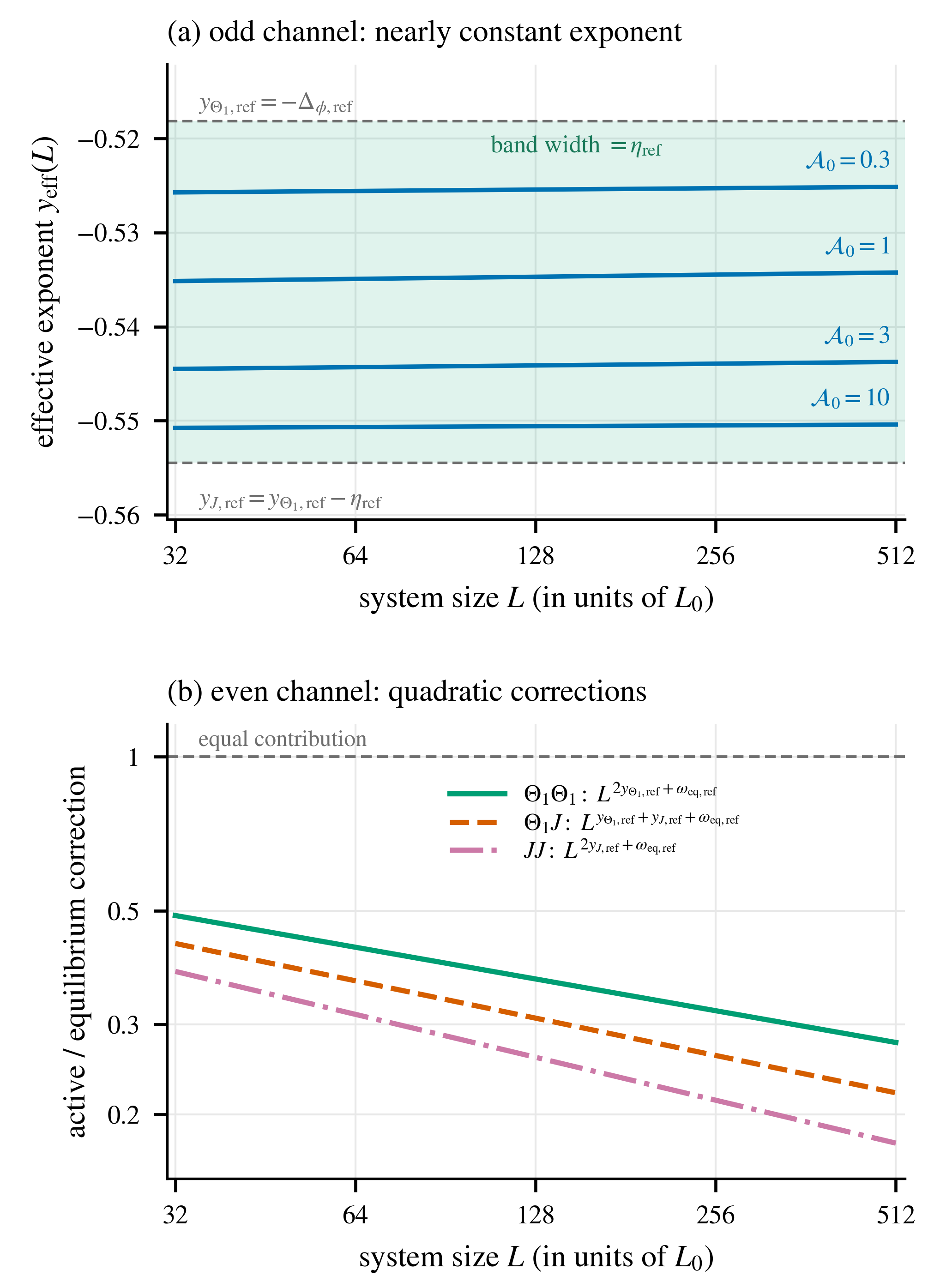}
\caption{Corrections at accessible system sizes.
The curves use the conditional estimates
$y_{\Theta_1,\rm ref}$ and $y_{J,\rm ref}$ of
Eq.~\eqref{eq:referenceImproved}, based on
$\eta_{\rm ref}$~\cite{Kos2016}, and the central reference value $\omega_{\rm eq,ref}=0.8303$~\cite{ElShowk2014}.
Sizes are measured in units of $L_0$.
(a) Effective exponent of Eq.~\eqref{eq:yeff} for four positive amplitude
ratios $\mathcal A_0$.  The curves change by less than $10^{-3}$ between
$L=32$ and $512$, while their offsets reflect the observable-dependent
mixing of two slow odd fields.  The shaded band has width $\eta_{\rm ref}$.
The illustration retains only these two contributions and assumes the
critical tuning fields vanish.
(b) Magnitude of each quadratic active correction in an even observable
divided by the leading equilibrium correction.  The three powers are
$L^{2y_{\Theta_1,\rm ref}+\omega_{\rm eq,ref}}$,
$L^{y_{\Theta_1,\rm ref}+y_{J,\rm ref}+\omega_{\rm eq,ref}}$, and
$L^{2y_{J,\rm ref}+\omega_{\rm eq,ref}}$.
The full prefactor of each term is set to unity, including the factor
two multiplying the mixed term in Eq.~\eqref{eq:FSSeven}.
The $\Theta\Theta$ contribution is generically leading among the active
terms; all three decay faster than the equilibrium power.  The dashed
horizontal line denotes equal individual contributions.}
\label{fig:accessible}
\end{figure}

With the conditional reference splitting
$\eta_{\rm ref}$~\cite{Kos2016}, amplitude ratios such as
$r_A=2$ or $3$ place the crossover far beyond the simulation
sizes considered here.  Finite systems still probe the blended odd exponent.
Assume that the two slow amplitudes have the same sign and dominate over
faster odd corrections throughout the fitted size window.
For nonnegative $\mathcal A(L)$, differentiating
Eq.~\eqref{eq:yeff} gives
\begin{equation}
 \frac{\dd y_{\rm eff}}{\dd\ln L}
 =(y_\Theta-y_J)^2
 \frac{\mathcal A(L)}{[1+\mathcal A(L)]^2}
 \le\frac{(y_\Theta-y_J)^2}{4}.
 \label{eq:driftBound}
\end{equation}
For the conditional reference estimates,
$y_{\Theta_1,\rm ref}-y_{J,\rm ref}=\eta_{\rm ref}$.
The bound per decade is therefore
$(\ln 10)\eta_{\rm ref}^2/4\simeq7.6\times10^{-4}$,
using $\eta_{\rm ref}$ from Ref.~\cite{Kos2016}.
Over $L=32$--$512$, Fig.~\ref{fig:accessible}(a) therefore predicts
little curvature for a two-mode odd signal evaluated with
$y_{\Theta_1,\rm ref}$ and $y_{J,\rm ref}$ from
Eq.~\eqref{eq:referenceImproved}, based on
Ref.~\cite{Kos2016}. A single-power fit can return an amplitude-dependent blend of these exponents rather than either individual asymptotic value. Whether the small drift can be resolved
depends on the simulation's statistical and tuning errors.

For these reference inputs, the two-mode odd effective exponent lies in
$[y_{J,\rm ref},y_{\Theta_1,\rm ref}]$.
Different observables can occupy different positions in this band
because their amplitude ratios differ.
In Fig.~\ref{fig:accessible}(a), evaluated using
Eq.~\eqref{eq:referenceImproved} and Ref.~\cite{Kos2016},
ratios from $0.3$ to $10$ produce offsets of order $0.02$
with little size dependence.  Measurements on
shared configurations can reduce common errors, but do not guarantee
cancellation of observable-dependent corrections or tuning errors.

Parity gives a separate prediction: active corrections are linear in an
odd observable and quadratic in an even one.  Generically the slowest
even active correction scales as $L^{2\yth}$; relative to an equilibrium
correction it carries $L^{2\yth+\omega_{\rm eq}}$.
Figure~\ref{fig:accessible}(b) evaluates these ratios using
$y_{\Theta_1,\rm ref}$ and $y_{J,\rm ref}$ from
Eq.~\eqref{eq:referenceImproved}, based on Ref.~\cite{Kos2016},
together with the central reference value
$\omega_{\rm eq,ref}=0.8303$~\cite{ElShowk2014}.
The full term prefactors are equal.
At $L=128$ their individual ratios are approximately $0.37$,
$0.31$, and $0.26$ for $\Theta\Theta$, $\Theta J$, and $JJ$,
respectively.  The $\Theta\Theta$ term is generically the slowest active
contribution; the mixed term is not singled out as leading.
These curves compare terms within the even channel.  In the tuned odd
channel there is no independent pure Wegner contribution; the allowed
equilibrium dressing is Eq.~\eqref{eq:oddWegnerDressing}.
The plotted ratios refer to individual magnitudes, not their sum.
Nonuniversal amplitudes and signs can change their relative weights and
produce cancellations; they are not fixed by the present calculation.

\section{Discussion and conclusions}
\label{sec:discussion}

Our principal result is a unified classification of the leading nonequilibrium corrections around the conserved Wilson--Fisher fixed point in three dimensions.  It explains how observables can display long-lived, amplitude-dependent effective powers while the flow still approaches passive Ising criticality.  A field-dependent noise-to-mobility ratio and
gradient activity are physically different operators and must not be
collapsed into one ``activity'' coupling.  Nevertheless their two slowest
odd scaling fields are separated only by the small Ising anomalous
dimension.

The status of these results is as follows.  Conservation fixes
the closure identity and the complete three-dimensional local $p^4$
vertex basis.  Conservation, locality, and It\^o causality also forbid
chemical-to-current mixing to all loop orders under the assumptions of
Sec.~\ref{sec:ward}.  By contrast, the block-triangular local-transport
flow and $y_{\Theta_1}=-\Dphi$ are established within the present
functional projection.  The absence of an additional current contact
counterterm is verified perturbatively through two loops.  The relation
$y_J-y_{\Theta_1}=-\eta$ beyond these calculations is conditional on
both contact anomalies vanishing; it is not an all-orders theorem.
Within the truncation the two slow exponents follow from $\eta_*$,
whereas $y_{\rm ch}$ is an independent numerical result.  The
two-dimensional nonequilibrium quotient retains this chemical mode and
the current mode; the one-dimensional current quotient used in the Ward
proof removes the whole chemical plane.

This focus distinguishes the present work from the recent analyses in
Refs.~\cite{PapanikolaouSpeck2024,FejosSzepYamamoto2026}.  Those works map
nonlinear Active-Model-B+ flows, operator generation, and active fixed
points, including regimes far from the passive basin.  Here we instead
resolve the local correction-to-scaling spectrum at the passive fixed point,
add the independent local-transport sector, quotient the full
$(\Lambda,\Xi,\zeta_J)$ gradient basis by its detailed-balance-preserving
tangent, and connect the resulting eigenoperators to observables.  The two
programs are complementary: the global-flow studies determine where active
trajectories may go, while the present spectrum determines how passive
criticality is approached when it is the infrared endpoint.

The local transport sector is controlled by an operator that differs from the fixed-$\eta$ potential operator only by the constant shift $d$.  Two features
make the leading eigenvalue unusually robust for a truncation result: the
eigenfunction $\vartheta_1\propto\vphi$ has vanishing second derivative and
therefore does not probe the threshold function at all, and the coupled
$(\delta\bar u,\vartheta)$ system is triangular, so the feedback of $\vartheta$
into the potential changes the scaling field but not its eigenvalue.
Equivalently, $\mathcal O_\Theta=\phi(\grad\phit)^2$ has the
dimension-counting value $y_\Theta^{(0)}=-\Delta_\phi$ and no additional
contact anomaly in the truncation.  The gradient sector requires the
complete active basis and a quotient by the equilibrium direction.  Its
current-like eigenvector is strongly mixed with the chemical direction, even
though its eigenvalue is fixed by the simpler operator
$\mathcal O_J=(\partial_i\phit)(\lap\phi)\partial_i\phi$.

Translation invariance represents $\mathcal O_J$ exactly by the stress tensor, modulo chemical gradients.  Conservation and causality then forbid the chemical sector from mixing into the current quotient.  Triangularity fixes only the off-diagonal zeros; the diagonal entry is fixed separately by the one-loop result $\gamma_{J,\rm ct}=0$.  At two loops
the nontrivial calculation is the causal dynamic-to-static reduction
$W_{\rm serial}=W_{\rm triangle}=1$.  Once it holds, the Ward-determined
stress pole must cancel against $Z_\phi^{-1}$, and the remaining result
$b_J=c_{J,\rm ct}=0$ states that no extra local quotient pole exists.  Such explicit
tests matter: driven diffusive systems can evade equilibrium expectations
\cite{SchmittmannZia1995}, and superficially exact nonrenormalization claims
in related conserved growth theories have required two-loop correction
\cite{Janssen1997}.
The $d=3$ functional estimates and the $4-\eps$ test are therefore not
combined into a single flow: the former quantify the spectrum within the
effective-average-action truncation, whereas the latter tests the
regulator-free current operator for an additional contact anomaly.

Numerically, the quotient modes are stable across the two regulators while
the removed equilibrium tangent varies by about $6\%$.  This contrast is a
useful truncation diagnostic.  The different background treatments are also
now explicit: the local sector resolves the full fixed function, whereas the
active three-leg sector is evaluated at $\phi_b=0$.  Symmetry removes its
linear background sensitivity, but the quadratic term mixes higher-field
active operators and should be included in a future enlarged basis.

The present paper is intentionally restricted to the three-dimensional
critical basin of the passive conserved fixed point.  In two dimensions
gradient activity is canonically marginal, a prediction whose logarithmic
signature has recently been looked for in deterministic simulations of Active
Model B and B+~\cite{BhowmickMohanty2026}, and requires a genuinely
nonlinear functional RG treatment; we leave that distinct strong-coupling
problem for separate work.

For applications, assuming no additional contact anomalies,
Eq.~\eqref{eq:referenceImproved} gives
$y_{\Theta_1,\rm ref}\simeq-0.518149$ and
$y_{J,\rm ref}\simeq-0.554447$.
These conditional estimates use the equilibrium input
$\eta_{\rm ref}$ from Ref.~\cite{Kos2016}.  Their small separation makes
the passive Ising fixed point compatible with very long active transients,
with a crossover scale that depends sensitively on the initial
current-to-local amplitude ratio.
The block skewness, odd part of the block-density histogram, and normalized
fifth moment in Eqs.~\eqref{eq:blockSkewness}--\eqref{eq:fifthMoment} provide
direct tests in fixed-density MIPS simulations.  Even observables see these
fields only at second order and can remain dominated by ordinary equilibrium
corrections.

\begin{acknowledgments}
P.Z. acknowledges salary support from the National Science Centre, Poland,
under OPUS 21 grant No.~\mbox{UMO-2021/41/B/ST8/04474}.  This work lies
outside the scientific scope of that project.
\end{acknowledgments}

\section*{Data availability}
The numerical data related to this work are openly available in
comma-separated form in the RepOD repository~\cite{ZdybelData2026}.

\appendix

\section{Diagrammatic notation and Feynman rules}
\label{app:diagrammatics}

This appendix fixes the diagrammatic conventions used in the main text and
in the subsequent appendices.  The conventions distinguish amputated
external legs from internal propagators and distinguish the two chemical
active vertices from the non-gradient current vertex.

\subsection{External legs and internal propagators}

The diagrams represent amputated 1PI vertex functions.  A plain single
external line denotes a physical field $\phi$; an arrowed single external
line denotes a response field $\phit$.  The arrow on an external segment is
only a field-type marker: it is not a propagator and carries no factor
$G_R$.

Internal lines have a different meaning.  A directed single line denotes
the retarded response propagator,
\begin{equation}
 G_R(\omega,q)=\frac{1}{-\ii\omega+\Omega_k(q)},\qquad
 \Omega_k(q)=M_0q^2P_k(q),
 \label{eq:diagGR}
\end{equation}
while the advanced propagator is
$G_A(\omega,q)=G_R(-\omega,q)$.  A double line denotes the correlation
propagator
\begin{equation}
 C(\omega,q)=2D_k(\phi)q^2G_R(\omega,q)G_A(\omega,q).
 \label{eq:diagC}
\end{equation}
There is no $\phit\phit$ propagator.  Consequently, every double line is
internal and connects two physical-field legs.  No amputated external leg
is ever drawn as a double line.

The arrow of an internal response line points from the vertex supplying the
internal response leg to the vertex at which it terminates on a physical
leg.  With the It\^o convention, every closed directed cycle of internal
response lines vanishes.

\begin{center}
\begin{tikzpicture}[x=1.0cm,y=0.65cm,font=\scriptsize]
 \draw[diagline] (0,1.0)--(1.1,1.0);
 \node[anchor=west] at (1.25,1.0) {external $\phi$ leg};
 \draw[response] (0,0.35)--(1.1,0.35);
 \node[anchor=west] at (1.25,0.35) {external $\phit$ leg};
 \draw[response] (0,-0.35)--(1.1,-0.35);
 \node[anchor=west] at (1.25,-0.35) {$G_R$ (internal)};
 \draw[correlation] (0,-1.05)--(1.1,-1.05);
 \node[anchor=west] at (1.25,-1.05) {$C$ (internal only)};
 \draw[diagline] (0,-1.75)--(1.1,-1.75)
   node[pos=0.5,regulatorinsertion] {};
 \node[anchor=west] at (1.25,-1.75)
   {single-scale regulator insertion};
\end{tikzpicture}
\end{center}

\subsection{Vertex dictionary}

All momenta are incoming and the overall momentum-conserving delta function
is suppressed.  We use
$\int_{\omega,q}\equiv\int\dd\omega/(2\pi)\int\dd^dq/(2\pi)^d$.
The nonnegative integers $m,n$ count physical-field legs on the
corresponding vertex.  The labels V1--V6 below are entries of this
dictionary and are not equation numbers.

\paragraph*{V1: Passive potential vertex.}
Its symbol is \passiveicon. At a uniform background,
\begin{equation}
 \Gamma_{U,k}^{(1,m)}(Q;p_1,\ldots,p_m)
 =M_0U_k^{(m+1)}(\phi_b)Q^2.
 \label{eq:diagPassiveVertex}
\end{equation}

\paragraph*{V2: Conserved-noise/local-transport vertex.}
Its symbol is \noiseicon. With two response legs and $n$ physical legs,
\begin{equation}
 \Gamma_{D,k}^{(2,n)}(Q_1,Q_2;p_1,\ldots,p_n)
 =2D_k^{(n)}(\phi_b)(Q_1\cdot Q_2).
 \label{eq:diagNoiseVertex}
\end{equation}

\paragraph*{V3: \texorpdfstring{$\Lambda$}{Lambda} chemical active vertex.}
In V3--V5, $a,b$ denote incoming physical momenta and $Q+a+b=0$.
Its symbol is \lambdaicon, and its three-leg factor is
\begin{equation}
 \Gamma_{\Lambda,k}^{(1,2)}(Q;a,b)
 =-2M_0\Lambda_kQ^2(a\cdot b).
 \label{eq:diagLambdaVertex}
\end{equation}

\paragraph*{V4: \texorpdfstring{$\Xi$}{Xi} chemical active vertex.}
Its symbol is \xiicon, and its three-leg factor is
\begin{equation}
 \Gamma_{\Xi,k}^{(1,2)}(Q;a,b)
 =-M_0\Xi_kQ^2(a^2+b^2).
 \label{eq:diagXiVertex}
\end{equation}

\paragraph*{V5: Current-like active vertex.}
Its symbol is \currenticon, and its three-leg factor is
\begin{equation*}
 \Gamma_{J,k}^{(1,2)}(Q;a,b)
 =M_0\zeta_{J,k}\big[(Q\cdot a)b^2+(Q\cdot b)a^2\big].
\end{equation*}

\paragraph*{V6: Single-scale insertion.}
An open circle on an internal line denotes the action of
$\widetilde\partial_\ell$ on that line.  If no regulator-insertion circle is shown in a compact main-text graph,
the kernel $\dot{\mathcal F}$ includes the sum over all allowed placements
of the single-scale derivative on the internal propagators.

The black circle, red square, blue diamond, cyan triangle, and violet
hexagon identify the passive, local-noise, $\Lambda$, $\Xi$, and current-like
vertices, respectively.  In the generic $T1$ and $T2$ graphs the neutral
pentagon labeled $j$ stands for any of the three active vertices selected by
$V_j$.

\subsection{Loop construction and projection}

For a specified 1PI vertex, one first enumerates all one-loop topologies
compatible with its external field content.  Internal response and
correlation lines are assigned according to Eqs.~\eqref{eq:diagGR} and
\eqref{eq:diagC}, and every graph containing a closed directed response cycle
is discarded.  The single-scale derivative acts only on the regulator, with
all running couplings held fixed.  Only after the complete vertex flow has
been assembled is the appropriate projector applied.  In particular, the
projector onto $\Lambda,\Xi$, and $\zeta_J$ acts on the quartic
external-momentum polynomial of $\partial_\ell\Gamma_k^{(1,2)}$, not on
individual vertices before the loop integration.

\section{Passive flow and anomalous dimension}
\label{app:passiveflow}

\subsection{Why the uniform-background trace must not be used}

It is tempting to obtain the potential flow by evaluating
Eq.~\eqref{eq:Wetterich} at $\phit=0$ on a uniform background.  That route
fails, and it is worth recording why.  Every term of Eq.~\eqref{eq:trunc}
carries at least one response field, so $\Gamma_k[\phi,\phit=0]=0$
identically and
$\Gamma^{(2)}_{k,\phi\phi}\big|_{\phit=0}=0$.  The regulated inverse
propagator is therefore off-diagonal up to its $\phit\phit$ entry, and its
inverse has components
\begin{equation*}
 \mathcal G=
 \begin{pmatrix}
 C&G_R\\ G_A&0
 \end{pmatrix}.
\end{equation*}
Since $\partial_{\tilde{t}}\mathcal R_k$ of Eq.~\eqref{eq:regMatrix} is purely off-diagonal, the trace samples $G_R+G_A$ and not the correlation function.
If the It\^o time-shift factors are temporarily suppressed, the symmetric
frequency integral of the displayed rational propagators gives
\begin{equation*}
 \int\frac{\dd\omega}{2\pi}\big[G_R+G_A\big]
 =\int\frac{\dd\omega}{2\pi}\frac{2\Omega_k}{\omega^2+\Omega_k^2}=1
\end{equation*}
independently of $\Omega_k$.  This field-independent equal-time contact is
removed by the causal normalization; with the strict It\^o prescription,
closed response traces vanish.  Neither expression carries information
about $U_k$, consistently with $\Gamma_k[\phi,\phit=0]=0$.  The
potential flow must instead be extracted either from the term linear in
$\phit$, or from the equilibrium argument given next, which is shorter.

\subsection{Equilibrium derivation of the potential flow}

By Eqs.~\eqref{eq:regMatrix}--\eqref{eq:staticRegulator} the regulator
deforms the static Hamiltonian only, leaving the noise term untouched.  The
regulated process is therefore Model-B relaxation towards
$\mathcal H_k=\mathcal H+\Delta\mathcal H_k$ with unchanged noise, and its
stationary measure is $\propto\exp(-\mathcal H_k/\Theta)$.  The equal-time
sector of the dynamic theory coincides with the static theory regulated by
$R_k$, whose effective average action obeys the static Wetterich equation.
On a uniform background this reads
\begin{equation}
 \partial_{\tilde{t}}U_k(\phi)
 =
 \frac{\Theta}{2}\int_{\mathbf q}
 \frac{\partial_{\tilde{t}}R_k(q)}
 {U_k''(\phi)+K_kq^2+R_k(q)} .
 \label{eq:BUdim}
\end{equation}
The explicit factor $\Theta$ is the temperature in energy units and is
the same object that appears in the equal-time correlation function,
\begin{equation}
 \int\frac{\dd\omega}{2\pi}C(\omega,q)
 =
 2D_0q^2\frac1{2M_0q^2P_k}
 =\frac{\Theta_0}{P_k},
 \label{eq:Bcorr}
\end{equation}
which is why Eq.~\eqref{eq:potFlow} carries $\Theta_k(\phi)$ once the local
transport perturbation is switched on.  We set $\Theta_0=1$ by absorbing
the temperature into the static functional and the mobility,
$U\to U/T$, $K\to K/T$, $M_0\to M_0T$; this leaves the coefficient of
$\phit\partial_t\phi$ equal to unity, as required by the delta-functional
representation.

Introduce
\begin{equation*}
 q^2=k^2y,\qquad
 R_k=K_kk^2yr(y),\qquad
 U_k''=K_kk^2w.
\end{equation*}
The radial measure becomes
\begin{equation*}
 \int_{\mathbf q}
 =
 2v_d k^d
 \int_0^\infty\dd y\,y^{d/2-1}.
\end{equation*}
At fixed $q$,
\begin{align}
 \partial_{\tilde{t}}R_k
 &=
 \partial_{\tilde{t}}[K_kq^2r(y)]
 =
 -\eta K_kq^2r
 -2yK_kq^2r'(y)
 \nonumber\\
 &=
 -K_kq^2[\eta r+2yr'].
 \label{eq:dtRk}
\end{align}
Substitution into Eq.~\eqref{eq:BUdim} gives Eq.~\eqref{eq:l0} and
\begin{equation*}
 \partial_{\tilde{t}}U_k
 =
 2v_dk^d\,\Theta\, l_0^d(w;\eta).
\end{equation*}

Now differentiate $u=k^{-d}U_k$ at fixed $\rho$.  Because
\begin{equation*}
 \rho=\frac12 K_kk^{2-d}\phi^2,
\end{equation*}
one has, at fixed dimensional field,
\begin{equation*}
 \partial_{\tilde{t}}\rho=-(d-2+\eta)\rho.
\end{equation*}
The chain rule then gives
\begin{align*}
 \partial_{\tilde{t}}u\big|_\rho
 &=
 -du
 +(d-2+\eta)\rho u'
 +k^{-d}\partial_{\tilde{t}}U_k\big|_\phi ,
\end{align*}
which is Eq.~\eqref{eq:potFlow}.

\subsection{Anomalous-dimension projection}

For an $O(N)$ field with $N$ components, temporarily introduce a
longitudinal background and
project $K_k$ from the transverse two-point function.  The only
momentum-dependent one-loop contribution contains two vertices with one
longitudinal and two transverse legs, denoted $LTT$.  The
result may be written
\begin{equation*}
 \eta
 =
 \frac{16v_d}{d}\kappa\lambda_2^2m_{22}^d(w_\kappa,\eta),
 \qquad w_\kappa=2\kappa\lambda_2,
\end{equation*}
with the threshold function defined compactly by
\begin{equation}
 m_{22}^d(w,\eta)
 =-\frac12\int_0^\infty\dd y\,y^{d/2}\,
 \widetilde\partial_{\tilde{t}}
 \Big[\partial_yP_w^{-1}\,\partial_yP_0^{-1}\Big],
 \label{eq:m22compact}
\end{equation}
where
\begin{equation*}
 P_0=y(1+r),\qquad P_w=P_0+w,
\end{equation*}
and $\widetilde\partial_{\tilde{t}}$ acts only on the regulator, i.e., through
$\widetilde\partial_{\tilde{t}}(yr)=-y(\eta r+2yr')$ as implied by
Eq.~\eqref{eq:dtRk}.  Carrying out the differentiation gives the explicit
smooth-regulator form
\begin{align}
m_{22}^d(w,\eta)
={}&-
\int_0^\infty\dd y\,y^{d/2}
\frac{1+r+yr'}{P_w^2P_0^2}
\nonumber\\
&\times\Bigg[
y(\eta r+2yr')(1+r+yr')
\left(\frac1{P_w}+\frac1{P_0}\right)
\nonumber\\
&\hspace{1cm}
-\eta r-(\eta+4)yr'-2y^2r''
\Bigg].
\label{eq:m22explicit}
\end{align}
The one-component limit is taken after the projection.  This is an analytic
continuation, because no transverse mode exists at $N=1$; it is the standard
choice in this truncation.  Its contribution to the offset from the Ising reference anomalous
dimension $\eta_{\rm ref}$~\cite{Kos2016} has not been separated
from the other derivative-expansion errors; see Sec.~\ref{sec:d3}.

\section{Local transport flow and the operator identity}
\label{app:thetaflow}

The local noise action is
\begin{equation*}
 \Gamma_{D,k}
 =
 -\int\dd t\,\dd^dx\,D_k(\phi)(\grad\phit)^2,
\end{equation*}
so that
\begin{equation*}
 \Gamma_k^{(2,0)}(q,-q;\phi)
 =
 -2D_k(\phi)q^2+\Order(q^4).
\end{equation*}

\subsection{Soft-momentum counting of the deterministic vertices}

Before classifying the one-loop contributions we record the momentum
counting of the deterministic response vertices, because the current vertex
is the only nonobvious case.  By Eqs.~\eqref{eq:diagPassiveVertex},
\eqref{eq:diagLambdaVertex} and \eqref{eq:diagXiVertex} the potential and
the two chemical vertices carry an explicit factor $Q^2$.  The current
vertex is naively $\Order(Q)$, but momentum conservation inside the loop
removes the linear term.  With loop momentum $q$, set $p_1=q$ and
$p_2=-q-Q$.  The coupling-stripped tensor of Eq.~\eqref{eq:activeVertices} is
\begin{align}
V_J(Q;q,-q-Q)
={}&(Q\cdot q)(q^2+2Q\cdot q+Q^2)
\nonumber\\
&\quad+(-Q\cdot q-Q^2)q^2
\nonumber\\
={}&2(Q\cdot q)^2-Q^2q^2
\nonumber\\
&\quad+Q^2(Q\cdot q)=\Order(Q^2).
\label{eq:VJsoft}
\end{align}
Every deterministic response vertex in the truncation is therefore at least
$\Order(Q^2)$ in the soft external response momentum.

\subsection{One-loop classification}

Write $\mathcal P=\Gamma_k^{(2)}+\mathcal R_k$ and
$\mathcal G=\mathcal P^{-1}$, and let
$\mathcal P_a=\delta_a\mathcal P$ and
$\mathcal P_{ab}=\delta_b\delta_a\mathcal P$ denote field derivatives.
The second derivative of the flow with respect to the response field gives a
tadpole and a bubble,
\begin{equation*}
 \partial_{\tilde{t}}\Gamma_k^{(2,0)}
 =\frac12\widetilde\partial_{\tilde{t}}
 \Big[\Tr\big(\mathcal G\,\mathcal P_{ab}\big)
 -\Tr\big(\mathcal G\mathcal P_a\mathcal G\mathcal P_b\big)\Big],
\end{equation*}
with $a=\phit(q)$, $b=\phit(-q)$.  In the tadpole the four-leg vertex
$\mathcal P_{ab}$ must carry two response legs; by Eq.~\eqref{eq:trunc} the
noise vertex is the only such vertex, so $\mathcal P_{ab}$ is nonzero only
in the $(\phi\phi)$ block and the trace closes on the correlation
propagator.  Diagrammatically,
\begin{center}
\begin{tikzpicture}[x=1.0cm,y=0.72cm,font=\scriptsize]
  \node[noisevertex] (d) at (0,0) {};
  \draw[response] (-1.0,0.42)--(d) node[pos=0,left] {$q_1$};
  \draw[response] (-1.0,-0.42)--(d) node[pos=0,left] {$q_2$};
  \draw[correlation] (d)
    .. controls +(1.35,0.95) and +(1.35,-0.95) ..
    node[pos=0.5,regulatorinsertion] {} (d);
  \node at (1.75,0) {$D_k''$};
\end{tikzpicture}
\end{center}
The two external response legs are arrowed single lines, while the closed
double line is the internal physical-field correlation.  The open circle on
that loop marks the single-scale regulator derivative.  Analytically,
using Eq.~\eqref{eq:diagNoiseVertex} with $Q_1=Q$ and $Q_2=-Q$,
\begin{align*}
 \partial_\ell\Gamma^{(2,0)}
 &=
 \frac12\widetilde\partial_\ell
 \int_{\omega,\mathbf q}
 2D_k''(\phi)(Q\cdot(-Q))\,C_k(\omega,q)
 \nonumber\\
 &=
 D_k''Q^2\Theta_k
 \int_{\mathbf q}
 \frac{\partial_\ell R_k}{P_k^2},
\end{align*}
where we used Eq.~\eqref{eq:Bcorr} and
$\widetilde\partial_\ell P_k^{-1}=-\partial_\ell R_k/P_k^2$.  Applying
Eq.~\eqref{eq:DprojMain} gives Eq.~\eqref{eq:Dflow}.

The remaining one-loop classes are
deterministic$\times$deterministic, deterministic$\times D'$, and
$D'\times D'$.  The first starts at $q^4$ by the counting of
Eq.~\eqref{eq:VJsoft}.  The second is $\Order(q^2)\times\Order(q)$, whose
leading part is odd in the loop momentum and vanishes upon angular
integration; the first surviving term is $\Order(q^4)$.  In the last class
each vertex supplies one internal response leg and one internal physical
leg, and since there is no $\phit\phit$ propagator the only admissible
pairing consists of two response lines forming a closed directed cycle,
which vanishes by It\^o causality.  Thus Eq.~\eqref{eq:Dflow} is the
complete local $\Order(q^2)$ one-loop projection, and the
$(\Lambda,\Xi,\zeta_J)$ sector contributes only at $\Order(q^4)$.

\subsection{Linearization}

Linearize with
$D_k=D_0[1+\vartheta(\vphi)]$ and
$\Theta_k=\Theta_0[1+\vartheta(\vphi)]$.  Since $D_k''$ is already first
order in $\vartheta$,
\begin{equation*}
 D_k''\Theta_k
 =
 D_0\Theta_0\partial_\phi^2\vartheta+\Order(\vartheta^2).
\end{equation*}
Using
\begin{align*}
 \partial_\phi^2
 &=K_kk^{2-d}\partial_\vphi^2,
 \nonumber\\
 K_kk^{2-d}\frac12
 \int_{\mathbf q}
 \frac{\partial_{\tilde{t}}R_k}{P_k^2}
 &=-2v_dl_0^{d\,\prime}(w;\eta),
\end{align*}
together with $\partial_\ell\vphi=+\Dphi\vphi$ at fixed dimensional field,
the chain rule gives Eq.~\eqref{eq:thetaFlow}.  Linearizing the passive
potential flow at fixed $\eta$ gives Eq.~\eqref{eq:LU0}, and subtracting the
two proves Eq.~\eqref{eq:operatorIdentity}.

\subsection{Exact zero mode of the coupled system}

The triangular structure of Eq.~\eqref{eq:triangularTheta} admits a check
that is independent of the regulator and of the threshold function.  Because
the physics depends only on $\mathcal H/\Theta$, the rescaling
$(U,K,\Theta)\mapsto\lambda(U,K,\Theta)$ is a pure change of energy units.
Since $\rho\propto K_k$, in dimensionless variables it acts as
\begin{equation*}
 \delta\bar u=u_*-\rho u_*',\qquad
 \delta\vartheta=1,
\end{equation*}
and must therefore be a redundant zero mode.  For the following operator
check we hold $\eta$ fixed and use $\mathcal{L}_U=\mathcal{L}_U^{(0)}$.
Writing $F$ for the right-hand
side of the fixed-point equation and
$\mathcal B_{U\Theta}\vartheta=-2v_d\vartheta\,l_0^d(w_*;\eta_*)$ as in
Eq.~\eqref{eq:triangularTheta}, a direct computation gives the identity
\begin{equation}
 \mathcal{L}_U\big[u-\rho u'\big]+\mathcal B_{U\Theta}[1]
 =\rho F'(\rho)-F(\rho),
 \label{eq:zeroModeIdentity}
\end{equation}
valid for arbitrary $u(\rho)$ and arbitrary threshold function.  At a fixed
point $F\equiv0$ for all $\rho$, hence also $F'\equiv0$, and the right-hand
side vanishes.  Equation~\eqref{eq:zeroModeIdentity} simultaneously
verifies the presence of the factor $\Theta_k$ in Eq.~\eqref{eq:potFlow},
the form of the coupling block $\mathcal B_{U\Theta}$, and the fact that $\mathcal{L}_\Theta$
annihilates constants.  The corresponding eigenvalue zero is the redundant
energy-unit direction and is not a physical marginal operator.

\section{Gradient-active vertex flow}
\label{app:activeflow}

\subsection{Third derivative of the Wetterich equation}

Let
\begin{equation*}
 \mathcal P=\Gamma_k^{(2)}+\mathcal R_k,
 \qquad
 \mathcal G=\mathcal P^{-1}.
\end{equation*}
The elementary identity
\begin{equation*}
 \delta\mathcal G=-\mathcal G(\delta\mathcal P)\mathcal G
\end{equation*}
gives the identity below~\cite{Kopietz2010}.  Here
$a=\phit(Q)$, $b=\phi(p_1)$, $c=\phi(p_2)$ label external fields,
$\delta_a$ denotes functional differentiation with respect to that field,
and subscripts on $\mathcal P$ denote successive derivatives:
\begin{align}
\delta_c\delta_b\delta_a\frac12\Tr\ln\mathcal P
={}&\frac12\Tr(\mathcal G\mathcal P_{abc})
\nonumber\\
&-\frac12\Tr(\mathcal G\mathcal P_c\mathcal G\mathcal P_{ab})
\nonumber\\
&-\frac12\Tr(\mathcal G\mathcal P_b\mathcal G\mathcal P_{ac})
\nonumber\\
&-\frac12\Tr(\mathcal G\mathcal P_{bc}\mathcal G\mathcal P_a)
\nonumber\\
&+\frac12\Tr(\mathcal G\mathcal P_c\mathcal G\mathcal P_b\mathcal G\mathcal P_a)
\nonumber\\
&+\frac12\Tr(\mathcal G\mathcal P_b\mathcal G\mathcal P_c\mathcal G\mathcal P_a).
\label{eq:Dthird}
\end{align}
Applying $\widetilde\partial_\ell$ to this identity gives
$\partial_\ell\Gamma_k^{(1,2)}$.
The topology is fixed by first listing the vertices that survive at the
symmetric background.  Among three-point vertices, the passive one is
proportional to $U'''(0)$ and the noise one to $D'(0)$, both of which vanish,
so the only surviving three-point vertex in the gradient-sector insertion
problem is the active one.  Here $D$ is held at its constant passive value;
the separate local-transport insertion was treated in Appendix~\ref{app:thetaflow}.  Among four-point
vertices, the noise vertex is proportional to $D''(0)=0$ and no four-leg
active vertex exists because $\Lambda_k$, $\Xi_k$ and $\zeta_{J,k}$ are field
independent, so the only surviving four-point vertex is the passive quartic
one.  Consequently the tadpole term of Eq.~\eqref{eq:Dthird} would require an
active five-leg vertex, which is absent, and the triangle terms would require
two surviving three-point vertices, which at linear order in activity would
have to be passive and therefore vanish.  Only the bubble terms remain, and
they are forced to consist of one active three-point vertex and one passive
quartic vertex.  Their three assignments,
$-\tfrac12\Tr(\mathcal G\mathcal P_c\mathcal G\mathcal P_{ab})$ and
$-\tfrac12\Tr(\mathcal G\mathcal P_b\mathcal G\mathcal P_{ac})$ for the two
terms of the sum over $i$ in Eq.~\eqref{eq:T1}, and
$-\tfrac12\Tr(\mathcal G\mathcal P_{bc}\mathcal G\mathcal P_a)$ for
Eq.~\eqref{eq:T2}, each carry a factor $\tfrac12$ that is cancelled by the two
admissible propagator assignments $(G_R,C)$ and $(C,G_A)$, which are equal
after the change of loop variable $q\to-(q+p_i)$.  This reproduces
Eqs.~\eqref{eq:T1} and \eqref{eq:T2} with unit coefficient.

The momentum routing also fixes the response-line directions.  Taking $q$
on the correlation line, in $T1$ the active vertex carries external $p_i$
and internal response momentum $-q-p_i$; its response line points to the
passive vertex, which carries external $Q$ and $p_{3-i}$.
In $T2$ the active vertex carries external $Q$ and the passive vertex
carries both physical legs.  The internal response line then points from
the passive to the active vertex and has shifted momentum $q+Q$.

The compact drawings in the main text suppress the placement of the
single-scale derivative.  In the Wetterich trace each topology contributes
twice: the open regulator-insertion circle lies either on the correlation
line or on the response line.  For the $i=1$ representative the four terms
used in the active-sector projection are
\begin{center}
\begin{tikzpicture}[x=0.72cm,y=0.66cm,font=\tiny]
  \node at (0.68,2.08) {$T1(C)$};
  \node[activevertex] (t1ca) at (0,1.35) {$j$};
  \node[passivevertex] (t1cb) at (1.35,1.35) {};
  \draw[diagline] (-0.68,1.35)--(t1ca) node[midway,above] {$p_1$};
  \draw[response] (t1cb)--(2.03,1.72) node[pos=1,right] {$Q$};
  \draw[diagline] (t1cb)--(2.03,0.98) node[pos=1,right] {$p_2$};
  \draw[response] (t1ca) to[out=35,in=145] (t1cb);
  \draw[correlation] (t1ca) to[out=-35,in=-145]
    node[pos=0.5,regulatorinsertion] {} (t1cb);

  \node at (4.78,2.08) {$T1(R)$};
  \node[activevertex] (t1ra) at (4.10,1.35) {$j$};
  \node[passivevertex] (t1rb) at (5.45,1.35) {};
  \draw[diagline] (3.42,1.35)--(t1ra) node[midway,above] {$p_1$};
  \draw[response] (t1rb)--(6.13,1.72) node[pos=1,right] {$Q$};
  \draw[diagline] (t1rb)--(6.13,0.98) node[pos=1,right] {$p_2$};
  \draw[response] (t1ra) to[out=35,in=145]
    node[pos=0.5,regulatorinsertion] {} (t1rb);
  \draw[correlation] (t1ra) to[out=-35,in=-145] (t1rb);

  \node at (0.68,0.28) {$T2(C)$};
  \node[activevertex] (t2ca) at (0,-0.45) {$j$};
  \node[passivevertex] (t2cb) at (1.35,-0.45) {};
  \draw[response] (-0.68,-0.45)--(t2ca) node[midway,above] {$Q$};
  \draw[diagline] (t2cb)--(2.03,-0.08) node[pos=1,right] {$p_1$};
  \draw[diagline] (t2cb)--(2.03,-0.82) node[pos=1,right] {$p_2$};
  \draw[response] (t2cb) to[out=145,in=35] (t2ca);
  \draw[correlation] (t2ca) to[out=-35,in=-145]
    node[pos=0.5,regulatorinsertion] {} (t2cb);

  \node at (4.78,0.28) {$T2(R)$};
  \node[activevertex] (t2ra) at (4.10,-0.45) {$j$};
  \node[passivevertex] (t2rb) at (5.45,-0.45) {};
  \draw[response] (3.42,-0.45)--(t2ra) node[midway,above] {$Q$};
  \draw[diagline] (t2rb)--(6.13,-0.08) node[pos=1,right] {$p_1$};
  \draw[diagline] (t2rb)--(6.13,-0.82) node[pos=1,right] {$p_2$};
  \draw[response] (t2rb) to[out=145,in=35]
    node[pos=0.5,regulatorinsertion] {} (t2ra);
  \draw[correlation] (t2ra) to[out=-35,in=-145] (t2rb);
\end{tikzpicture}
\end{center}
Here $C$ and $R$ specify whether
$\widetilde\partial_\ell$ acts on the correlation or response propagator.
Their sum is exactly the derivative $\dot{\mathcal F}(q,q')$ used in
Eqs.~\eqref{eq:T1} and \eqref{eq:T2}; the four drawings introduce no extra
topology or approximation.  The $i=2$ pair follows by
$p_1\leftrightarrow p_2$.

\subsection{Frequency integral}

For a correlation line of momentum $q$ and response line of momentum $q'$,
\begin{equation*}
 \mathcal F(q,q')
 =
 2q^2\int\frac{\dd\omega}{2\pi}
 \frac{1}{(\omega^2+E_q^2)(E_{q'}-\ii\omega)}.
\end{equation*}
Closing the contour in the upper half plane, only $\omega=\ii E_q$
contributes.  Its residue is
\begin{align*}
 \mathcal F(q,q')
 &=
 2q^2\ii
 \frac{1}{(2\ii E_q)(E_{q'}+E_q)}
 \nonumber\\
 &=
 \frac{q^2}{E_q(E_q+E_{q'})}.
\end{align*}
Differentiating with respect to the regulated energies,
\begin{align*}
 \frac{\partial\mathcal F}{\partial E_q}
 &=
 -q^2\frac{2E_q+E_{q'}}{E_q^2(E_q+E_{q'})^2},\\
 \frac{\partial\mathcal F}{\partial E_{q'}}
 &=
 -q^2\frac{1}{E_q(E_q+E_{q'})^2},
\end{align*}
which yields Eq.~\eqref{eq:Fdot}.  For
\begin{equation*}
 R_k=K_kq^2r(y),
\end{equation*}
the infrared-time single-scale derivative is
\begin{equation*}
 \widetilde\partial_\ell R_k
 =
 K_kq^2[\eta r+2yr'].
\end{equation*}
Both the unshifted and shifted internal lines must be differentiated.

\section{Translation Ward identity and the all-orders current quotient}
\label{app:wardproof}

Introduce a nondynamical metric $g_{ij}(x)$, with determinant
$g=\det(g_{ij})$.  Indices are raised and lowered with this metric and
$\nabla_i$ denotes its covariant derivative.  Define
\begin{equation*}
 T^{ij}
 =
 \frac{2}{\sqrt g}
 \frac{\delta\Gamma_{\rm stat}}{\delta g_{ij}}.
\end{equation*}
Under an infinitesimal local translation with displacement $\alpha^i(x)$,
\begin{equation*}
 \delta\phi=-\alpha^k\nabla_k\phi,\qquad
 \delta g_{ij}=\nabla_i\alpha_j+\nabla_j\alpha_i,
\end{equation*}
translation-preserving renormalization gives $\delta\Gamma_{\rm stat}=0$:
\begin{align*}
0
={}&
\int\dd^dx\sqrt g
\left[
-\EOM\alpha^k\nabla_k\phi
+T^{ij}\nabla_i\alpha_j
\right]
\\
={}&
-\int\dd^dx\sqrt g\,\alpha_j
\left[
\nabla_iT^{ij}+\EOM\nabla^j\phi
\right].
\end{align*}
Since $\alpha_j(x)$ is arbitrary,
\begin{equation*}
 \nabla_iT^{ij}=-\EOM\nabla^j\phi.
\end{equation*}
This gives Eqs.~\eqref{eq:WardMain}--\eqref{eq:stressRep}.

The stress tensor is defined only up to improvement
terms~\cite{CallanColemanJackiw1970}
$\Delta T_{ij}=(\partial_i\partial_j-\delta_{ij}\lap)f_{\rm imp}(\phi)$,
where $f_{\rm imp}$ is an arbitrary local scalar.  These terms have
vanishing divergence and therefore do not affect Eq.~\eqref{eq:WardMain}.
They also drop out of the doubly contracted insertion used in the two-loop
test of Sec.~\ref{sec:ward}: in momentum space the improvement contributes
$-Q_iQ_j+\delta_{ij}Q^2$, and
\begin{equation}
 Q_iQ_j\big[-Q_iQ_j+\delta_{ij}Q^2\big]=-Q^4+Q^4=0 .
 \label{eq:improvementDrops}
\end{equation}
The cancellation established below is consequently free of any
improvement-scheme ambiguity.

For the momentum quotient, use
\begin{equation*}
 Q^2=\kinS+2\kinT.
\end{equation*}
The chemical subspace at fourth order is generated by
\begin{equation*}
 Q^2\kinS=\kinS^2+2\kinS\kinT,\qquad
 Q^2\kinT=\kinS\kinT+2\kinT^2.
\end{equation*}
The current tensor is
\begin{align*}
V_J
&=(Q\cdot p_1)p_2^2+(Q\cdot p_2)p_1^2\\
&=-(p_1^2+\kinT)p_2^2-(p_2^2+\kinT)p_1^2\\
&=-\kinS\kinT-2\kinV.
\end{align*}
Therefore $\Pi_J[V_J]=1$ while $\Pi_J$ annihilates the chemical subspace.
Combining this with the counting below Eq.~\eqref{eq:closure} makes the
quotient basis independent: the conserving subspace is three dimensional, the
chemical subspace $\Span\{Q^2\kinS,Q^2\kinT\}$ is two dimensional and contained in
it, and $V_J$ is conserving but not chemical.  The current quotient $\mathcal V_{\rm grad}/\mathcal V_{\rm chem}$
at order $p^4$ is therefore one dimensional, represented by $V_J$.
The nonequilibrium quotient $\mathcal V_{\rm grad}/\mathcal V_{\rm eq}$
used for the spectrum remains two dimensional.

Now consider an arbitrary 1PI graph with one external response leg.
If that leg is attached to a deterministic chemical vertex, the vertex
contains an explicit $Q^2$.  Local UV counterterms are Taylor polynomials
in external momenta and cannot remove that factor.  If instead the external
response leg is attached to the additive-noise vertex, the second response
leg is internal.  Following retarded propagators through the finite graph
must eventually close a directed response cycle, which vanishes by It\^o
causality.  These are the only two possibilities in the purely chemical
theory, proving Eq.~\eqref{eq:triangularExact}.

At linear order in $\zeta_J$ there is a third possibility: the external
response field is attached directly to the current insertion.  Hence every
graph surviving $\Pi_J$ is rooted at the current insertion.  This fact is
used in the two-loop reduction below.

\section{Explicit two-loop calculation of the current counterterm}
\label{app:twoloop}

This appendix works entirely with regulator-free, renormalized 1PI vertices
at $k=0$.  The infrared cutoff $R_k$ and the Wetterich flow do not enter.
Instead, $\mu_R$ denotes the ultraviolet renormalization scale used in
dimensional regularization and minimal subtraction.  The calculation is an
independent perturbative test of the current-operator statement in
Sec.~\ref{sec:ward}, not a perturbative evaluation of the Wetterich equation.
To avoid confusing the RG scale $k$ with integration variables, loop momenta
in this appendix are denoted by $q$ and $r$.

We give the calculation in sufficient detail to make the cancellation
reproducible.

\subsection{Definition of the counterterm}

We follow the standard conventions of renormalized perturbation theory and of
composite-operator renormalization~\cite{Collins1984}.
Throughout this appendix renormalized quantities are written as
renormalization constants times bare ones, $\phi=Z_\phi^{1/2}\phi_0$ and
$\phit=Z_{\phit}^{1/2}\phit_0$, the two-point vertex is defined with
$\Gamma^{(2)}=p^2+\Sigma$, so that $\Gamma_R^{(2)}=Z_\phi^{-1}\Gamma_0^{(2)}$,
and the calculation is performed in the massless theory with minimal
subtraction at generic external momenta.  This convention is the inverse of
the more common one and fixes the signs of the exponents below.
Let $\mu_R$ be the dimensional-regularization scale and write the bare
static quartic coupling as
\begin{equation*}
 \lambda_{{\rm MS},0}=8\pi^2\mu_R^\eps Z_g\,\widetilde g_4.
\end{equation*}
Here $Z_g$ is the quartic-coupling renormalization factor in minimal subtraction.
The ultraviolet beta function
$\widetilde\beta_4\equiv\mu_R\dd \widetilde g_4/\dd\mu_R$ satisfies
\begin{align*}
 \widetilde\beta_4&=-\eps \widetilde g_4+\frac32\widetilde g_4^2+\Order(\widetilde g_4^3),\\
 \widetilde g_{4,*}&=\frac23\eps+\Order(\eps^2).
\end{align*}
Let $Z_\phi,Z_{\phit}$, and $Z_M$ be the physical-field, response-field,
and mobility renormalization factors.  Because the coefficient of
$\phit\partial_t\phi$ is not renormalized and equilibrium Model B satisfies
the fluctuation--dissipation relation~\cite{Tauber2014},
\begin{equation*}
 Z_{\phit}Z_\phi=1,\qquad Z_M=Z_\phi.
\end{equation*}
Writing
\begin{equation*}
 \zeta_{J,0}
 =
 \mu_R^{-1+\eps/2}
 Z_\phi^{3/2}Z_{J,\rm ct}\zeta_J,
\end{equation*}
we define
\begin{equation*}
 Z_{J,\rm ct}
 =
 1+\frac{b_J\widetilde g_4^2}{\eps}+\Order(\widetilde g_4^3).
\end{equation*}
The ultraviolet anomalous dimension associated with this additional factor
is
\begin{align*}
 \gamma_{J,\mu}
 &\equiv\mu_R\frac{\dd}{\dd\mu_R}\ln Z_{J,\rm ct}
 \nonumber\\
 &=\widetilde\beta_4\partial_{\widetilde g_4}\left(\frac{b_J\widetilde g_4^2}{\eps}\right)
 +\Order(\widetilde g_4^3)
 \nonumber\\
 &=-2b_J\widetilde g_4^2+\Order(\widetilde g_4^3).
\end{align*}
At $\widetilde g_{4,*}=2\eps/3+\Order(\eps^2)$,
\begin{equation*}
 \gamma_{J,\mu,*}=-\frac89b_J\eps^2
 +\Order(\eps^3).
\end{equation*}
Our infrared-RG convention is
$\gamma_{J,\rm ct}=-\gamma_{J,\mu,*}$; therefore
\begin{equation*}
 c_{J,\rm ct}=\frac89b_J.
\end{equation*}
The desired two-loop test is the determination of $b_J$; in particular,
$b_J=0$ implies $c_{J,\rm ct}=0$ independently of the sign convention.

\subsection{Two-loop topologies and dynamic-to-static reduction}
\label{sec:causalReduction}

After the current quotient the external response leg is rooted at the
current insertion.  With two passive quartic vertices, the connected 1PI
graphs contain two response lines and two correlation lines.  The classes are
fixed by counting.  Let $n_{XY}$ be the number of internal lines joining
vertices $X$ and $Y$.  Let $e_X$ be the number of external physical legs on
vertex $X$, with $e_J+e_A+e_B=2$; the current insertion has $2-e_J$ internal
physical legs and each passive vertex has $3-e_X$ physical plus one response
leg, so the number of internal line ends is $10-2=8$ and there are four
internal lines, hence two loops.  Solving
$n_{JA}+n_{JB}=2-e_J$, $n_{JA}+n_{AB}=4-e_A$ and $n_{JB}+n_{AB}=4-e_B$ gives
six cases: $e_J=2$ disconnects the current insertion, $e_J=1$ leaves a bridge
and is one-particle reducible, and $e_J=0$ leaves
$(n_{JA},n_{JB},n_{AB})=(2,0,2)$, $(1,1,2)$ and $(0,2,2)$.  There are
therefore two inequivalent multigraph classes: a serial graph, which occurs
twice as a mirror pair, and a triangle graph.

\begin{center}
\begin{tikzpicture}[x=0.92cm,y=0.72cm,font=\scriptsize]
  \node[currentvertex] (js) at (0,0.8) {};
  \node[passivevertex] (as) at (1.25,0.8) {};
  \node[passivevertex] (bs) at (2.5,0.8) {};
  \node[anchor=south] at (0,1.18) {$J$};
  \node[anchor=south] at (1.25,1.33) {$A$};
  \node[anchor=south] at (2.5,1.33) {$B$};
  \draw[response] (js)--(-0.75,0.8);
  \draw[response] (as) to[out=135,in=45] (js);
  \draw[correlation] (js) to[out=-35,in=-145] (as);
  \draw[response] (bs) to[out=135,in=45] (as);
  \draw[correlation] (as) to[out=-35,in=-145] (bs);
  \draw[diagline] (bs)--(3.25,1.2);
  \draw[diagline] (bs)--(3.25,0.4);
  \node at (1.25,-0.05) {serial};

  \node[currentvertex] (jt) at (4.55,0.8) {};
  \node[passivevertex] (at) at (5.9,1.35) {};
  \node[passivevertex] (bt) at (5.9,0.25) {};
  \node[anchor=south] at (4.55,1.18) {$J$};
  \node[anchor=south west] at (5.98,1.35) {$A$};
  \node[anchor=north west] at (5.98,0.25) {$B$};
  \draw[response] (jt)--(3.8,0.8);
  \draw[response] (at)--(jt);
  \draw[correlation] (jt)--(bt);
  \draw[response] (bt) to[bend right=60] (at);
  \draw[correlation] (at)--(bt);
  \draw[diagline] (at)--(6.10,2.05);
  \draw[diagline] (bt)--(6.10,-0.45);
  \node at (5.55,-0.55) {triangle};
\end{tikzpicture}
\end{center}
The arrowed single line attached to the violet current insertion is the
amputated external response leg.  The two external physical legs are plain
single lines.  Each topology contains exactly two directed internal response
propagators and two internal double-line correlations; neither topology
contains a directed response cycle.

Use $K=M_0=\Theta_0=1$ and write the renormalized static interaction
as $\mu_R^\eps\widetilde{\lambda}\phi^4/4!$.
The frequency-domain lines are
$G_R(\omega,q)=(-\ii\omega+E_q)^{-1}$ and
$C(\omega,q)=2q^2/(\omega^2+E_q^2)$.
Their Fourier transforms are
\begin{align*}
 G_R(t,q)&=\theta(t)e^{-E_qt},\\
 C(t,q)&=\frac{1}{P(q)}e^{-E_q|t|},
 \qquad E_q=q^2P(q).
\end{align*}
Here $J$ is the current insertion and $A,B$ are the two passive vertices,
as labeled in the diagrams.  For an internal line $r$ joining vertices $X$
and $Y$, we write
$E_{XY,r}\equiv E_{q_{XY,r}}=q_{XY,r}^2P(q_{XY,r})>0$; the line label $r$
is omitted when only one line joins the pair.  These quantities are positive
relaxation rates of propagator lines, not running couplings.
A passive response vertex contributes $q^2=E_q/P(q)$ on the response edge.
Thus each internal edge carries the same static factor $1/P(q)$; response
edges carry additional relaxation rates and causal step functions.  The
current insertion contributes no such factor, because in both topologies its
response leg is external and supplies no internal relaxation rate.  In the stress representation
the external contraction is $Q_iQ_j$; the full momentum numerator of the
stress/current insertion must still be retained in the spatial integral.

In the serial graph the $J$--$A$ and $A$--$B$ pairs each carry two parallel
internal lines.  In this topology we define explicitly
\begin{equation}
 \Omega_{JA}\equiv E_{JA,1}+E_{JA,2},\qquad
 \Omega_{AB}\equiv E_{AB,1}+E_{AB,2}.
 \label{eq:serialRates}
\end{equation}
After summing the allowed response/correlation assignment within each bundle,
the corresponding rate numerator is $\Omega_{JA}\Omega_{AB}$.  Define positive backward time differences $x=t_J-t_A$ and
$y=t_J-t_B$, where $t_X$ is the time assigned to vertex $X$.
For the serial orientation $B\to A\to J$, causality
requires $t_B<t_A<t_J$, hence $0<x<y$.  Therefore
\begin{align*}
W_{\rm serial}
&=
\Omega_{JA}\Omega_{AB}
\int_0^\infty\dd x
\int_x^\infty\dd y\,
e^{-\Omega_{JA} x-\Omega_{AB}(y-x)}
\nonumber\\
&=
\Omega_{JA}\Omega_{AB}
\int_0^\infty\dd x\,e^{-\Omega_{JA} x}
\frac1{\Omega_{AB}}
=1.
\end{align*}

In the triangle graph there is one line on each of the $J$--$A$ and
$J$--$B$ edges and two parallel lines on the $A$--$B$ edge.  Here the
topology-specific definitions are
\begin{align}
 \Omega_{JA}&\equiv E_{JA},\qquad
 \Omega_{JB}\equiv E_{JB},\nonumber\\
 \Omega_{AB}&\equiv E_{AB,1}+E_{AB,2}.
 \label{eq:triangleRates}
\end{align}
Consequently the response-rate numerators are $\Omega_{JA}\Omega_{JB}$ for the star
orientation $A\to J\leftarrow B$, $\Omega_{JB}\Omega_{AB}$ for $A\to B\to J$, and $\Omega_{JA}\Omega_{AB}$ for
$B\to A\to J$.  For the star orientation, $x,y>0$ are independent, and the common
exponential is
$\exp[-\Omega_{JA}x-\Omega_{JB}y-\Omega_{AB}|x-y|]$.
Splitting the integration quadrant into $x>y$ and $y>x$ gives
\begin{align*}
 &\int_0^\infty\dd x\int_0^x\dd y\,
 e^{-\Omega_{JA}x-\Omega_{JB}y-\Omega_{AB}(x-y)}\\
 &\qquad=\frac{1}{(\Omega_{JA}+\Omega_{AB})(\Omega_{JA}+\Omega_{JB})},
\end{align*}
and its $JA\leftrightarrow JB$ partner.  The first domain is also the
causal domain of $A\to B\to J$, whereas the second belongs to
$B\to A\to J$.  Multiplying by the respective response-rate numerators
therefore gives
\begin{align*}
W_\star
&=
\frac{\Omega_{JA}\Omega_{JB}}{\Omega_{JA}+\Omega_{JB}}
\left[
\frac1{\Omega_{JA}+\Omega_{AB}}+\frac1{\Omega_{JB}+\Omega_{AB}}
\right],\\
W_{A\to B\to J}
&=
\frac{\Omega_{JB}\Omega_{AB}}{(\Omega_{JA}+\Omega_{AB})(\Omega_{JA}+\Omega_{JB})},\\
W_{B\to A\to J}
&=
\frac{\Omega_{JA}\Omega_{AB}}{(\Omega_{JB}+\Omega_{AB})(\Omega_{JA}+\Omega_{JB})}.
\end{align*}
Summing,
\begin{align*}
W_{\rm triangle}
={}&
\frac{\Omega_{JB}(\Omega_{JA}+\Omega_{AB})}
{(\Omega_{JA}+\Omega_{JB})(\Omega_{JA}+\Omega_{AB})}
\nonumber\\
&+\frac{\Omega_{JA}(\Omega_{JB}+\Omega_{AB})}
{(\Omega_{JA}+\Omega_{JB})(\Omega_{JB}+\Omega_{AB})}
\nonumber\\
={}&
\frac{\Omega_{JB}+\Omega_{JA}}{\Omega_{JA}+\Omega_{JB}}
=1.
\end{align*}
Thus the sum over all noncyclic dynamical orientations equals the
corresponding static graph topology by topology.
The identities $W_{\rm serial}=W_{\rm triangle}=1$ are the genuinely
dynamic content of the test: they require the full causal orientation sum
and are not consequences of the static translation Ward identity.  Once
they hold, the fully contracted stress insertion below is fixed by
$\Gamma^{(2)}$, so its cancellation against ordinary field
renormalization is kinematically enforced.  The only remaining independent
question is whether a local quotient contact pole survives.

\subsection{Explicit reduction to the sunset integral}
\label{sec:sunsetReduction}

The causal sum leaves four static propagators.
The reduction to three propagators uses the current-vertex numerator.
We show it before evaluating the scalar integral.
In the triangle choose momenta $q$ on $JA$, $q+Q$ on $JB$, and
$r$, $q-p_1-r$ on the two $AB$ lines, with physical external legs
$p_1$ at $A$ and $p_2$ at $B$ and $Q+p_1+p_2=0$.
Let $\mathcal V_{J,\triangle}^{(2)}$ denote the amputated two-loop
vertex with its overall current coupling removed.  In the massless
theory the result of the causal sum is
\begin{align}
 \mathcal V_{J,\triangle}^{(2)}
 &=\frac{\widetilde{\lambda}^2}{2}\mu_R^{2\eps}
 \int_q\int_r
 \frac{V_J(Q;q,-Q-q)}
 {q^2(q+Q)^2r^2(q-p_1-r)^2}.
 \label{eq:triangleFourPropagators}
\end{align}
The coefficient $1/2$ includes the external-leg assignments and the
symmetry under exchange of the two parallel $AB$ lines.  Causal weights
are already summed and must not be included again.
Using the same vertex convention as Eq.~\eqref{eq:activeVertices},
\begin{align}
 V_J(Q;q,-Q-q)
 &=(Q\cdot q)(q+Q)^2\nonumber\\
 &\quad-[Q\cdot(q+Q)]q^2,\nonumber\\
 \frac{V_J(Q;q,-Q-q)}{q^2(q+Q)^2}
 &=\frac{Q\cdot q}{q^2}
 -\frac{Q\cdot(q+Q)}{(q+Q)^2}.
 \label{eq:currentNumeratorCancellation}
\end{align}
Each term now has three propagators.  Write $I(p)$ for the scalar
sunset integral in Eq.~\eqref{eq:sunsetIntegralG} below and define
\begin{equation*}
 F_i(p)=\mu_R^{2\eps}\int_q\int_r
 \frac{(q)_i}{q^2r^2(p-q-r)^2}.
\end{equation*}
The three internal momenta $q,r,p-q-r$ are interchangeable in this
integral.  Adding the corresponding vector components gives $(p)_i$ and hence
$3F_i(p)=(p)_iI(p)$.  In the first term of
Eq.~\eqref{eq:currentNumeratorCancellation}, $r\mapsto-r$ makes the
external sunset momentum $p_1$.  In the second, set $q'=q+Q$ and
$r\mapsto-r$; the external sunset momentum is then $-p_2$.
Since $I(-p)=I(p)$, the result is
\begin{equation}
 \mathcal V_{J,\triangle}^{(2)}
 =\frac{\widetilde{\lambda}^2}{6}
 \big[(Q\cdot p_1)I(p_1)+(Q\cdot p_2)I(p_2)\big].
 \label{eq:triangleSunsetBridge}
\end{equation}
This identity explicitly connects the dynamic three-point calculation
to two static sunset integrals.  Momentum conservation also gives
\begin{align*}
 &(Q\cdot p_1)I(p_1)+(Q\cdot p_2)I(p_2)\\
 &\quad=-(Q\cdot p_1)I(p_2)-(Q\cdot p_2)I(p_1)\\
 &\qquad\quad-Q^2[I(p_1)+I(p_2)].
\end{align*}
Together with $\Sigma_{\rm sun}^{(2)}=-\widetilde{\lambda}^2 I/6$, this
has precisely the Ward form plus a $Q^2$-proportional term.
The latter is discarded only after extracting its local ultraviolet
pole: $\Pi_J$ is defined on local quartic polynomials, not arbitrary
nonlocal momentum functions.  The serial graphs depend on external
momenta only through $Q$, so their local quartic poles are proportional
to $Q^4$ and vanish under the same projection.

At criticality, in the massless theory and with the convention
\begin{equation}
 \Gamma_{\rm stat}^{(2)}(p)=p^2+\Sigma(p),
 \label{eq:staticTwoPointG}
\end{equation}
one-loop tadpoles and two-loop double-tadpole graphs are momentum independent;
in dimensional regularization they are scaleless and vanish.  The first graph
that can generate a $p^2/\eps$ pole is therefore the two-loop sunset built
from two conventional $\widetilde{\lambda}\phi^4/4!$ vertices:
\begin{center}
\begin{tikzpicture}[x=1.0cm,y=0.72cm,font=\scriptsize]
  \node[passivevertex] (a) at (0,0) {};
  \node[passivevertex] (b) at (2.7,0) {};
  \draw[diagline] (-1.0,0)--(a)
    node[pos=0.36,above] {$p$};
  \draw[diagline] (b)--(3.7,0);
  \draw[diagline] (a) to[bend left=42]
    node[midway,above] {$q$} (b);
  \draw[diagline] (a)--(b)
    node[midway,above] {$p-q-r$};
  \draw[diagline] (a) to[bend right=42]
    node[midway,below] {$r$} (b);
\end{tikzpicture}
\end{center}
Assigning independent loop momenta $q$ and $r$ to two internal lines fixes
the third momentum to $p-q-r$.  Each critical static propagator contributes
the inverse square of the momentum carried by that line.  Removing the two
vertex factors and the graph symmetry factor leaves the scalar integral
\begin{align}
 I(p)
 &=
 \mu_R^{2\eps}\int_q\int_r
 \frac1{q^2r^2(p-q-r)^2},
 \label{eq:sunsetIntegralG}
 \\
 \int_q
 &\equiv
 \int\frac{\dd^dq}{(2\pi)^d},
 \qquad d=4-\eps.
 \nonumber
\end{align}
The factor $\mu_R^{2\eps}$ accompanies the two quartic vertices because
$\lambda_{{\rm MS},0}\propto\mu_R^\eps\widetilde{\lambda}$.  The external momentum must be kept
nonzero while extracting the ultraviolet pole: setting $p=0$ prematurely
would turn Eq.~\eqref{eq:sunsetIntegralG} into a scaleless integral and erase
the UV information together with the infrared contribution.

For completeness, the integral can be evaluated by applying twice the
standard massless convolution
\begin{align*}
&\int\frac{\dd^dr}{(2\pi)^d}
\frac1{(r^2)^a[(\mathcal Q-r)^2]^b}
\nonumber\\
&\quad=
\frac{(\mathcal Q^2)^{d/2-a-b}}{(4\pi)^{d/2}}
\frac{\Gamma(a+b-d/2)}{\Gamma(a)\Gamma(b)}
\frac{\Gamma(d/2-a)\Gamma(d/2-b)}
{\Gamma(d-a-b)}.
\end{align*}
For fixed $q$, the $r$ integration corresponds to $a=b=1$ and
$\mathcal Q=p-q$.  It gives the one-loop bubble
\begin{align}
 B(p-q)
 &\equiv
 \mu_R^\eps\int_r\frac1{r^2(p-q-r)^2}
\nonumber\\
 &=
 \mu_R^\eps
 \frac{[(p-q)^2]^{-\eps/2}}
 {(4\pi)^{2-\eps/2}}
 \frac{\Gamma(\eps/2)\Gamma(1-\eps/2)^2}
 {\Gamma(2-\eps)}.
 \label{eq:bubbleIntegralG}
\end{align}
Equation~\eqref{eq:sunsetIntegralG} is then
$I(p)=\mu_R^\eps\int_q B(p-q)/q^2$.  The remaining convolution has
$a=1$ and $b=\eps/2$, yielding
\begin{equation*}
 I(p)
 =
 \mu_R^{2\eps}(p^2)^{1-\eps}
 (4\pi)^{-4+\eps}
 \frac{\Gamma(1-\eps/2)^3\Gamma(\eps-1)}
 {\Gamma(3-3\eps/2)}.
\end{equation*}
Since
$\Gamma(\eps-1)=-1/\eps+\gamma_E-1+\Order(\eps)$, where $\gamma_E$
is the Euler--Mascheroni constant, its pole part is
\begin{equation}
 I(p)\big|_{1/\eps}
 =
 -\frac{p^2}{2(16\pi^2)^2}\frac1\eps.
 \label{eq:sunsetPoleG}
\end{equation}
The bubble in Eq.~\eqref{eq:bubbleIntegralG} contains a logarithmic
four-point subdivergence.  Its minimal-subtraction counterterm is independent
of the external momentum of that subgraph; inserted into the remaining
$q$ integral, it is proportional to
$\eps^{-1}\int_q1/q^2$, which is a scaleless tadpole and vanishes in
dimensional regularization.  Therefore the $p^2/\eps$ residue in
Eq.~\eqref{eq:sunsetPoleG} is unchanged by the incomplete subtraction
operation $\bar R_{\rm BPHZ}$ used in Sec.~\ref{sec:wardCancellation}.

For two identical $\widetilde{\lambda}\phi^4/4!$ vertices, the sunset symmetry factor is
$1/6$.  The normalization used here, the resulting $Z_\phi$ and the value
$\eta=\eps^2/54$ agree with the standard tabulations of the $\eps$
expansion~\cite{KleinertSchulteFrohlinde2001}.  The order-$\widetilde{\lambda}^2$ connected propagator correction and the
corresponding term in the inverse propagator have opposite signs; with the
definition in Eq.~\eqref{eq:staticTwoPointG},
\begin{equation*}
 \Sigma^{(2)}_{\rm sun}(p)
 =
 -\frac{\widetilde{\lambda}^2}{6}I(p).
\end{equation*}
Using $\widetilde{\lambda}=8\pi^2\widetilde g_4$ therefore gives
\begin{align}
 \Sigma^{(2)}_{\rm sun}(p)\big|_{1/\eps}
 &=
 \frac{\widetilde g_4^2}{48\eps}p^2,\nonumber\\
 Z_\phi&=1+\frac{\widetilde g_4^2}{48\eps}+\Order(\widetilde g_4^3).
 \label{eq:staticPoleInputG}
\end{align}
In our convention $\phi=Z_\phi^{1/2}\phi_0$, the coefficient of the bare
kinetic term expressed through the renormalized field is $Z_\phi^{-1}$.
Consequently $Z_\phi^{-1}-1=-\widetilde g_4^2/(48\eps)+\Order(\widetilde g_4^3)$ cancels the
two-point pole in Eq.~\eqref{eq:staticPoleInputG}.  The residue
$\widetilde g_4^2/(48\eps)$ is the two-point input needed below.  Direct substitution into
Eq.~\eqref{eq:triangleSunsetBridge} gives the independent local check
\begin{equation}
 \mathcal V_{J,\triangle}^{(2)}\big|_{1/\eps}
 =\frac{\widetilde g_4^2}{48\eps}(V_J+Q^2\kinS).
 \label{eq:triangleLocalPole}
\end{equation}
Thus its current projection is $\widetilde g_4^2/(48\eps)$, while the
chemical polynomial $Q^2\kinS$ is annihilated by $\Pi_J$.

\subsection{Ward conversion and cancellation}
\label{sec:wardCancellation}

The explicit numerator reduction in Sec.~\ref{sec:sunsetReduction}
can now be compared with the translation Ward identity.  The contracted
stress insertion differs from the current representative by chemical
terms, which do not affect its local current projection.
Let $\Gamma_{T_{ij}\phi\phi}$ denote the 1PI vertex with one stress-tensor
insertion and two physical-field legs.  Differentiating the Ward identity
twice with respect to $\phi$ and contracting with $Q_i$ gives
\begin{equation*}
 Q_iQ_j\Gamma_{T_{ij}\phi\phi}
 =
 (Q\cdot p_1)\Gamma^{(2)}(p_2)
 +(Q\cdot p_2)\Gamma^{(2)}(p_1).
\end{equation*}
The self-energy pole therefore fixes, rather than merely approximates, the
fully contracted stress insertion:
\begin{align*}
 Q_iQ_j\Gamma_{T_{ij}\phi\phi}\big|_{\rm pole}
 &=
 \frac{\widetilde g_4^2}{48\eps}
 \left[
 (Q\cdot p_1)p_2^2+(Q\cdot p_2)p_1^2
 \right]\\
 &=
 \frac{\widetilde g_4^2}{48\eps}V_J.
\end{align*}
Meanwhile
$Z_\phi^{-1}=1-\widetilde g_4^2/(48\eps)+\Order(\widetilde g_4^3)$
supplies the counterterm
$-\widetilde g_4^2V_J/(48\eps)$.  Let $\bar R_{\rm BPHZ}$ denote the incomplete Bogoliubov--Parasiuk--Hepp--Zimmermann (BPHZ) operation
that subtracts all proper ultraviolet subdivergences but not the final
overall pole.  Writing $\Gamma_J^{(1,2)}$ for the complete current-insertion
vertex at the indicated order, the projected overall pole is therefore zero:
\begin{equation*}
 \Pi_J\bar R_{\rm BPHZ}\,\Gamma_J^{(1,2)}
 \big|_{\widetilde g_4^2\zeta_J,1/\eps}
 =0.
\end{equation*}
The zero in this equation is consequently not presented as an independent
cancellation miracle.  It confirms that, after the nontrivial
dynamic-to-static reduction, no additional local quotient counterterm is
left.  There is no one-loop $Z_{J,\rm ct}$ counterterm, because the one-loop current
quotient vanishes for the same reason as in the functional calculation: the
external response leg sits either on the quartic vertex, which supplies an
explicit $Q^2$ and hence a chemical structure, or on the current insertion, in
which case the loop closes on the quartic vertex and the integral depends on
the external momenta only through $Q$, so that its quartic part is
proportional to $Q^4$.  In both cases $c_{\kinV}=0$.  Consequently the
quartic-coupling subdivergence multiplies a one-loop current graph whose
quotient already vanishes; mass subdivergences are lower derivative or $Q^2$-divisible; and
$Z_\phi-1$ begins only at $\widetilde g_4^2$.  These observations exhaust the forest
formula~\cite{Collins1984} at the
order-$\widetilde g_4^2\zeta_J$ forests.  Hence
\begin{equation*}
 b_J=0,\qquad c_{J,\rm ct}=0.
\end{equation*}

\section{Numerical extraction and convergence checks}
\label{app:numerics}

\begin{figure*}[t]
\centering
\includegraphics[width=0.98\textwidth]{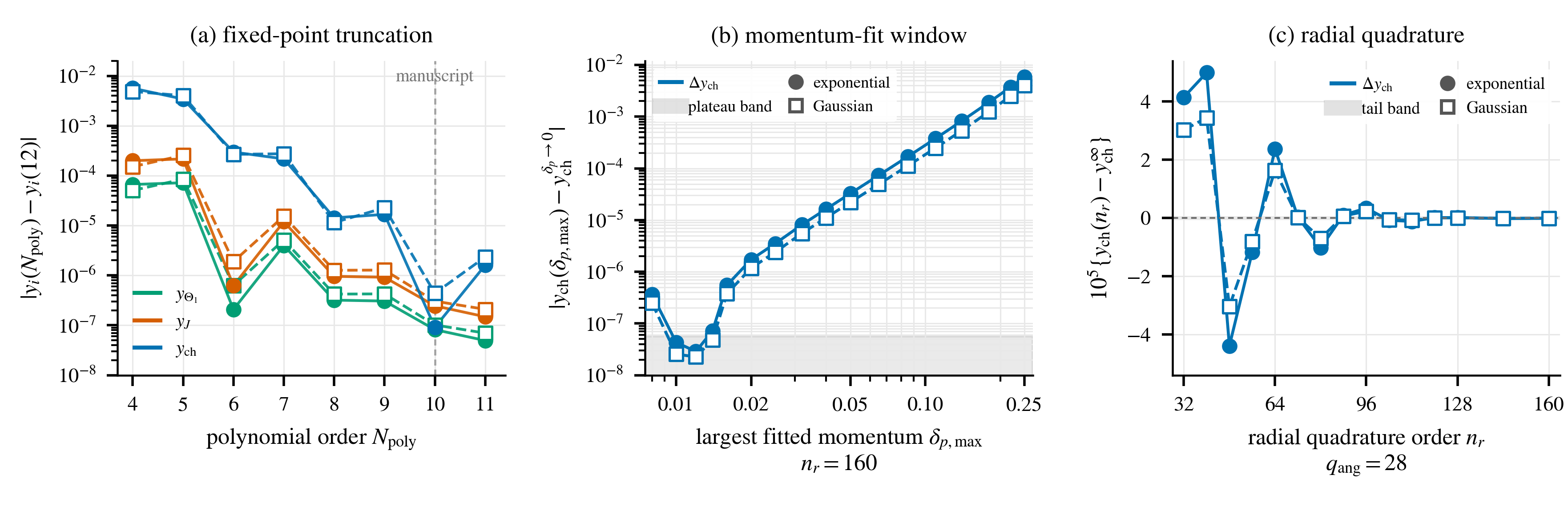}
\caption{Numerical convergence of the three-dimensional FRG projection for
the exponential (filled circles) and Gaussian (open squares) regulators.
(a) Deviation of the three active eigenvalues from their value at
$N_{\rm poly}=12$, the highest order reached, as the polynomial order is
increased from $N_{\rm poly}=4$; the vertical line marks $N_{\rm poly}=10$,
the order used in the main tables.  Panel (a) holds the momentum
window at $\delta_{p,\max}=0.04$ and the quadrature orders at
$n_r=72$, $q_{\rm ang}=36$; the plotted differences isolate polynomial
convergence at those settings.  The eigenvalues themselves are spread over
$0.4$ while the truncation effect ranges from $5\times10^{-8}$ to
$6\times10^{-3}$, so the deviation rather than the eigenvalue is plotted, on
the same logarithmic presentation as panel (b).  The sequence is not monotone
below $N_{\rm poly}=8$.  (b) Small-momentum convergence at $n_r=160$ and
$q_{\rm ang}=28$.  The reference is the mean over
$\delta_{p,\max}=0.010,0.012,0.014$, and the shaded band is half their range.
Extending the window to $\delta_{p,\max}=0.25$ displays contamination by higher
momentum powers, while the upturn below $\delta_{p,\max}\simeq0.01$ marks numerical
loss of significance.  (c) Signed radial-quadrature residual at fixed
$q_{\rm ang}=28$ and $\delta_{p,\max}=0.04$.  Here $y_{\rm ch}^{\infty}$ is the
mean over $n_r=120,128,144,160$, and the shaded band is half the range of
those four values.}
\label{fig:numericalConvergence}
\end{figure*}

For each fixed external momentum shape $v_1,v_2$ and active insertion,
we set
\begin{equation*}
 p_1=\delta_p v_1,\qquad p_2=\delta_p v_2,\qquad Q=-\delta_p(v_1+v_2)
\end{equation*}
and fit the integrated loop result at several small $\delta_p$ to
\begin{equation*}
 X(\delta_p)=A_2\delta_p^2+A_4\delta_p^4+A_6\delta_p^6+A_8\delta_p^8.
\end{equation*}
Here $A_{2n}$ is the fitted coefficient of $\delta_p^{2n}$ for
that momentum configuration and active insertion. We use the following
six pairs of dimensionless vectors:
\[
\begin{array}{c|c|c}
 & v_1 & v_2 \\ \hline
1 & (1,0,0)          & (0,0,1) \\
2 & (1,0,0)          & (0.6,0,0.8) \\
3 & (1,0,0.2)        & (-0.4,0.3,0.9) \\
4 & (0.7,-0.2,0.4)   & (0.2,0.8,-0.3) \\
5 & (1.1,0.3,-0.2)   & (-0.5,0.4,0.7) \\
6 & (0.5,-0.6,0.9)   & (0.8,0.1,0.3)
\end{array}
\]
The vectors need not be normalized or mutually orthogonal.
The requirement for a unique reconstruction is that these
configurations distinguish the four quartic invariants in
Eq.~\eqref{eq:p4fit}. To make this explicit, define
\[
 \tilde S=|v_1|^2+|v_2|^2,\qquad
 \tilde T=v_1\cdot v_2,\qquad
 \tilde V=|v_1|^2|v_2|^2.
\]
Then $\kinS=\delta_p^2 \tilde S$, $\kinT=\delta_p^2 \tilde T$, and
$\kinV=\delta_p^4 \tilde V$. Each configuration therefore supplies
one row $(\tilde S^2,\tilde S \tilde T,\tilde T^2,\tilde V)$ of the reconstruction matrix.
For the six pairs listed above, this $6\times4$ matrix has
rank four. For each active insertion, we use the six fitted
$A_4$ values to determine
$(c_{\kinS^2},c_{\kinS\kinT},c_{\kinT^2},c_{\kinV})$
by least squares, with the normalization of
Eq.~\eqref{eq:p4fit}.
Equation~\eqref{eq:activeProj} then gives the three beta functions.
The basis change in Eq.~\eqref{eq:Tmatrix} yields
$\mathsf M_{\rm act}'=\mathsf S_{\rm eq}^{-1}\mathsf M_{\rm act}\mathsf S_{\rm eq}$;
the prime denotes transformed coordinates, whose lower $2\times2$ block
is the quotient matrix.

Figure~\ref{fig:numericalConvergence} separately tests polynomial order,
external-momentum window, and quadrature; its caption specifies the fixed
settings and reference values for each panel.  The plotted data are
deposited in Ref.~\cite{ZdybelData2026}.

The polynomial sequence in panel (a) is nonmonotone, notably at
$N_{\rm poly}=7$, but stabilizes over $N_{\rm poly}=10$--12.
The full ranges of the chemical eigenvalue over these three orders are
$1.7\times10^{-6}$ and $2.4\times10^{-6}$ for the exponential and Gaussian
regulators, respectively, about three orders of magnitude below their
$3.1\times10^{-3}$ spread.  From order 10 to 12 the two slow exponents
change by about $10^{-7}$.  This direct comparison tests the derivatives
at $\rho=0$ needed for the gradient projection, evaluated from the
potential expanded about $\kappa\simeq0.041$.

For panel (b), the mean of the three plateau points
$\delta_{p,\max}=0.010,0.012,0.014$ gives
\begin{align}
 y_{\rm ch}^{\delta_p\to0,E}&=-0.89601881, \nonumber
 \\
 y_{\rm ch}^{\delta_p\to0,G}&=-0.89296341,
 \label{eq:ychHlimit}
\end{align}
where $E$ and $G$ denote the exponential and Gaussian regulators.
The matrices and eigenvalues in the main text use this plateau average.
For $\delta_{p,\max}\gtrsim0.02$, terms beyond $A_8\delta_p^8$ contaminate
the fit; below $\delta_{p,\max}\simeq0.01$, cancellation in extracting
the small quartic coefficient causes loss of significance.

The radial-quadrature half-ranges in panel (c) are
$1.7\times10^{-7}$ and $1.1\times10^{-7}$ for the exponential and Gaussian
regulators, respectively.  Increasing the angular order beyond 28 at
fixed radial order changes the eigenvalues by less than these bands,
identifying the visible oscillation as a radial-quadrature effect.

Two further diagnostics are independent of these convergence bands.
At the momentum plateau the closure residual of Eq.~\eqref{eq:closure}
is below $2.1\times10^{-7}$; it measures violation of conservation, not
completeness of the operator basis.  The norm of the raw loop contribution
to the third row is below $5.8\times10^{-6}$ and decreases with
$\delta_{p,\max}$, consistently approaching the analytic one-loop result
$\Delta M_{33}=0$.  In contrast,
$(\mathsf M_{\rm act})_{31}=(\mathsf M_{\rm act})_{32}=0$ follows exactly
from Sec.~\ref{sec:ward}, not from numerical smallness.

Finally, before setting $\phi_b=0$, we test linear background sensitivity
using the odd finite-difference part
\begin{equation*}
 \delta_{\rm odd}\mathsf M(\delta_b)=
 \frac{\mathsf M_{\rm act}(\delta_b)-
       \mathsf M_{\rm act}(-\delta_b)}{2\delta_b},
\end{equation*}
where $\delta_b$ is a small background-field displacement.
It vanishes algebraically in the retained vertex basis before quadrature,
verifying Eq.~\eqref{eq:backgroundSensitivity}.  The generally nonzero
symmetric curvature generates higher-field active vertices and is not
a closed-truncation shift of the three retained eigenvalues.


\end{document}